\documentclass[usenatbib]{mn2e}
\usepackage{graphicx}
\title{The GEMS project: X-ray analysis and statistical properties of the group
       sample}

\author[J. P. F. Osmond \& T. J. Ponman]
  {John P. F. Osmond\thanks{E-mail: jpfo@star.sr.bham.ac.uk} and Trevor J.
   Ponman\\ School of Physics and Astronomy, The University of Birmingham,
   Edgbaston, Birmingham B15 2TT, UK\\}

\date{Accepted 2004 ??. Received 2004 ??; in original form 2004 ??}

\pubyear{2004}

\pagerange{\pageref{firstpage}--\pageref{lastpage}}

\begin{document}

%%%%%%%%%%%%%%%%%%%%%%%%%%%%%%%%%%%%%%%%%%%%%%%%%%%%%%%%%%%%%%%%%%%%%%%%%%%%%%%%%
%%%%%%%%%%%%%%%%%%%%%%%%%%%%%%%%%%%%%%%%%%%%%%%%%%%%%%%%%%%%%%%%%%%%%%%%%%%%%%%%%

% Definitions:

\newcommand{\betafit}{\ensuremath{\beta_{\mathrm{fit}}}}
\newcommand{\betaspec}{\ensuremath{\beta_{\mathrm{spec}}}}
\newcommand{\D}{\ensuremath{D}}
\newcommand{\Dec}{\ensuremath{Dec}}
\newcommand{\dengal}{\ensuremath{\bar{\rho}_{\mathrm{gal}}}}
\newcommand{\dom}{\ensuremath{L_{\mathrm{12}}}}
\newcommand{\e}{\ensuremath{e}}
\newcommand{\fgas}{\ensuremath{f_{\mathrm{gas}}}}
\newcommand{\fsp}{\ensuremath{f_{\mathrm{sp}}}}
\newcommand{\hfifty}{\ensuremath{h_{50}}}
\newcommand{\Hone}{\ensuremath{HI}}
\newcommand{\Hzero}{\ensuremath{H_{\mathrm{0}}}}
\newcommand{\K}{\ensuremath{K}}
\newcommand{\LB}{\ensuremath{L_{\mathrm{B}}}}
\newcommand{\LBGG}{\ensuremath{L_{\mathrm{BGG}}}}
\newcommand{\Lcut}{\ensuremath{L_{\mathrm{cut}}}}
\newcommand{\Lstar}{\ensuremath{L_{\star}}}
\newcommand{\LX}{\ensuremath{L_{\mathrm{X}}}}
\newcommand{\MB}{\ensuremath{M_{\mathrm{B}}}} 
\newcommand{\Ngal}{\ensuremath{N_{\mathrm{gal}}}}
\newcommand{\NH}{\ensuremath{N_{\mathrm{H}}}}
\newcommand{\omegam}{\ensuremath{\Omega_{\mathrm{m}}}}
\newcommand{\phistar}{\ensuremath{\phi_{\star}}}
\newcommand{\rcore}{\ensuremath{r_{\mathrm{core}}}}
\newcommand{\rcut}{\ensuremath{r_{\mathrm{cut}}}}  
\newcommand{\rext}{\ensuremath{r_{\mathrm{ext}}}}  
\newcommand{\rfh}{\ensuremath{r_{\mathrm{500}}}}
\newcommand{\rth}{\ensuremath{r_{\mathrm{200}}}}
\newcommand{\RA}{\ensuremath{RA}}
\newcommand{\sigmav}{\ensuremath{\sigma_{\mathrm{v}}}}
\newcommand{\Szero}{\ensuremath{S_{\mathrm{0}}}}
\newcommand{\TBGG}{\ensuremath{T_{\mathrm{BGG}}}}
\newcommand{\TX}{\ensuremath{T_{\mathrm{X}}}}
\newcommand{\vel}{\ensuremath{v}}
\newcommand{\Vfh}{\ensuremath{V_{\mathrm{500}}}}
\newcommand{\Z}{\ensuremath{Z}}

\newcommand{\LXrfh}{\ensuremath{L_{\mathrm{X}}(r_{\mathrm{500}})}}
\newcommand{\LXMDMB}{\ensuremath{L_{\mathrm{X}}\mathrm{(MDMB)}}}
\newcommand{\LXpLB}{\ensuremath{L_{\mathrm{X}}/L_{\mathrm{B}}}}
\newcommand{\LXpLBGG}{\ensuremath{L_{\mathrm{X}}/L_{\mathrm{BGG}}}}
\newcommand{\MpL}{\ensuremath{M/L}}
\newcommand{\rfhsigma}{\ensuremath{\rfh(\sigmav)}}
\newcommand{\rfhLB}{\ensuremath{r_{\mathrm{500}}(L_{\mathrm{B}})}}
\newcommand{\rfhTX}{\ensuremath{r_{\mathrm{500}}(T_{\mathrm{X}})}}

\newcommand{\ltsim}{\,\rlap{\raise 0.5ex\hbox{$<$}}{\lower 1.0ex\hbox{$\sim$}}\,}
\newcommand{\gtsim}{\,\rlap{\raise 0.5ex\hbox{$>$}}{\lower 1.0ex\hbox{$\sim$}}\,}

\newcommand{\arcm}{\ensuremath{^\prime}}
\newcommand{\degree}{\ensuremath{^\circ}}
\newcommand{\pcmsq}{\ensuremath{\mbox{cm}^{-2}}}
\newcommand{\pMpc}{\ensuremath{\mbox{Mpc}^{-1}}}
\newcommand{\pMpccu}{\ensuremath{\mbox{Mpc}^{-3}}}
\newcommand{\kev}{\ensuremath{\mbox{keV}}}
\newcommand{\erg}{\ensuremath{\mbox{erg}}}
\newcommand{\cm}{\ensuremath{\mbox{cm}}}
\newcommand{\km}{\ensuremath{\mbox{km}}}
\newcommand{\kpc}{\ensuremath{\mbox{kpc}}}
\newcommand{\m}{\ensuremath{\mbox{m}}}
\newcommand{\Mpc}{\ensuremath{\mbox{Mpc}}}
\newcommand{\Mpccu}{\ensuremath{\mbox{Mpc}^{3}}}
\newcommand{\kmps}{\ensuremath{\mbox{km~s}^{-1}}}
\newcommand{\ps}{\ensuremath{\s^{-1}}}
\newcommand{\s}{\ensuremath{\mathrm{s}}}
\newcommand{\ergps}{\ensuremath{\mbox{erg~s}^{-1}}}
\newcommand{\ergpspLsol}{\ensuremath{\mbox{erg~s}^{-1}~\mathrm{L_{\odot}^{-1}}}}
\newcommand{\kmpspMpc}{\ensuremath{\mbox{km~s}^{-1}~\mbox{Mpc}^{-1}}}
\newcommand{\Lsol}{\ensuremath{\mathrm{L_{\odot}}}}
\newcommand{\Msol}{\ensuremath{\mathrm{M_{\odot}}}}
\newcommand{\Zsol}{\ensuremath{\mathrm{Z_{\odot}}}}

\newcommand{\ASCA}{\emph{ASCA}}
\newcommand{\Chandra}{\emph{Chandra}}
\newcommand{\GEMS}{\emph{GEMS}}
\newcommand{\IPAC}{\emph{IPAC}}
\newcommand{\LEDA}{\emph{LEDA}}
\newcommand{\LEDAS}{\emph{LEDAS}}
\newcommand{\MEKAL}{\textsc{MEKAL}}
\newcommand{\NASA}{\emph{NASA}}
\newcommand{\NED}{\emph{NED}}
\newcommand{\PSPC}{\emph{PSPC}}
\newcommand{\ROSAT}{\emph{ROSAT}}
\newcommand{\XMM}{\emph{XMM-Newton}}

%%%%%%%%%%%%%%%%%%%%%%%%%%%%%%%%%%%%%%%%%%%%%%%%%%%%%%%%%%%%%%%%%%%%%%%%%%%%%%%%%
%%%%%%%%%%%%%%%%%%%%%%%%%%%%%%%%%%%%%%%%%%%%%%%%%%%%%%%%%%%%%%%%%%%%%%%%%%%%%%%%%

\maketitle

\label{firstpage}

%%%%%%%%%%%%%%%%%%%%%%%%%%%%%%%%%%%%%%%%%%%%%%%%%%%%%%%%%%%%%%%%%%%%%%%%%%%%%%%%%
%%%%%%%%%%%%%%%%%%%%%%%%%%%%%%%%%%%%%%%%%%%%%%%%%%%%%%%%%%%%%%%%%%%%%%%%%%%%%%%%%

\begin{abstract}

The \GEMS\ project involves a multi-wavelength study of a sample of 60
galaxy groups, chosen to span a wide range of group properties. Substantial
\ROSAT\ \PSPC\ observations, available for all of these groups, are used to
characterise the state of the intergalactic medium in each.  We present the
results of a uniform analysis of these \ROSAT\ data, and a statistical
investigation of the relationship between X-ray and optical properties across the
sample.  Our analysis improves in several respects on previous work: (a) we
distinguish between systems in which the hot gas is a group-scale medium, and
those in which it appears to be just a hot halo associated  a central galaxy,
(b) we extrapolate X-ray luminosities to a fixed overdensity radius (\rfh) using
fitted surface brightness models, in order to avoid biases arising from the fact
that cooler systems are detectable to smaller radii, and (c) optical properties
have been rederived in a uniform manner from the \NASA\ Extragalactic Database,
rather than relying on the data in the disparate collection of group catalogues
from which our systems are drawn.

The steepening of the \LX-\TX\ relation in the group regime reported previously
is not seen in our sample, which fits well onto the cluster trend, albeit with
large non-statistical scatter.  A number of biases affect the fitting of
regression lines under these circumstances, and until the impact of these has
been thoroughly investigated it seems best to regard the slope of the group
\LX-\TX\ relation as being poorly determined. A significant problem in comparing
the properties of groups and clusters is the derivation of system radii, to allow
different systems to be compared within regions having the same overdensity. We
find evidence that group velocity dispersion (\sigmav) provides a very unreliable
measure of system mass (and hence radius), with a number of groups having
remarkably low values of \sigmav, given that they appear from their X-ray
properties to be collapsed systems. We confirm that the surface brightness
profiles of groups are significantly flatter than those of clusters -- the {\it
maximum} value of the \betafit\ parameter for our sample is 0.58, lower than the
typical value of 0.67 seen in clusters -- however, we find no significant
tendency within our sample for cooler groups to show flatter profiles. This
result is inconsistent with simple universal preheating models. The morphology of
the galaxies in the \GEMS\ groups is correlated to their X-ray properties in a
number of ways: we confirm the very strong relationship between X-ray emission
and a dominant early-type central galaxy which has been noted since the early
X-ray studies of groups, and also find that spiral fraction is correlated with
the temperature of the hot gas, and hence the depth of the gravitational
potential. A class of spiral-rich groups with little or no X-ray emission,
probably corresponds to groups which have not yet fully collapsed.
\end{abstract}

%%%%%%%%%%%%%%%%%%%%%%%%%%%%%%%%%%%%%%%%%%%%%%%%%%%%%%%%%%%%%%%%%%%%%%%%%%%%%%%%%
%%%%%%%%%%%%%%%%%%%%%%%%%%%%%%%%%%%%%%%%%%%%%%%%%%%%%%%%%%%%%%%%%%%%%%%%%%%%%%%%%

\begin{keywords}
X-rays: galaxies: clusters -
galaxies: clusters: general -
galaxies: general -
galaxies: intergalactic medium -
galaxies: formation -
galaxies: evolution
\end{keywords}

\clearpage

%%%%%%%%%%%%%%%%%%%%%%%%%%%%%%%%%%%%%%%%%%%%%%%%%%%%%%%%%%%%%%%%%%%%%%%%%%%%%%%%%
%%%%%%%%%%%%%%%%%%%%%%%%%%%%%%%%%%%%%%%%%%%%%%%%%%%%%%%%%%%%%%%%%%%%%%%%%%%%%%%%%

\section{Introduction}
\label{sec_intro}

The principle that {\it we are nowhere special}, which is fundamental to
cosmology, also applies to galaxies.  The majority of galaxies are, like our own,
located within bound systems, mostly containing just a handful of bright galaxies
\citep{tully87}. These are characterised as {\it galaxy groups}, which are
distinguished rather arbitrarily from richer and rarer {\it galaxy clusters}.
These systems are evolving, as they turn round from the Hubble expansion,
virialise, and grow through mergers and accretion.  This dynamical evolution
modifies the environment of their constituent galaxies, and can in turn have
profound effects on the evolution of the galaxies themselves. On the other hand,
energetic galaxy winds can have a substantial impact on the surrounding
intergalactic medium (IGM) within groups and clusters \citep*[e.g][]{ponman99}, so
that there is a two-way interaction between the structure of galaxies and galaxy
systems.  The picture which emerges is that galaxies and the systems in which
most of them are located {\it co-evolve}, and a full understanding of the
evolution of either galaxies or galaxy clusters must take into account the
two-way interactions which couple the development of both.

Galaxy groups have received far less attention from astronomers than either
galaxies or galaxy clusters, and their properties are clearly very diverse, in
terms of structure, dynamics and the types of galaxies they contain
\citep{hickson97,zabludoff98,mulchaey00}. Any meaningful study of the
relationship between groups and galaxies needs to acknowledge this fact. We have
therefore commenced a study of the properties of a substantial sample of 60
galaxy groups, and the galaxies they contain, with a view to clarifying the
different stages of group evolution, and the ways in which this is related to
galaxy properties. This \GEMS\ (Group Evolution Multi-wavelength Study) project
involves optical photometry and spectroscopy to study the galaxies, radio
observations to explore the HI content of galaxies and to look for cool
intergalactic gas, and X-ray studies to probe the hot gas which dominates the
baryonic content of at least some galaxy groups, and also provides a valuable
indicator that a group is truly a dense system in 3-dimensions.

Given the value of X-ray data, all groups in our sample have been selected to
have high quality \ROSAT\ observations available -- though we have {\it not}
selected only groups which are detected in the X-ray. The present paper describes
the analysis of these \ROSAT\ \PSPC\ data, and the properties derived from them,
and combines these with other properties of these systems and their galaxies
drawn from the literature, and in particular from the \NASA-\IPAC\ Extragalactic
Database (\NED).  There have been a number of previous studies of samples of
galaxy groups based primarily on pointed \ROSAT\ observations
(e.g. \citealt*{pildis95};
\citealt{mulchaey96,ponman96,mulchaey98,helsdon00a,helsdon00b,mulchaey03}). 

The present work improves on these in a number of respects: 

\begin{itemize}
\item it is one of the largest samples for which the X-ray data have been
analysed in a uniform manner, 
\item it includes systems with low X-ray luminosity, and some which are entirely
undetected in the X-ray,
\item systems showing intergalactic X-ray emission have been distinguished from
those in which the X-ray emitting gas appears to constitute only hot halo
associated with the central galaxy,
\item galaxy membership and internal velocity dispersion of the groups have been
rederived in a consistent way, using NED data and a sigma-clipping approach,
within a projected overdensity radius,
\item fitted models have been used to extrapolate X-ray luminosity to a fixed
overdensity radius, to compensate for systematic trends with temperature in the
radial extent to which X-ray data are available.
\end{itemize}

The only other studies which share some (but not all) of these features, are
those of \citet{helsdon00a,helsdon00b} and \citet{mulchaey03}, with which we make
a number of comparisons below.

Throughout this paper we use \Hzero\ = 70 \kmpspMpc, and all errors correspond
to 1$\sigma$.

%%%%%%%%%%%%%%%%%%%%%%%%%%%%%%%%%%%%%%%%%%%%%%%%%%%%%%%%%%%%%%%%%%%%%%%%%%%%%%%%%
%%%%%%%%%%%%%%%%%%%%%%%%%%%%%%%%%%%%%%%%%%%%%%%%%%%%%%%%%%%%%%%%%%%%%%%%%%%%%%%%%

\section{The Sample}
\label{sec_sample}

We have sought to assemble the largest sample of galaxy groups for which an X-ray
analysis can be performed.  We have therefore compiled a list of 4320 groups from
10 optical catalogues
(\citealt{hickson82,huchra82,geller83,fouque92,garcia93,nolthenius93,barton96};
\citealt*{ramella97}; \citealt{giudice99,white99})
and compared it to the \ROSAT\ \PSPC\ observation log.  Groups with a recessional
velocity in the range 1000 $<$ \vel\ $<$ 3000 \kmps were then selected from this
list if there was a \PSPC\ pointing with t $>$ 10000 s, within 20\arcm\ of the
group position.  This was to ensure the availability of good quality \ROSAT\ data
and that the system was neither so close as to overfill the \PSPC\ field of view
nor so distant as to make an X-ray detection unlikely.

Duplicate entries resulting from the overlap between catalogues were removed,
along with 7 groups in or around the Hydra and Virgo clusters.  One further
system, NGC\,7552 (drawn from \citet{huchra82}), was excluded after a calculation
of the optical membership described in Section~\ref{sec_optical} revealed that,
although the catalogued group position adhered to all of the above criteria, the
group galaxies are all located at radii $>$ 20\arcm\ from the \ROSAT\ \PSPC\
pointing, so that none of them actually lie within the mirror shell support ring
of the \PSPC. 

To the resulting sample of 45 selected groups, we added a further 13 which had
previously been studied with the \PSPC\ by \citet{helsdon00a}, who in turn
assembled their sample from those of \citet{nolthenius93,ledlow96,mulchaey98},
and two additional Hickson compact groups (HCG\,4 and HCG\,40) for which we had
collected useful optical data. The resulting ensemble of groups is clearly not a
true statistical sample, but is chosen to represent a wide range of group
properties.

Details of the full sample of 60 groups can be found in Table~\ref{tab_sample},
where the names given are taken from the optical catalogues which have been used,
and generally (apart from the Hickson compact groups) correspond to the name of a
prominent galaxy within the group.  The group positions given in
Table~\ref{tab_sample}, were defined in the following way: (a) where X-ray
emission is present (i.e. most cases) the position is that of the galaxy most
centrally located within this emission, (b) where no X-rays are detected, the
catalogued group position was used.  Note that, following these rules, group
positions do {\it not} always correspond to the location of the galaxy whose name
appears in the first column of Table~\ref{tab_sample}.

We have further segregated our sample into 3 subsamples according to the presence
and nature of their X-ray emission, since it is important to distinguish between
emission which is genuinely intergalactic, and that which appears to be
associated with the halo of an individual galaxy.  It has been shown that the
emission from X-ray bright groups is typically characterised by a two-component
surface brightness profile \citep{mulchaey98}, where the extended component
corresponds to the group emission, and the central component to either the
central galaxy, or a bright group core.  Therefore, the presence of two such
components confirms that a group contains intergalactic hot gas.  Unfortunately,
poor statistics can often make fitting a two-component model difficult, even if
the distribution is truly two-component, and in such cases a one-component model
may be all that is available. Hence some other criterion is required, which can
discriminate between group-scale and galaxy halo emission, even in the case of
poor quality data.  Two simple criteria were investigated: the detectable extent
of group emission (\rext, Section \ref{sec_spatial}), and the ratio of X-ray
luminosity, to the luminosity of the central BGG (\LXpLBGG). The former was found
to give more satisfactory results, in that a simple threshold in \rext\ of
60~\kpc\ was found to result in the classification all two-component systems as
X-ray groups.  We therefore assumed that any systems with one-component fits,
which had emission more extensive than 60~\kpc, also possessed intergalactic gas,
but that poor quality data did not permit a two-component fit.

The implication of our extension threshold, is that individual galaxies should
not have X-ray halos extending to more than 60~\kpc\ in radius. To check this, it
would be useful to compare our threshold value of \rext\ with the radii derived
from a sample of isolated early-type galaxies (late-type galaxies have much less
extended hot gas halos).  Unfortunately, isolated early-type galaxies are rare,
and very few have been studied with X-ray instruments.  \citet{osullivan03}
studied \ROSAT\ data from a sample of 39 early-type galaxies, of which 8 were
neither BGGs, nor brightest cluster galaxies.  The X-ray radii of these galaxies
ranged from $\sim$ 3 to 9 \kpc\ - much smaller than our threshold.  A recent
study, by \citet{osullivan04}, of a rather X-ray bright isolated elliptical
galaxy, NGC\,4555, detected emission extending to just 60~\kpc.  Hence this would
have (just) been correctly classified by our extension criterion.

As a further check, we also examined \LXpLBGG\ for all groups, and found one case
in which we felt that our extension criterion had failed.  In HCG\,22, the X-ray
emission is centred on a rather faint elliptical (\LB\ = 10$^{9.29}$ \Lsol), and
the BGG lies outside the main X-ray emitting region.  Using this dimmer central
galaxy to calculate \LXpLB, results in a value of 10$^{31.39}$ \ergpspLsol, which
is significantly higher than the maximum value for any other galaxy halo system
(\LXpLBGG\ = 10$^{30.75}$ \ergpspLsol).  We have therefore altered the
classification of this system from galaxy halo to group emission, to reflect its
high value of \LX, relative to its optically dim central galaxies.

Finally, groups with an X-ray flux (\LX, Section~\ref{sec_spectral}) less than
3$\sigma$ above the background level, have been classed as X-ray undetected
groups. We therefore have the following 3 subsamples:

\begin{itemize}
\item G-sample: 37 groups (36 with \rext\ $>$ 60 kpc + HCG 22)
\item H-sample: 15 groups with \rext\ $\leq$ 60 kpc
\item U-sample: 8 groups with \LX\ $<$ bg + 3$\sigma$(bg)
\end{itemize}

Selection effects in our sample originate from the requirement to have
\ROSAT\ archive data available, the velocity cut we have used, and the sample of
\citet{helsdon00a} from which a large fraction of our groups have been taken.  It
is not clear what biasing is inherent in the \ROSAT\ pointing agenda. Some of our
targets were observed serendipitously by \ROSAT, which reduces any bias involved,
but most were the subject of direct pointed observations. Our sample should
therefore be viewed as diverse (and in particular it is not restricted to
X-ray bright systems) rather than statistically representative. Details of the
60 groups in our full sample are given in Table~\ref{tab_sample}.

\begin{table*}
\begin{center}
\scriptsize
\begin{tabular}{@{}lcccccr@{}}
\hline

%%%%%%%%%%%%%%%%%%%%%%%%%%%%%%%%%%%%%%%%%%%%%%%%%%%%%%%%%%%%%%%%%%%%%%%%%%%%%%%%%

Group     &  \RA         &  \Dec       &  \vel      &  \sigmav    &  \Ngal    &  Catalogue             \\
Name      &  (J2000)     &  (J2000)    &  (\kmps)   &  (\kmps)    &           &                        \\

\hline
     
HCG 4     &  00 34 13.8  &  -21 26 21  &  8394      &  n/a        &  5        &  \citet{hickson82}     \\
NGC 315   &  00 57 48.9  &  +30 21 09  &  4920      &  122        &  4        &  *\citet{nolthenius93} \\
NGC 383   &  01 07 24.9  &  +32 24 45  &  5190      &  466        &  29       &  *\citet{ledlow96}     \\
NGC 524   &  01 24 47.8  &  +09 32 19  &  2632      &  167        &  9        &  \citet{garcia93}      \\
NGC 533   &  01 25 31.3  &  +01 45 33  &  5430      &  464        &  36       &  *\citet{mulchaey98}   \\
HCG 10    &  01 25 40.4  &  +34 42 48  &  4827      &  240        &  4        &  \citet{hickson82}     \\
NGC 720   &  01 53 00.4  &  -13 44 18  &  1760      &  162        &  4        &  \citet{garcia93}      \\
NGC 741   &  01 56 21.0  &  +05 37 44  &  5370      &  432        &  41       &  *\citet{mulchaey98}   \\
HCG 15    &  02 07 37.5  &  +02 10 50  &  6835      &  457        &  6        &  \citet{hickson82}     \\
HCG 16    &  02 09 24.7  &  -10 08 11  &  3957      &  135        &  4        &  \citet{hickson82}     \\
NGC 1052  &  02 41 04.8  &  -08 15 21  &  1477      &  93         &  14       &  \citet{garcia93}      \\
HCG 22    &  03 03 31.0  &  -15 41 10  &  2700      &  13         &  5        &  \citet{hickson82}     \\
NGC 1332  &  03 26 17.1  &  -21 20 05  &  1499      &  n/a        &  n        &  \citet{barton96}      \\
NGC 1407  &  03 40 11.8  &  -18 34 48  &  1695      &  151        &  8        &  \citet{garcia93}      \\
NGC 1566  &  04 20 00.6  &  -54 56 17  &  1292      &  99         &  6        &  \citet{garcia93}      \\
NGC 1587  &  04 30 39.9  &  +00 39 43  &  3660      &  106        &  4        &  *\citet{nolthenius93} \\
NGC 1808  &  05 07 42.3  &  -37 30 46  &  1141      &  213        &  6        &  \citet{giudice99}     \\
NGC 2563  &  08 20 35.7  &  +21 04 04  &  4890      &  336        &  29       &  *\citet{mulchaey98}   \\
HCG 40    &  09 38 54.5  &  -04 51 07  &  6685      &  162        &  6        &  \citet{hickson82}     \\
HCG 42    &  10 00 14.2  &  -19 38 03  &  3840      &  240        &  4        &  \citet{hickson82}     \\
NGC 3227  &  10 23 30.6  &  +19 51 54  &  1407      &  118        &  4        &  \citet{ramella97}     \\
HCG 48    &  10 37 49.5  &  -27 07 18  &  2818      &  355        &  4        &  \citet{hickson82}     \\
NGC 3396  &  10 49 55.2  &  +32 59 27  &  1578      &  96         &  6        &  \citet{garcia93}      \\
NGC 3557  &  11 09 57.4  &  -37 32 17  &  2635      &  377        &  11       &  \citet{garcia93}      \\
NGC 3607  &  11 16 54.7  &  +18 03 06  &  1232      &  245        &  10       &  \citet{ramella97}     \\
NGC 3640  &  11 21 06.9  &  +03 14 06  &  1260      &  178        &  6        &  \citet{garcia93}      \\
NGC 3665  &  11 24 43.4  &  +38 45 44  &  2076      &  65         &  5        &  \citet{garcia93}      \\
NGC 3783  &  11 39 01.8  &  -37 44 19  &  2819      &  169        &  4        &  \citet{giudice99}     \\
HCG 58    &  11 42 23.7  &  +10 15 51  &  6206      &  178        &  5        &  \citet{hickson82}     \\
NGC 3923  &  11 51 02.1  &  -28 48 23  &  1376      &  103        &  5        &  \citet{garcia93}      \\
NGC 4065  &  12 04 06.2  &  +20 14 06  &  7050      &  495        &  9        &  *\citet{ledlow96}     \\
NGC 4073  &  12 04 27.0  &  +01 53 48  &  6120      &  607        &  22       &  *\citet{ledlow96}     \\
NGC 4151  &  12 10 32.6  &  +39 24 21  &  1358      &  95         &  3        &  \citet{ramella97}     \\
NGC 4193  &  12 13 53.6  &  +13 10 22  &  2695      &  168        &  4        &  \citet{nolthenius93}  \\
NGC 4261  &  12 19 23.2  &  +05 49 31  &  2355      &  120        &  18       &  \citet{garcia93}      \\
NGC 4325  &  12 23 06.7  &  +10 37 16  &  7560      &  256        &  18       &  *\citet{mulchaey98}   \\
NGC 4589  &  12 21 45.0  &  +75 18 43  &  2027      &  147        &  11       &  \citet{garcia93}      \\
NGC 4565  &  12 36 20.8  &  +25 59 16  &  1245      &  62         &  3        &  \citet{giudice99}     \\
NGC 4636  &  12 42 50.4  &  +02 41 24  &  1696      &  475        &  12       &  \citet{nolthenius93}  \\
NGC 4697  &  12 48 35.7  &  -05 48 03  &  1363      &  241        &  7        &  \citet{giudice99}     \\
NGC 4725  &  12 50 26.6  &  +25 30 06  &  1495      &  17         &  4        &  \citet{ramella97}     \\
HCG 62    &  12 53 05.8  &  -09 12 16  &  4380      &  324        &  4        &  \citet{hickson82}     \\
NGC 5044  &  13 15 24.0  &  -16 23 06  &  2460      &  129        &  9        &  \citet{garcia93}      \\
NGC 5129  &  13 24 10.0  &  +13 58 36  &  6960      &  294        &  33       &  *\citet{mulchaey98}   \\
NGC 5171  &  13 29 21.6  &  +11 44 07  &  6960      &  424        &  8        &  *\citet{ledlow96}     \\
HCG 67    &  13 49 11.4  &  -07 13 28  &  7345      &  240        &  4        &  \citet{hickson82}     \\
NGC 5322  &  13 49 15.5  &  +60 11 28  &  2106      &  176        &  8        &  \citet{garcia93}      \\
HCG 68    &  13 53 26.7  &  +40 16 59  &  2400      &  170        &  5        &  \citet{hickson82}     \\
NGC 5689  &  14 34 52.0  &  +48 39 36  &  2226      &  n/a        &  3        &  \citet{white99}       \\
NGC 5846  &  15 06 29.2  &  +01 36 21  &  1890      &  368        &  20       &  *\citet{mulchaey98}   \\
NGC 5907  &  15 15 53.9  &  +56 19 46  &  1055      &  56         &  4        &  \citet{geller83}      \\
NGC 5930  &  15 26 07.9  &  +41 40 34  &  2906      &  70         &  3        &  \citet{ramella97}     \\
NGC 6338  &  17 15 22.9  &  +57 24 40  &  8490      &  589        &  7        &  *\citet{ledlow96}     \\
NGC 6574  &  18 12 00.7  &  +14 02 44  &  2435      &  34         &  4        &  \citet{garcia93}      \\
NGC 7144  &  21 52 42.9  &  -48 15 16  &  1855      &  105        &  5        &  \citet{garcia93}      \\
HCG 90    &  22 02 08.4  &  -31 59 30  &  2640      &  110        &  4        &  \citet{hickson82}     \\
HCG 92    &  22 35 58.4  &  +33 57 57  &  6446      &  447        &  5        &  \citet{hickson82}     \\
IC  1459  &  22 57 10.6  &  -36 27 44  &  1707      &  144        &  5        &  \citet{garcia93}      \\
NGC 7714  &  23 36 14.1  &  +02 09 19  &  2908      &  152        &  n/a      &  \citet{fouque92}      \\
HCG 97    &  23 47 22.9  &  -02 18 02  &  6535      &  407        &  5        &  \citet{hickson82}     \\

%%%%%%%%%%%%%%%%%%%%%%%%%%%%%%%%%%%%%%%%%%%%%%%%%%%%%%%%%%%%%%%%%%%%%%%%%%%%%%%%%

\hline
\end{tabular}
\end{center}

%%%%%%%%%%%%%%%%%%%%%%%%%%%%%%%%%%%%%%%%%%%%%%%%%%%%%%%%%%%%%%%%%%%%%%%%%%%%%%%%%

\caption
{\label{tab_sample}
The sample listed in order of right ascension (Section \ref{sec_sample}).  \RA\
and \Dec\ are defined as discussed in the text.  \vel, \sigmav\ and \Ngal\ are
taken from the respective catalogues, and are rederived in Table \ref{tab_optical}
for use in the present work.  Groups where the catalogue is marked with * have
been included from the \citet{helsdon00a} sample and parameters were taken from
that paper. Velocity dispersions of Hickson compact groups are taken from
\citet{ponman96}.}

%%%%%%%%%%%%%%%%%%%%%%%%%%%%%%%%%%%%%%%%%%%%%%%%%%%%%%%%%%%%%%%%%%%%%%%%%%%%%%%%%

\end{table*}

%%%%%%%%%%%%%%%%%%%%%%%%%%%%%%%%%%%%%%%%%%%%%%%%%%%%%%%%%%%%%%%%%%%%%%%%%%%%%%%%%
%%%%%%%%%%%%%%%%%%%%%%%%%%%%%%%%%%%%%%%%%%%%%%%%%%%%%%%%%%%%%%%%%%%%%%%%%%%%%%%%%

\section{Optical Properties}
\label{sec_optical}

Our groups have been assembled from a large number of catalogues, and in
order to reduce the inhomogeneity in our optical data, we have rederived their
galaxy membership in a uniform manner.  Group galaxies were selected from the
\NASA-\IPAC\ Extragalactic Database (\NED) using the algorithm described below,
and optical properties such as total B-band luminosity, morphological type,
velocity and position were extracted.  There will however remain a degree of
inhomogeneity in the \NED\ data, due to the range of sources from which 
the NED data have been compiled. We discuss some checks on the
effects of this heterogeneity, later in this section.

For each group we have searched \NED\ for galaxies within a projected radius
\rfh\ of the group position, defined above, and in a velocity range of \vel\
$\pm$ 3\sigmav.  Values of \rfh\ were calculated from temperature
(\TX, Section~\ref{sec_spectral}) using a relation derived from simulations by
\citet*{evrard96},

\begin{equation}
\rfhTX ~ = ~ \frac{124}{\Hzero} \sqrt{\frac{\TX}{10~\kev}} ~ \Mpc,
\label{eqn_r500_TX}
\end{equation}

\noindent where \TX\ is the temperature in \kev\ and \Hzero\ is the Hubble
constant in \kmpspMpc.  Where no value of \TX\ was available, a relation
between \rfh\ and \LB\ was derived using Equation~\ref{eqn_r500_TX} and a
regression fit in the \LB-\TX\ plane for the systems with group-scale emission
(Section~\ref{sec_relations}),

\begin{equation}
\rfhLB ~ = ~ \frac{124}{\Hzero} \sqrt{\frac{1}{10}\left(\frac{\LB}{10^{11.33}}\right)^{\frac{1}{1.28}}} ~ \Mpc.
\label{eqn_r500_LB}
\end{equation}

\noindent A value of \rfh\ can also be estimated from galaxy velocity dispersion,
using the virial theorem.  Assuming energy equipartition between gas and galaxies
(i.e. \betaspec\ = 1, see Section~\ref{sec_spectral}), Equation~\ref{eqn_r500_TX}
can be rewritten in terms of \sigmav,

\begin{equation}
\rfhsigma ~ = ~ \frac{0.096\sigmav}{\Hzero} ~ \Mpc,
\label{eqn_r500_sig}
\end{equation}

\noindent where \sigmav\ is the velocity dispersion in \kmps.  However we find
evidence that this method is unreliable at low values of \sigmav, as discussed in
Section~\ref{sec_radius}.

Starting values of \vel\ and \sigmav\ were taken from the respective catalogues.
It has been shown that a virialised group should have \sigmav\ \gtsim\ 100 \kmps\
\citep{mamon94}, and as such we have constrained our starting value of \sigmav\
to be no less than this.  In cases where no value of velocity dispersion was
available from the source catalogue we have used 200 \kmps.

Optical data resulting from the galaxy extraction were used to recalculate \vel\
and \sigmav, where 

\begin{equation}
\sigmav ~ = ~ \sqrt{\frac{\sum(v-\bar{v})^{2}}{N-\frac{3}{2}}} ~ \pm ~ \frac{\sigmav}{\sqrt{2(N-\frac{3}{2})}} ~ \kmps,
\label{eqn_sigmav}
\end{equation}

\noindent and the updated values used to redefine our search criteria.  The
denominator includes a correction for the effects of biasing in systems with a
small number of galaxies \citep{helsdon04b}.  The selection and recalculation
were then repeated until the values of \vel\ and \sigmav\ became stable.  If
the final number of galaxies within a group was less than 3, then the membership
calculation was repeated with the starting value of \sigmav\ set to 200 \kmps.
Distances ($\D$) were calculated from velocities after correcting for infall into
Virgo and the Great Attractor.

For two of our systems it was necessary to adjust the membership calculation in
order to reduce contamination from nearby clusters.  NGC\,4261 is in the vicinity
of two clusters (WBL\,392 \& WBL\,397) and to prevent the \sigmav\ from increasing to
include cluster galaxies, we have used only one iteration of the membership
calculation.  HCG\,48 is falling into the cluster Abell\,1060, and to reduce
contamination we have used a group radius equivalent to the distance away of the
minimum in the galaxy density distribution between the centre of the two systems.

We improve the completeness of our sample by applying a luminosity cut to the
optical selection.  The value of luminosity was chosen so as to include 90\% of
the B-band luminosity of the galaxy population, as described by a
Schechter function of the form:

\begin{equation}
\phi(L)dL ~ = ~ \phistar \left( \frac{L}{\Lstar} \right)^{\alpha} exp \left( - \frac{L}{\Lstar} \right) \frac{dL}{\Lstar},
\label{eqn_lumfun}
\end{equation}

\noindent  where $L$ is the galaxy luminosity and $\phi$($L$)$dL$ is the number of
galaxies with a luminosity between $L$ and $L$+$dL$ per \Mpccu.  Free parameters are
the slope at the faint end ($\alpha$), the characteristic Schechter luminosity
(\Lstar) and the normalisation (\phistar).  We have taken values of $\alpha$ and
\Lstar\ from \citet{zabludoff00}, and applied a correction of B-R = 1.57 to
convert from R-band to B-band magnitudes \citep*{fukugita95} giving $\alpha$ =
-1.3, \Lstar\ = 1.60$\times$10$^{10}$ \Lsol, and a minimum luminosity of \Lcut\ =
5.28$\times$10$^{8}$ \Lsol\ (corresponding to \MB\ = -16.32).  We investigate the
completeness obtained by applying this cut in our comparison with \citet{miles04}
below.  The assumed value of B-R is appropriate for early-type
galaxies, and in the case of late types, it will typically result in a luminosity
cut which is too low by about 0.5 magnitudes.  However, no correction has been
applied for the effects of extinction on our galaxy magnitudes, and for typical
spirals this amounts to $\sim$ 0.5$^{\m}$ in the B-band. Hence the two effects
approximately cancel, and our galaxy membership cut should be reasonably
accurate.

We have applied this cut following the membership calculation, and as such it 
does not affect the calculated values of \vel\ and \sigmav, which are based on
the full sample of galaxies associated with each group.  Data surviving the cut
were used to calculate total optical luminosity \LB, corrected for the effect of the
magnitude cut, spiral fraction by number, \fsp, and mean galaxy density, \dengal,
assuming a spherical volume of radius \rfh.  The brightest galaxy within 0.25\rfh\
of the the group
position was designated as the brightest group galaxy (BGG) and its luminosity
(\LBGG) was divided by the luminosity of the second brightest galaxy to define
the dominance of the BGG (\dom).  The results of the membership calculation are
shown in Table~\ref{tab_optical}.  The number of galaxies (\Ngal) is quoted both
before and after the luminosity cut. Systems with \Ngal\ $<$ 4 before the
luminosity cut have been excluded from the statistical analysis, on the grounds
that (a) many of the optical properties of interest to us are poorly defined for
these systems, and (b) such very poor systems are quite likely to be line-of-sight
projections, rather than genuinely overdense in 3-dimensions \citep{frederic95a}.
This richness cut excludes 6 groups, two of which have group-scale X-ray
emission, but with statistical quality too poor to derive an X-ray temperature.
These 6 systems are included in the main data tables, but are flagged with
daggers in Tables~\ref{tab_optical}, \ref{tab_spatial} and \ref{tab_spectral}.

Hence our sample for {\it statistical} analysis in the present paper consists of
54 \GEMS\ groups, which are divided into three subsets according to their X-ray
properties. We refer to these below as the {\it G-sample} (35 systems) which
have group-scale emission, the {\it H-sample} (13 systems) with galaxy halo
emission, and the {\it U-sample} (6 systems) which are undetected in X-rays.

\begin{figure}
  \includegraphics[height=\linewidth,angle=270]{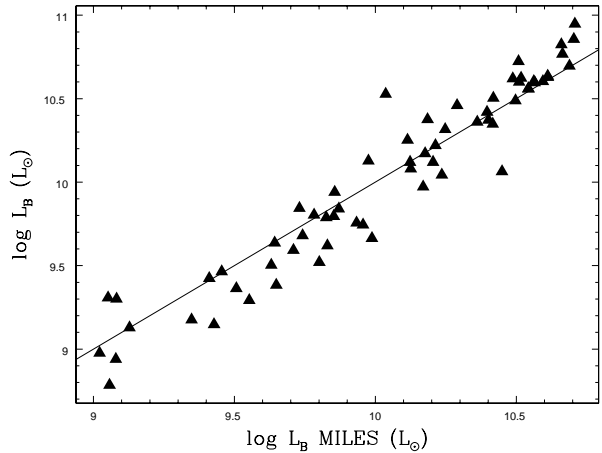}
  \caption{A comparison between our galaxy luminosities, and those of
           \citet{miles04}.  The solid line represents equality.}
  \label{fig_LB_LBMILES}
\end{figure}

As discussed above, the optical data drwan from NED are inevitably inhomogeneous.
To invesitigate the effects of this on our optical luminosities, we have compared
our optical results to those of \citet{miles04}, who obtained B-band photometry
for a subset of 25 of the \GEMS\ groups.  Figure~ \ref{fig_LB_LBMILES} shows a
comparison between our galaxy luminosities and those of \citeauthor{miles04} for
each galaxy that appears in both samples.  Assuming the \citeauthor{miles04} data
to be accurate, we find that our luminosities appear to be biased high (by
$\approx$ 20\%) for the brightest galaxies, and low (by $\approx$ 50\%) for faint
ones.  However, agreement for \LB\ $\sim$ \Lstar\ = 10.20, where most of the total
optical luminosity resides, is good, so that our estimates of total optical
luminosity for groups should be substantially unbiased.

\begin{figure}
  \includegraphics[height=\linewidth,angle=270]{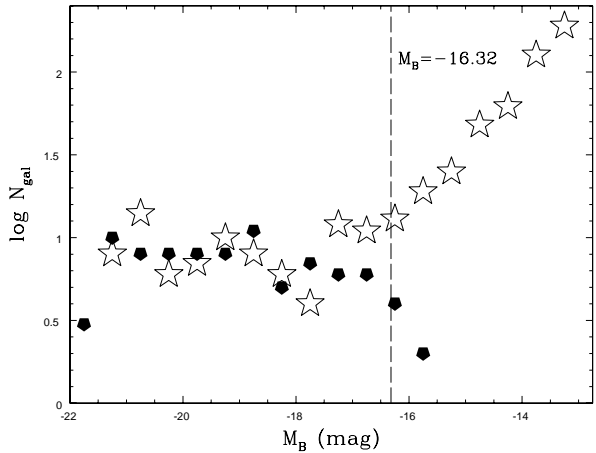}
  \caption{A comparison between our galaxy luminosity function (filled pentagons)
           and that of \citet{miles04} (open stars), for the subset of \GEMS\
           groups covered by the latter study.  The dashed line represents the
           luminosity cut at \MB\ = -16.32.}
  \label{fig_lumfun}
\end{figure}

We also derive a luminosity function from our data using galaxies that are
associated with groups in the \citeauthor{miles04} subsample, and are situated
within their extraction radii of r = 0.3\rfh.  Figure~\ref{fig_lumfun} shows a
comparison of this luminosity function with that derived by \citeauthor{miles04}
We find 81 galaxies above our luminosity cut of \MB\ = -16.32 compared to the 90
found by \citeauthor{miles04}  Since it is clear from Figure~\ref{fig_lumfun}
that the Miles et al. luminosity function is complete to a magnitude much fainter
than our cut, we conclude that our sample is approximately 90\% (81/90) complete,
down to our cut.

A further check on the completeness of our sample close to the cut is obtained by
comparing the total light in galaxies above two different cuts.  Using
Equation~\ref{eqn_lumfun} we calculate the luminosity above which 50\% of the
optical light should lie, to be 1.34$\times$10$^{43}$ \ergps\ (\MB\ = -20.55). If
our sample were complete to the 90\% cut then we would expect the ratio of total
light above the two cuts to be $\sum$\LB(90\%)/$\sum$\LB(50\%) = 1.8.  In
practice this ratio is found to be 1.9 for our data, suggesting that our
completeness is still very high down to the 90\% cut, under the assumption that
our galaxy luminosity function is well represented by the adopted Schechter
function.

\begin{table*}
\begin{center}
\scriptsize
\begin{tabular}{@{}lr@{\hspace{0.1cm}}lllccl@{\hspace{0.2cm}}ccc@{\hspace{0.2cm}}c@{\hspace{0.2cm}}r@{}}
\hline

%%%%%%%%%%%%%%%%%%%%%%%%%%%%%%%%%%%%%%%%%%%%%%%%%%%%%%%%%%%%%%%%%%%%%%%%%%%%%%%%%

Group                 &  \multicolumn{2}{c}{\Ngal}    &  \vel            &  \sigmav        &  \D        &  \rfh      &  \dengal        &  \LB            &  \fsp  &  \LBGG          &  \dom   &  \TBGG    \\
Name                  &      &                        &  (\kmps)         &  (\kmps)        &  (\Mpc)    &  (\Mpc)    &  (\pMpccu)      &  (log \Lsol)    &        &  (log \Lsol)    &         &           \\

\hline

HCG 4$^{\dagger}$     &  2   &  2                     &  8138 $\pm$ 146  &  207 $\pm$ 207  &  115       &  0.36      &  10   $\pm$ 7   &  10.85          &  0.50  &  10.73          &  5.45   &  Late     \\
NGC 315               &  5   &  4                     &  5141 $\pm$ 173  &  387 $\pm$ 146  &  72        &  0.55      &  6    $\pm$ 3   &  11.19          &  0.33  &  11.03          &  5.25   &  Early    \\
NGC 383               &  33  &  27                    &  5174 $\pm$ 8    &  450 $\pm$ 57   &  73        &  0.69      &  20   $\pm$ 4   &  11.54          &  0.20  &  10.56          &  1.22   &  Early    \\
NGC 524               &  10  &  10                    &  2470 $\pm$ 55   &  175 $\pm$ 42   &  35        &  0.45      &  26   $\pm$ 8   &  11.01          &  0.40  &  10.77          &  9.46   &  Early    \\
NGC 533               &  28  &  21                    &  5413 $\pm$ 3    &  439 $\pm$ 60   &  76        &  0.58      &  25   $\pm$ 6   &  11.52          &  0.55  &  10.99          &  1.14   &  Early    \\
HCG 10                &  5   &  5                     &  4843 $\pm$ 103  &  231 $\pm$ 87   &  68        &  0.24      &  84   $\pm$ 38  &  11.01          &  0.50  &  10.60          &  1.06   &  Early    \\
NGC 720               &  4   &  4                     &  1640 $\pm$ 136  &  273 $\pm$ 122  &  23        &  0.40      &  15   $\pm$ 7   &  10.55          &  0.50  &  10.46          &  23.55  &  Early    \\
NGC 741               &  33  &  15                    &  5595 $\pm$ 79   &  453 $\pm$ 57   &  79        &  0.62      &  15   $\pm$ 4   &  11.41          &  0.46  &  11.11          &  8.09   &  Early    \\
HCG 15                &  7   &  7                     &  6742 $\pm$ 153  &  404 $\pm$ 122  &  95        &  0.54      &  11   $\pm$ 4   &  10.85          &  0.43  &  10.26          &  1.15   &  Early    \\
HCG 16                &  7   &  6                     &  3956 $\pm$ 30   &  80  $\pm$ 24   &  57        &  0.32      &  45   $\pm$ 18  &  10.95          &  0.83  &  10.44          &  1.58   &  Late     \\
NGC 1052              &  5   &  4                     &  1366 $\pm$ 41   &  91  $\pm$ 35   &  20        &  0.36      &  21   $\pm$ 10  &  10.37          &  0.50  &  9.97           &  1.47   &  Early    \\
HCG 22                &  4   &  4                     &  2599 $\pm$ 13   &  25  $\pm$ 11   &  39        &  0.29      &  40   $\pm$ 20  &  10.57          &  0.25  &  10.42          &  9.12   &  Early    \\
NGC 1332              &  10  &  9                     &  1489 $\pm$ 59   &  186 $\pm$ 45   &  23        &  0.42      &  29   $\pm$ 10  &  10.55          &  0.56  &  10.06          &  1.42   &  Early    \\
NGC 1407              &  20  &  18                    &  1682 $\pm$ 71   &  319 $\pm$ 52   &  26        &  0.57      &  24   $\pm$ 6   &  11.05          &  0.33  &  10.74          &  5.55   &  Early    \\
NGC 1566              &  9   &  9                     &  1402 $\pm$ 61   &  184 $\pm$ 47   &  21        &  0.47      &  21   $\pm$ 7   &  11.27          &  0.33  &  10.70          &  1.43   &  Late     \\
NGC 1587              &  7   &  6                     &  3671 $\pm$ 43   &  115 $\pm$ 35   &  55        &  0.55      &  9    $\pm$ 4   &  11.07          &  0.40  &  10.60          &  1.10   &  Early    \\
NGC 1808              &  4   &  4                     &  1071 $\pm$ 52   &  104 $\pm$ 47   &  17        &  0.32      &  29   $\pm$ 15  &  10.73          &  1.00  &  10.35          &  1.11   &  Late     \\
NGC 2563              &  32  &  31                    &  4688 $\pm$ 68   &  384 $\pm$ 49   &  73        &  0.57      &  39   $\pm$ 7   &  11.45          &  0.53  &  10.63          &  1.80   &  Early    \\
HCG 40                &  6   &  6                     &  6596 $\pm$ 64   &  157 $\pm$ 52   &  102       &  0.45      &  16   $\pm$ 6   &  11.08          &  0.33  &  10.71          &  3.10   &  Early    \\
HCG 42                &  23  &  19                    &  3801 $\pm$ 59   &  282 $\pm$ 43   &  64        &  0.48      &  40   $\pm$ 9   &  11.33          &  0.36  &  10.95          &  5.40   &  Early    \\
NGC 3227              &  6   &  5                     &  1265 $\pm$ 69   &  169 $\pm$ 56   &  27        &  0.34      &  29   $\pm$ 13  &  10.80          &  0.80  &  10.60          &  3.02   &  Late     \\
HCG 48                &  4   &  2                     &  2587 $\pm$ 158  &  316 $\pm$ 141  &  41        &  0.23      &  39   $\pm$ 28  &  10.39          &  0.50  &  9.44           &  n/a    &  Late     \\
NGC 3396              &  12  &  11                    &  1595 $\pm$ 31   &  106 $\pm$ 23   &  31        &  0.48      &  24   $\pm$ 7   &  10.89          &  1.00  &  10.22          &  1.24   &  Late     \\
NGC 3557              &  14  &  11                    &  2858 $\pm$ 80   &  300 $\pm$ 60   &  39        &  0.27      &  132  $\pm$ 40  &  11.12          &  0.40  &  10.82          &  4.33   &  Early    \\
NGC 3607              &  13  &  11                    &  1099 $\pm$ 78   &  280 $\pm$ 58   &  23        &  0.33      &  72   $\pm$ 22  &  11.02          &  0.33  &  10.61          &  2.25   &  Early    \\
NGC 3640              &  8   &  7                     &  1509 $\pm$ 75   &  211 $\pm$ 59   &  29        &  0.35      &  37   $\pm$ 14  &  10.83          &  0.43  &  10.56          &  4.17   &  Early    \\
NGC 3665              &  4   &  3                     &  2043 $\pm$ 43   &  87  $\pm$ 39   &  37        &  0.38      &  13   $\pm$ 7   &  10.81          &  0.00  &  10.62          &  3.40   &  Early    \\
NGC 3783$^{\dagger}$  &  1   &  1                     &  2917            &  n/a            &  36        &  0.25      &  n/a            &  10.29          &  1.00  &  10.25          &  n/a    &  Late     \\
HCG 58                &  7   &  7                     &  6269 $\pm$ 70   &  184 $\pm$ 55   &  98        &  0.51      &  12   $\pm$ 5   &  11.23          &  0.67  &  10.59          &  1.18   &  Late     \\
NGC 3923              &  8   &  4                     &  1764 $\pm$ 85   &  239 $\pm$ 66   &  22        &  0.40      &  15   $\pm$ 7   &  10.80          &  0.50  &  10.54          &  2.58   &  Early    \\
NGC 4065              &  13  &  13                    &  6880 $\pm$ 125  &  450 $\pm$ 94   &  106       &  0.62      &  13   $\pm$ 4   &  11.57          &  0.31  &  10.81          &  1.41   &  Early    \\
NGC 4073              &  32  &  31                    &  6042 $\pm$ 100  &  565 $\pm$ 72   &  96        &  0.69      &  22   $\pm$ 4   &  11.70          &  0.13  &  11.19          &  4.57   &  Early    \\
NGC 4151              &  6   &  4                     &  1023 $\pm$ 42   &  102 $\pm$ 34   &  23        &  0.29      &  38   $\pm$ 19  &  10.62          &  1.00  &  10.31          &  1.29   &  Late     \\
NGC 4193              &  7   &  6                     &  2159 $\pm$ 76   &  202 $\pm$ 61   &  39        &  0.39      &  24   $\pm$ 10  &  10.93          &  0.83  &  10.11          &  3.02   &  Late     \\
NGC 4261              &  29  &  25                    &  2332 $\pm$ 37   &  197 $\pm$ 27   &  41        &  0.64      &  23   $\pm$ 5   &  11.47          &  0.21  &  10.86          &  2.17   &  Early    \\
NGC 4325              &  16  &  16                    &  7632 $\pm$ 94   &  376 $\pm$ 70   &  117       &  0.51      &  29   $\pm$ 7   &  11.06          &  0.20  &  10.61          &  1.53   &  Early    \\
NGC 4589              &  10  &  9                     &  1640 $\pm$ 90   &  284 $\pm$ 69   &  29        &  0.43      &  26   $\pm$ 9   &  10.69          &  0.67  &  10.15          &  1.25   &  Early    \\
NGC 4565$^{\dagger}$  &  2   &  2                     &  1318 $\pm$ 50   &  71  $\pm$ 71   &  27        &  0.34      &  13   $\pm$ 9   &  10.93          &  1.00  &  10.87          &  27.04  &  Late     \\
NGC 4636              &  9   &  4                     &  936  $\pm$ 95   &  284 $\pm$ 73   &  10        &  0.51      &  7    $\pm$ 4   &  10.45          &  0.00  &  10.04          &  1.07   &  Early    \\
NGC 4697              &  6   &  5                     &  1404 $\pm$ 49   &  120 $\pm$ 40   &  20        &  0.32      &  38   $\pm$ 17  &  10.88          &  0.75  &  10.74          &  5.06   &  Early    \\
NGC 4725              &  4   &  2                     &  1228 $\pm$ 25   &  49  $\pm$ 22   &  25        &  0.40      &  8    $\pm$ 5   &  11.02          &  1.00  &  10.95          &  13.80  &  Late     \\
HCG 62                &  35  &  33                    &  4291 $\pm$ 71   &  418 $\pm$ 51   &  74        &  0.67      &  26   $\pm$ 5   &  11.50          &  0.33  &  10.54          &  1.20   &  Early    \\
NGC 5044              &  18  &  18                    &  2518 $\pm$ 100  &  426 $\pm$ 74   &  33        &  0.62      &  18   $\pm$ 4   &  11.18          &  0.31  &  10.50          &  1.43   &  Early    \\
NGC 5129              &  23  &  23                    &  7012 $\pm$ 71   &  342 $\pm$ 52   &  108       &  0.51      &  40   $\pm$ 8   &  11.54          &  0.60  &  11.05          &  2.31   &  Early    \\
NGC 5171              &  14  &  12                    &  6924 $\pm$ 132  &  494 $\pm$ 99   &  107       &  0.58      &  15   $\pm$ 4   &  11.28          &  0.00  &  10.76          &  2.65   &  Early    \\
HCG 67                &  10  &  10                    &  7455 $\pm$ 83   &  261 $\pm$ 63   &  115       &  0.46      &  24   $\pm$ 8   &  11.32          &  0.60  &  10.94          &  4.29   &  Early    \\
NGC 5322              &  5   &  3                     &  2032 $\pm$ 74   &  166 $\pm$ 63   &  35        &  0.27      &  37   $\pm$ 22  &  10.90          &  0.33  &  10.82          &  22.08  &  Early    \\
HCG 68                &  17  &  16                    &  2407 $\pm$ 46   &  191 $\pm$ 34   &  41        &  0.43      &  50   $\pm$ 12  &  11.41          &  0.67  &  10.64          &  1.19   &  Early    \\
NGC 5689              &  5   &  4                     &  2240 $\pm$ 36   &  80  $\pm$ 30   &  38        &  0.26      &  57   $\pm$ 29  &  10.48          &  0.75  &  9.51           &  1.74   &  Late     \\
NGC 5846              &  25  &  14                    &  1866 $\pm$ 69   &  346 $\pm$ 51   &  30        &  0.48      &  30   $\pm$ 8   &  11.24          &  0.27  &  10.72          &  1.57   &  Early    \\
NGC 5907              &  6   &  3                     &  768  $\pm$ 29   &  72  $\pm$ 24   &  17        &  0.24      &  50   $\pm$ 29  &  10.42          &  1.00  &  10.23          &  2.75   &  Late     \\
NGC 5930              &  4   &  4                     &  2500 $\pm$ 75   &  150 $\pm$ 67   &  41        &  0.55      &  6    $\pm$ 2   &  10.32          &  1.00  &  9.98           &  1.58   &  Late     \\
NGC 6338              &  37  &  36                    &  8789 $\pm$ 107  &  651 $\pm$ 77   &  127       &  0.88      &  13   $\pm$ 2   &  11.80          &  0.44  &  11.05          &  1.37   &  Early    \\
NGC 6574$^{\dagger}$  &  2   &  1                     &  2266 $\pm$ 21   &  29  $\pm$ 29   &  35        &  0.16      &  56   $\pm$ 56  &  10.00          &  1.00  &  9.96           &  n/a    &  Late     \\
NGC 7144$^{\dagger}$  &  2   &  2                     &  1912 $\pm$ 29   &  41  $\pm$ 41   &  27        &  0.41      &  7    $\pm$ 5   &  10.65          &  0.00  &  10.36          &  1.37   &  Early    \\
HCG 90                &  15  &  9                     &  2559 $\pm$ 34   &  131 $\pm$ 25   &  36        &  0.38      &  39   $\pm$ 13  &  10.87          &  0.62  &  10.37          &  1.60   &  Early    \\
HCG 92                &  5   &  5                     &  6347 $\pm$ 209  &  467 $\pm$ 176  &  88        &  0.47      &  11   $\pm$ 5   &  11.06          &  0.40  &  10.52          &  1.19   &  Late     \\
IC  1459              &  8   &  7                     &  1835 $\pm$ 79   &  223 $\pm$ 62   &  26        &  0.35      &  39   $\pm$ 15  &  10.93          &  0.86  &  10.62          &  3.40   &  Early    \\
NGC 7714$^{\dagger}$  &  2   &  2                     &  2784 $\pm$ 20   &  28  $\pm$ 28   &  39        &  0.22      &  48   $\pm$ 34  &  10.30          &  1.00  &  10.17          &  4.70   &  Late     \\
HCG 97                &  14  &  14                    &  6638 $\pm$ 114  &  425 $\pm$ 85   &  92        &  0.51      &  26   $\pm$ 7   &  11.07          &  0.50  &  10.39          &  1.15   &  Early    \\

%%%%%%%%%%%%%%%%%%%%%%%%%%%%%%%%%%%%%%%%%%%%%%%%%%%%%%%%%%%%%%%%%%%%%%%%%%%%%%%%%

\hline
\end{tabular}
\end{center}

%%%%%%%%%%%%%%%%%%%%%%%%%%%%%%%%%%%%%%%%%%%%%%%%%%%%%%%%%%%%%%%%%%%%%%%%%%%%%%%%%

\caption
{\label{tab_optical}
The optical data (section \ref{sec_optical}).  Group membership was calculated
using a position/velocity search for each group, and a luminosity cut of \Lcut\
= 2.73$\times$10$^{41}$ \ergps. The number of galaxies is given before and after
the cut. Groups marked with $\dagger$ have \Ngal\ $<$ 4 before the cut, and have
been excluded from the statistical analysis.}

%%%%%%%%%%%%%%%%%%%%%%%%%%%%%%%%%%%%%%%%%%%%%%%%%%%%%%%%%%%%%%%%%%%%%%%%%%%%%%%%%

\end{table*}

%%%%%%%%%%%%%%%%%%%%%%%%%%%%%%%%%%%%%%%%%%%%%%%%%%%%%%%%%%%%%%%%%%%%%%%%%%%%%%%%%
%%%%%%%%%%%%%%%%%%%%%%%%%%%%%%%%%%%%%%%%%%%%%%%%%%%%%%%%%%%%%%%%%%%%%%%%%%%%%%%%%

\section{X-ray Data Analysis}
\label{sec_reduction}

\ROSAT\ \PSPC\ datasets were prepared for analysis by first eliminating sources
of contamination such as particle emission and solar X-ray emission scattered
from the Earth's atmosphere.  Detectors aboard the spacecraft identify and
exclude over 99\% of these events and record them in a master veto file.  Times
for which this master veto rate exceeded 170 counts \ps\ were considered
significantly contaminated and excluded from our analysis.  Further contamination
from reflected solar X-rays can be identified by an increase in the total X-ray
event rate.  To remove this contamination we have excluded all times for which
the total event rate exceeded the mean by greater than 2$\sigma$.  The remaining
counts were binned into a 3-dimensional x,y,energy data cube.  Images were
created by projecting the data cube along its energy axis, and smoothed images
generated by convolving with a 2-dimensional Gaussian with $\sigma$ = 0.05\arcm.

The background for each dataset was estimated from an annulus, the radius of
which was chosen so as to place it approximately in the region of lowest flux.
Diffuse emission was removed from this annulus by extracting an azimuthal profile
and masking regions with a number of counts greater than 4$\sigma$ above the
mean.  A background model could then be constructed and subtracted from the
datasets.

Point sources within the image were found using maximum likelihood searching and
removed to 1.2 times the 95\% radius for 0.5 \kev\ photons.  The background
subtraction and point source searching were then repeated until the same number
of sources were found upon successive iterations.  Extended sources such as
background clusters were manually identified and removed to the radius at which
their contribution became approximately equal to the surrounding emission.
Extended emission co-incident with the BGG has been shown to exhibit properties
which correlate with the properties of the surrounding group emission
\citep{helsdon00a}, and is hence best identified with the group rather than the
central galaxy.  We have therefore included any such emission in our analysis.
Each exposure was further corrected for dead time effects, vignetting and the
shadow formed by the mirror shell support ring, and finally divided by the
exposure time to produce a map of spectral flux.

%%%%%%%%%%%%%%%%%%%%%%%%%%%%%%%%%%%%%%%%%%%%%%%%%%%%%%%%%%%%%%%%%%%%%%%%%%%%%%%%%

\subsection{Spatial Analysis}
\label{sec_spatial}

On completion of the data reduction, a radial profile centred on the group
position was examined, and the radius at which the group emission fell to the
background level was used to define an extent radius (\rext), and hence a radius
(\rcut) from within which X-ray data are extracted for analysis.  The \rcut\
radii are given in Table~\ref{tab_spectral}.  In the case of HCG\,48, \rext\
included emission from the nearby cluster Abell\,1060 and we have therefore
reduced \rcut\ to a value which only includes regions in which the group emission
dominates over the cluster emission.  In all other detected groups \rcut\ =
\rext.  In cases where no emission was evident, an \rcut\ value of 50 kpc was
used to calculate upper limits. The number of source counts within this region
was then calculated by subtracting the background contribution, as predicted by
the background model.  In cases where the number of source counts was greater
than 3$\sigma$ above background, the dataset was deemed to contain detected
X-ray emission.

\begin{figure*}
  \begin{minipage}[t]{241pt}
    \centering

    \includegraphics[width=0.9\linewidth]{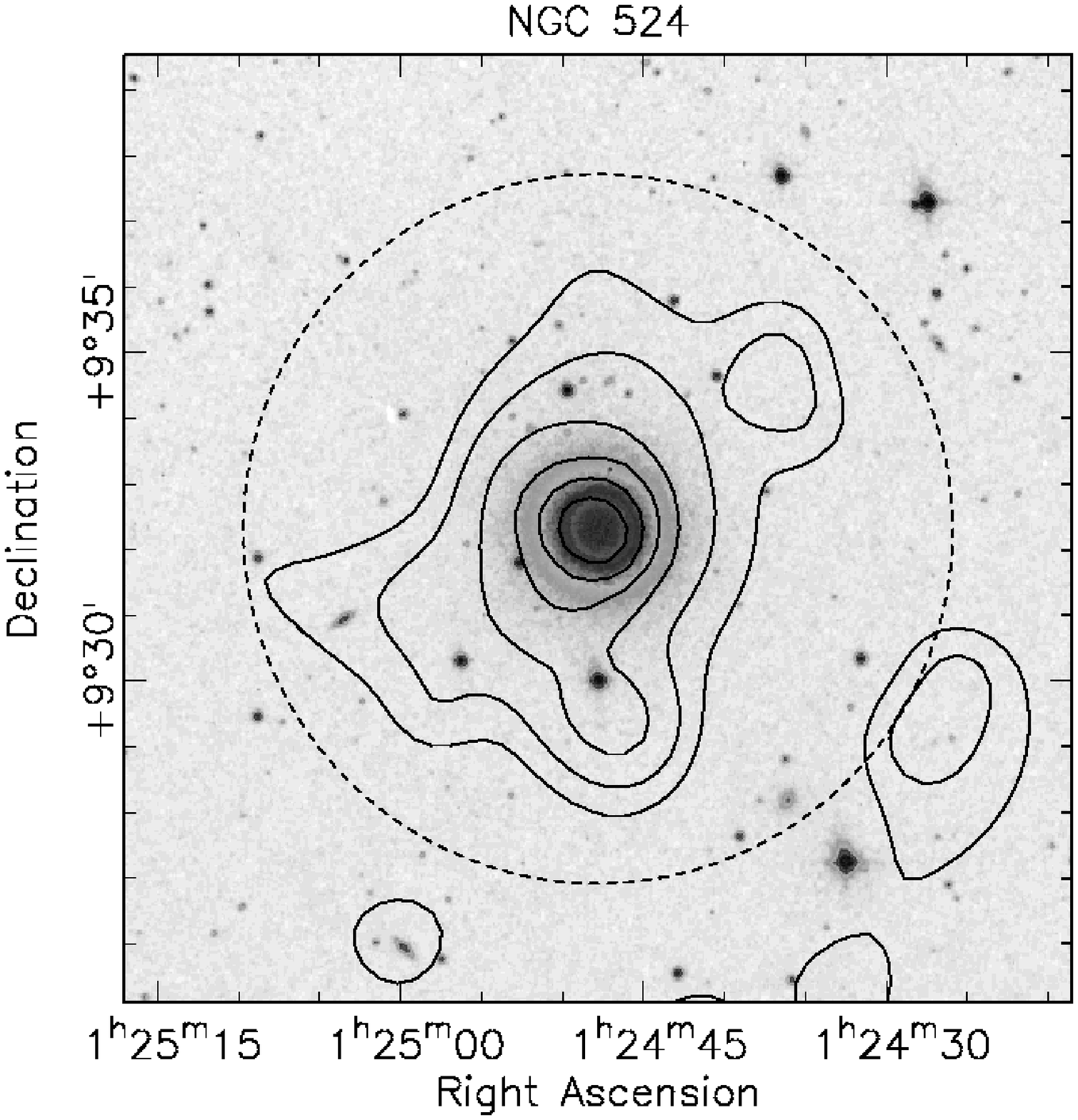}
    \caption{\ROSAT\ \PSPC\ contours for NGC\,524 overlaid on an optical DSS
             image.  The dashed circle represents \rcut\ = 56 kpc.  This system
             has rather compact X-ray emission which we classify as a galaxy halo
             (H-sample).}
    \label{fig_ngc524_ovly}
  
  \end{minipage}\hspace{18pt}
  \begin{minipage}[t]{241pt}
  \centering

    \includegraphics[width=0.9\linewidth]{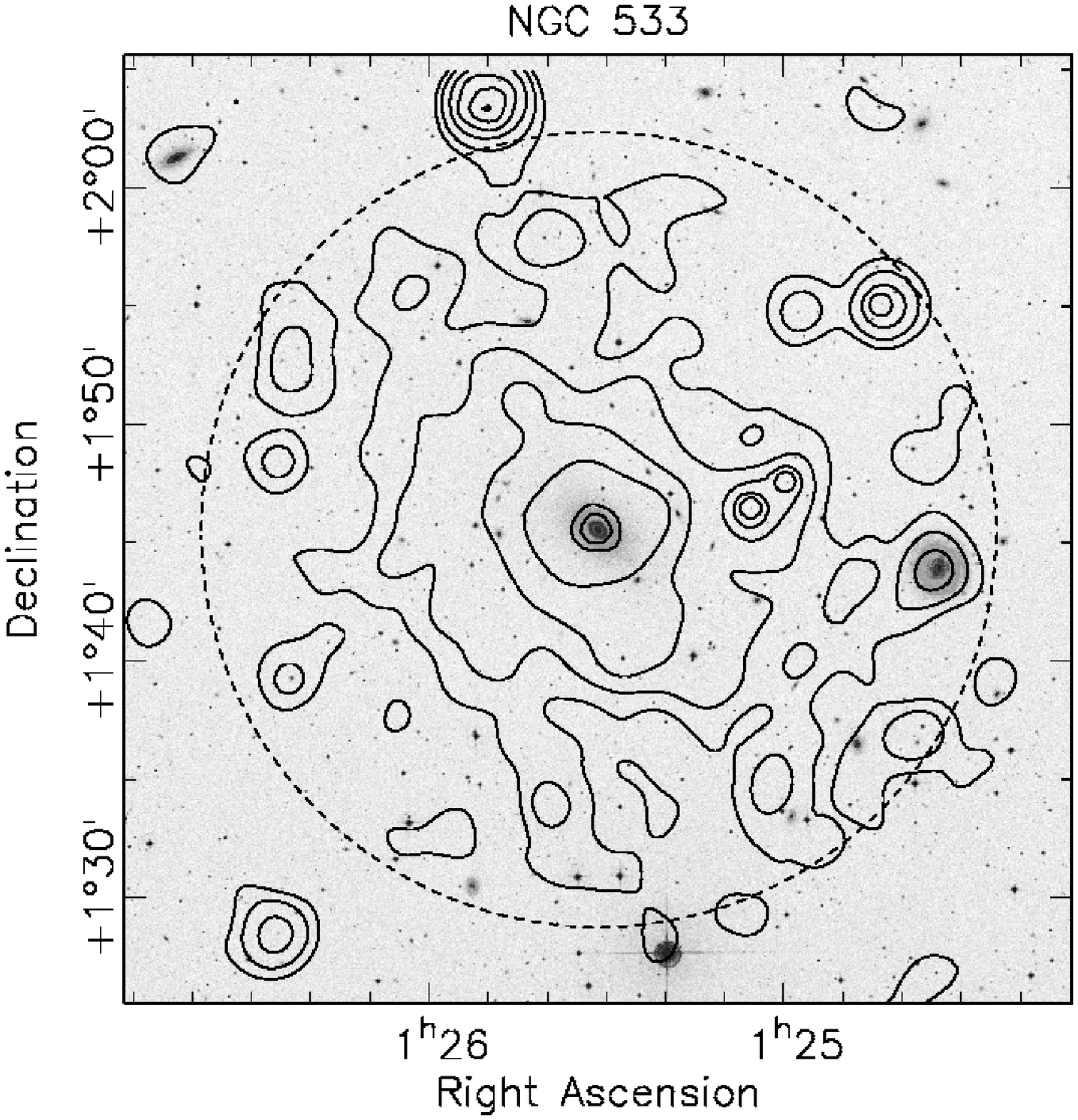}
    \caption{\ROSAT\ \PSPC\ contours for NGC\,533 overlaid on an optical DSS image.
             The dashed circle represents \rcut\ = 372 kpc. This extensive hot
             halo is clearly associated with the group as a whole (G-sample).}
    \label{fig_ngc533_ovly}

  \end{minipage}
\end{figure*}

It was often useful during the course of this data reduction to examine images of
the groups in question.  Quantitative results, such as the emission radii
calculated in Section~\ref{sec_reduction}, could be examined and confirmed using
such images.  We have therefore produced optical images, with X-ray contours
overlaid, for all of the groups in our sample.  Background variance maps were
created assuming Poissonian statistics, smoothed in the usual way and divided
into smoothed images to produce significance maps.  Contours were drawn on at
5$\times2^{n}$ sigma above the background (n = 0,1..10) and overlaid onto optical
images taken from the Digitised Sky Survey (DSS).  Figure~\ref{fig_ngc524_ovly}
shows an X-ray/optical overlay of NGC\,524 which has \rcut\ = 56 kpc (represented
by the dashed circle) and as such is identified with emission from a galactic
halo. NGC\,533 (Figure~\ref{fig_ngc533_ovly}) has \rcut\ = 372 kpc indicating
group-scale emission.

We have sought to characterise the surface brightness properties of our sample of
groups by fitting their emission with a 2-dimensional $\beta$-model, of the form

\begin{equation}
S(r) ~ = ~ \Szero \left(1+ \left(\frac{r}{\rcore}\right)^2\right)^{ -3 \betafit + 0.5}.
\label{eqn_sr}
\end{equation}

\noindent It has been shown that fitting such a profile in one dimension can lead
to an overestimate of the \betafit\ parameter in systems with particularly
elongated or offset components \citep{helsdon00a}.  An image in the band 0.5 to 2
\kev\ was extracted from all datasets containing a detection, and data outside
\rcut\ removed.  Remaining data were then fitted with a single component
$\beta$-model.

Models were convolved with the PSF at an energy determined from the mean photon
energy of the data, and free parameters were the central surface brightness
\Szero, core radius (\rcore), slope (\betafit) and the co-ordinates of
the centre of emission. We also allowed our fits to be elliptical by introducing
the axis ratio ($e$) and position angle as additional free parameters.

In cases where a single component $\beta$-model was inadequate in describing the
emission, a second component was added to the model.  Such an inadequacy was
identified by examining a radial profile for each group in the G-sample, and
looking for a shoulder in which the single $\beta$-model significantly departed
from the data.  In marginal cases a fit using the two component model was
attempted.  In order to limit the number of free parameters in our two-component
fits, we have fixed the axis ratio and position angle of the central component,
thus constraining it to be circular.

Surface brightness models were used to correct bolometric fluxes for the removal
of point sources and other contamination.  For each group the fitted
$\beta$-model (two-component where available) was taken and used to generate a
model image from which a count rate was extracted.  Regions of contamination, as
defined in Section~\ref{sec_reduction}, were then removed, and the reduced count
rate combined with the original count rate to derive a correction factor for the
luminosity.  Groups with no fitted $\beta$-model were corrected by taking an
image with the regions removed and patching over the holes using a local mean.  A
correction factor was obtained and applied in the same way.  Results of the
spatial analysis are presented in Table~\ref{tab_spatial}.

\begin{table*}
\begin{center}
\scriptsize
\begin{tabular}{@{}l@{}c@{\hspace{0.6cm}}ccc@{}c@{\hspace{0.6cm}}cc@{}}
\hline

%%%%%%%%%%%%%%%%%%%%%%%%%%%%%%%%%%%%%%%%%%%%%%%%%%%%%%%%%%%%%%%%%%%%%%%%%%%%%%%%%

Group                 & &  \multicolumn{3}{c}{Extended}                              & &  \multicolumn{2}{c}{Central}           \\
\cline{3-5}  \cline{7-8}
Name                  & &  \rcore             &  \betafit         &  \e              & &  \rcore            &  \betafit         \\
                      & &  (\kpc)             &                   &                  & &  (\kpc)            &                   \\

\hline

HCG 4$^{\dagger}$     & &  n/a               &  n/a              &  n/a              & &  n/a               &  n/a              \\
NGC 315               & &  n/a               &  n/a              &  n/a              & &  n/a               &  n/a              \\
NGC 383               & &  2.11  $\pm$ 0.21  &  0.36 $\pm$ 0.00  &  1.08 $\pm$ 0.03  & &  0.41  $\pm$ 0.43  &  0.60 $\pm$ 0.15  \\
NGC 524               & &  n/a               &  n/a              &  n/a              & &  n/a               &  n/a              \\
NGC 533               & &  2.21  $\pm$ 1.68  &  0.42 $\pm$ 0.01  &  1.50 $\pm$ 0.03  & &  2.52  $\pm$ 0.83  &  0.59 $\pm$ 0.06  \\
HCG 10                & &  n/a               &  n/a              &  n/a              & &  n/a               &  n/a              \\
NGC 720               & &  1.15  $\pm$ 0.20  &  0.47 $\pm$ 0.01  &  1.21 $\pm$ 0.06  & &  n/a               &  n/a              \\
NGC 741               & &  2.30  $\pm$ 0.18  &  0.44 $\pm$ 0.01  &  1.30 $\pm$ 0.09  & &  n/a               &  n/a              \\
HCG 15                & &  n/a               &  n/a              &  n/a              & &  n/a               &  n/a              \\
HCG 16                & &  n/a               &  n/a              &  n/a              & &  n/a               &  n/a              \\
NGC 1052              & &  n/a               &  n/a              &  n/a              & &  n/a               &  n/a              \\
HCG 22                & &  1.34  $\pm$ 4.75  &  0.44 $\pm$ 0.20  &  1.37 $\pm$ 0.77  & &  n/a               &  n/a              \\
NGC 1332              & &  0.07  $\pm$ 0.16  &  0.52 $\pm$ 0.01  &  1.19 $\pm$ 0.12  & &  n/a               &  n/a              \\
NGC 1407              & &  0.08  $\pm$ 0.15  &  0.46 $\pm$ 0.01  &  1.22 $\pm$ 0.06  & &  n/a               &  n/a              \\
NGC 1566              & &  n/a               &  n/a              &  n/a              & &  n/a               &  n/a              \\
NGC 1587              & &  4.34  $\pm$ 4.34  &  0.46 $\pm$ 0.09  &  1.26 $\pm$ 0.45  & &  n/a               &  n/a              \\
NGC 1808              & &  n/a               &  n/a              &  n/a              & &  n/a               &  n/a              \\
NGC 2563              & &  2.14  $\pm$ 0.12  &  0.37 $\pm$ 0.01  &  1.26 $\pm$ 0.06. & &  n/a               &  n/a              \\
HCG 40                & &  n/a               &  n/a              &  n/a              & &  n/a               &  n/a              \\
HCG 42                & &  4.69  $\pm$ 0.72  &  0.56 $\pm$ 0.02  &  1.29 $\pm$ 0.08  & &  n/a               &  n/a              \\
NGC 3227              & &  0.77  $\pm$ 0.60  &  0.57 $\pm$ 0.02  &  1.21 $\pm$ 0.09  & &  n/a               &  n/a              \\
HCG 48                & &  1.20  $\pm$ 1.56  &  0.48 $\pm$ 0.02  &  1.62 $\pm$ 0.30  & &  n/a               &  n/a              \\
NGC 3396              & &  n/a               &  n/a              &  n/a              & &  n/a               &  n/a              \\
NGC 3557              & &  1.13  $\pm$ 0.21  &  0.52 $\pm$ 0.03  &  1.93 $\pm$ 0.34  & &  n/a               &  n/a              \\
NGC 3607              & &  1.98  $\pm$ 0.93  &  0.39 $\pm$ 0.02  &  2.06 $\pm$ 0.18  & &  5.16  $\pm$ 2.73  &  0.60 $\pm$ 0.21  \\
NGC 3640              & &  0.08  $\pm$ 0.19  &  0.43 $\pm$ 0.05  &  2.31 $\pm$ 0.84  & &  n/a               &  n/a              \\
NGC 3665              & &  1.08  $\pm$ 1.31  &  0.47 $\pm$ 0.03  &  1.67 $\pm$ 0.40  & &  n/a               &  n/a              \\
NGC 3783$^{\dagger}$  & &  n/a               &  n/a              &  n/a              & &  n/a               &  n/a              \\
HCG 58                & &  n/a               &  n/a              &  n/a              & &  n/a               &  n/a              \\
NGC 3923              & &  0.63  $\pm$ 0.06  &  0.55 $\pm$ 0.01  &  1.18 $\pm$ 0.08  & &  n/a               &  n/a              \\
NGC 4065              & &  3.08  $\pm$ 0.51  &  0.36 $\pm$ 0.01  &  2.75 $\pm$ 0.35  & &  6.68  $\pm$ 7.86  &  0.37 $\pm$ 0.03  \\
NGC 4073              & &  9.42  $\pm$ 2.89  &  0.43 $\pm$ 0.01  &  1.20 $\pm$ 0.03  & &  3.72  $\pm$ 1.50  &  0.53 $\pm$ 0.07  \\
NGC 4151              & &  n/a               &  n/a              &  n/a              & &  n/a               &  n/a              \\
NGC 4193              & &  n/a               &  n/a              &  n/a              & &  n/a               &  n/a              \\
NGC 4261              & &  40.08 $\pm$ 12.01 &  0.44 $\pm$ 0.09  &  1.17 $\pm$ 0.12  & &  3.31  $\pm$ 1.26  &  1.17 $\pm$ 0.44  \\
NGC 4325              & &  27.56 $\pm$ 4.97  &  0.58 $\pm$ 0.01  &  1.16 $\pm$ 0.05  & &  n/a               &  0.49 $\pm$ 0.03  \\
NGC 4589              & &  9.33  $\pm$ 0.83  &  0.52 $\pm$ 0.07  &  2.65 $\pm$ 0.39  & &  3.41  $\pm$ 2.04  &  n/a              \\
NGC 4565$^{\dagger}$  & &  n/a               &  n/a              &  n/a              & &  n/a               &  n/a              \\
NGC 4636              & &  0.30  $\pm$ 0.06  &  0.47 $\pm$ 0.00  &  1.08 $\pm$ 0.02  & &  2.67  $\pm$ 0.25  &  0.79 $\pm$ 0.04  \\
NGC 4697              & &  1.25  $\pm$ 0.29  &  0.46 $\pm$ 0.02  &  1.37 $\pm$ 0.10  & &  n/a               &  n/a              \\
NGC 4725              & &  n/a               &  n/a              &  n/a              & &  n/a               &  n/a              \\
HCG 62                & &  2.44  $\pm$ 0.26  &  0.48 $\pm$ 0.01  &  1.12 $\pm$ 0.03  & &  10.75 $\pm$ 0.60  &  1.00 $\pm$ 0.05  \\
NGC 5044              & &  5.96  $\pm$ 0.16  &  0.51 $\pm$ 0.00  &  1.04 $\pm$ 0.01  & &  11.04 $\pm$ 0.66  &  0.80 $\pm$ 0.06  \\
NGC 5129              & &  3.14  $\pm$ 1.71  &  0.43 $\pm$ 0.02  &  1.18 $\pm$ 0.18  & &  n/a               &  n/a              \\
NGC 5171              & &  n/a               &  n/a              &  n/a              & &  n/a               &  n/a              \\
HCG 67                & &  4.77  $\pm$ 1.57  &  0.54 $\pm$ 0.07  &  3.16 $\pm$ 0.05  & &  n/a               &  n/a              \\
NGC 5322              & &  n/a               &  n/a              &  n/a              & &  n/a               &  n/a              \\
HCG 68                & &  5.97  $\pm$ 3.43  &  0.45 $\pm$ 0.05  &  1.63 $\pm$ 0.07  & &  n/a               &  n/a              \\
NGC 5689              & &  n/a               &  n/a              &  n/a              & &  n/a               &  n/a              \\
NGC 5846              & &  2.19  $\pm$ 0.26  &  0.51 $\pm$ 0.01  &  1.13 $\pm$ 0.04  & &  n/a               &  n/a              \\
NGC 5907              & &  n/a               &  n/a              &  n/a              & &  n/a               &  n/a              \\
NGC 5930              & &  n/a               &  n/a              &  n/a              & &  n/a               &  n/a              \\
NGC 6338              & &  3.72  $\pm$ 0.98  &  0.44 $\pm$ 0.01  &  1.30 $\pm$ 0.05  & &  10.32 $\pm$ 4.35  &  0.86 $\pm$ 0.34  \\
NGC 6574$^{\dagger}$  & &  n/a               &  n/a              &  n/a              & &  n/a               &  n/a              \\
NGC 7144$^{\dagger}$  & &  0.77  $\pm$ 1.41  &  0.45 $\pm$ 0.03  &  5.03 $\pm$ 3.65  & &  n/a               &  n/a              \\
HCG 90                & &  0.91  $\pm$ 1.54  &  0.41 $\pm$ 0.03  &  1.69 $\pm$ 0.26  & &  3.89  $\pm$ 1.12  &  1.00 $\pm$ 0.20  \\
HCG 92                & &  n/a               &  n/a              &  n/a              & &  n/a               &  n/a              \\
IC  1459              & &  0.74  $\pm$ 2.26  &  0.45 $\pm$ 0.02  &  1.26 $\pm$ 0.07  & &  n/a               &  n/a              \\
NGC 7714$^{\dagger}$  & &  n/a               &  n/a              &  n/a              & &  n/a               &  n/a              \\
HCG 97                & &  2.73  $\pm$ 3.06  &  0.44 $\pm$ 0.01  &  1.53 $\pm$ 0.13  & &  4.31  $\pm$ 1.22  &  0.50 $\pm$ 0.03  \\

%%%%%%%%%%%%%%%%%%%%%%%%%%%%%%%%%%%%%%%%%%%%%%%%%%%%%%%%%%%%%%%%%%%%%%%%%%%%%%%%%

\hline
\end{tabular}
\end{center}

%%%%%%%%%%%%%%%%%%%%%%%%%%%%%%%%%%%%%%%%%%%%%%%%%%%%%%%%%%%%%%%%%%%%%%%%%%%%%%%%%

\caption
{\label{tab_spatial}
Results of the spatial analysis (Section \ref{sec_spatial}). Groups marked with
$\dagger$ have \Ngal\ $<$ 4 before the luminosity cut and have been excluded from
the statistical analysis.}

%%%%%%%%%%%%%%%%%%%%%%%%%%%%%%%%%%%%%%%%%%%%%%%%%%%%%%%%%%%%%%%%%%%%%%%%%%%%%%%%%

\end{table*}

%%%%%%%%%%%%%%%%%%%%%%%%%%%%%%%%%%%%%%%%%%%%%%%%%%%%%%%%%%%%%%%%%%%%%%%%%%%%%%%%%

\subsection{Spectral Analysis}
\label{sec_spectral}

We have performed a spectral analysis for all pointings containing a detected
galaxy group.  A spectrum was obtained by removing all data outside \rcut, and
projecting the cube along its spatial axes.  Each spectrum was then
fitted with a single component \MEKAL\ hot plasma model \citep*{mewe86}, and a
multiplicative absorption component with the neutral hydrogen column density
fixed at a value taken from \Hone\ radio observations \citep{dickey90}.

A spectral fit was considered reliable if the error on the temperature was less
than the fitted value of temperature.  Where this was not the case, the value of
metal abundance was fixed at 0.3 \Zsol\ and the fitting repeated.  If the fit
remained unreliable then the value of temperature was fixed at 1 \kev\ and the
normalisation fitted.  Unabsorbed bolometric fluxes were obtained from all
spectral models by setting the neutral hydrogen column density to zero.  We have
calculated an upper limit on the flux from undetected groups, by taking the same
fixed model and fitting the normalisation to 3$\sigma$ above the background
level.  Values of flux were then converted to luminosities, \LX, using the
optically derived distance, \D\ (Section~\ref{sec_optical}). Results of
the spectral analysis are presented in Table~\ref{tab_spectral}.

The poor spectral resolution of the \ROSAT\ \PSPC\ means that values of metal
abundance (\Z) can often be misleading, even when the value of temperature is
deemed reliable.  However simulations have shown us that fixing this value can
bias the fitted temperature by up to approximately 20\%.  We found that fitting a
one-component spectral model to variable temperature emission results in a
greatly underestimated metal abundance, whilst still producing a reliable value
for temperature.

We have produced simple projected temperature profiles for all groups with
sufficiently good statistics.  For each group we extracted spectra from
concentric annuli of increasing radius from the group position, and fitted \MEKAL\
hot plasma models to them.  The neutral hydrogen column density was fixed as
before, and \Z\ was allowed to vary in cases where the data quality had allowed a
global value to be fitted.  Spectral profiles including more than 3 bins, and
showing a significant drop in temperature in the centre were deemed to
demonstrate a cool core. In these cases (9 systems), data within the central cool
region were removed and the global spectrum refitted, to derive a
``cooling-corrected'' temperature.  This correction was found to be small -- the
average drop in \TX\ being only 4\%,  and lying well within the statistical
error on \TX.  In these cases the X-ray luminosity was corrected for any central
data excised, using the fitted surface brightness model.

Spectral data were combined with optical data to calculate two compound
parameters:  the ratio of X-ray luminosity to optical luminosity (\LXpLB), and the
spectral index (\betaspec) defined by

\begin{equation}
\betaspec ~ = ~ \frac{\mu\sigmav^{2}}{kT} ~ = ~ 6.26 \times 10^{-6} \left(\frac{\sigmav^{2}}{\TX}\right),
\label{eqn_betaspec}
\end{equation}

\noindent where \sigmav\ is the line-of-sight velocity dispersion in \kmps, and
\TX\ the temperature in \kev.

\begin{figure}
  \includegraphics[height=\linewidth,angle=270]{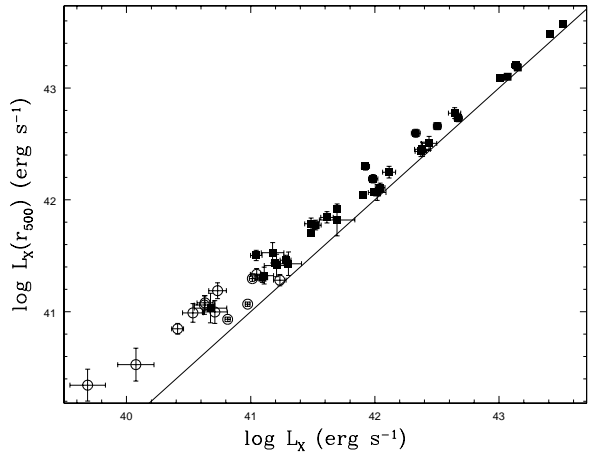}
  \caption{A plot of the luminosity within \rfh\ against that within \rcut.
           Filled squares represent the G-sample and open circles the H-sample.
           The solid line represents equality.}
  \label{fig_LXrfh_LX}
\end{figure}

We have used the fitted $\beta$-models to calculate a luminosity within \rfh, and
Figure~\ref{fig_LXrfh_LX} shows this extrapolated luminosity (\LXrfh) plotted
against \LX.  Errors in \LXrfh\ plotted in the Figure, and listed in 
Table~\ref{tab_spectral}, are extrapolated from errors in \LX, ignoring any
errors arising from uncertainties in \rcore\ or \betafit.  In cases where no
fitted $\beta$-model was available, a standard model with the average values of
\rcore\ = 6 \kpc\ and \betafit\ = 0.5 (Table \ref{tab_statistics}), was used
instead.  As expected, the greatest correction to the luminosity occurs within
the lowest luminosity systems, where it can be as large as a factor of $\sim$ 3.

To investigate the impact on  \LXrfh\ of errors in \rcore\ and \betafit, we
peformed a full Monte-Carlo analysis, incorporating a Gaussian spread in
normalisation, \rcore\ and \betafit, for the system (NGC\,720), which has fairly
typical parameter values. The total derived error on \LXrfh\ was 8\%, a factor of
2 greater than the value of 4\% based on the normalisation error alone.

\begin{table*}
\begin{center}
\scriptsize
\begin{tabular}{@{}lcccccccccl@{}}
\hline

%%%%%%%%%%%%%%%%%%%%%%%%%%%%%%%%%%%%%%%%%%%%%%%%%%%%%%%%%%%%%%%%%%%%%%%%%%%%%%%%%

Group                &  \TX              &  \Z               &  log \LX           &  log \LXrfh         &  log \LXpLB        &  \betaspec        &  \multicolumn{2}{c}{\rcut}    &  \NH                 &  Sample \\
Name                 &  (\kev)           &  (\Zsol)          &  (\ergps)          &  (\ergps)           &  (\ergpspLsol)     &                   &  ($\degree$)  &  (\kpc)       &  (10$^{21}$ \pcmsq)  &         \\

\hline                                                                                                                                

HCG 4$^{\dagger}$    &  n/a              &  n/a              &  41.48 $\pm$ 0.19  &  41.62  $\pm$ 0.19  &  30.64 $\pm$ 0.19  &  n/a              &  0.06         &  120          &  0.15                &  G     \\
NGC 315              &  0.97 $\pm$ 0.22  &  0.30 $\pm$ 0.98  &  41.21 $\pm$ 0.10  &  41.41  $\pm$ 0.10  &  30.02 $\pm$ 0.10  &  0.46 $\pm$ 0.43  &  0.08         &  98           &  0.59                &  G     \\
NGC 383              &  1.51 $\pm$ 0.06  &  0.42 $\pm$ 0.08  &  43.07 $\pm$ 0.01  &  43.10  $\pm$ 0.01  &  31.53 $\pm$ 0.01  &  0.84 $\pm$ 0.22  &  0.50         &  633          &  0.54                &  G     \\
NGC 524              &  0.65 $\pm$ 0.07  &  0.22 $\pm$ 0.15  &  41.05 $\pm$ 0.05  &  41.33  $\pm$ 0.05  &  30.03 $\pm$ 0.05  &  0.30 $\pm$ 0.15  &  0.09         &  56           &  0.48                &  H     \\
NGC 533              &  1.08 $\pm$ 0.05  &  0.68 $\pm$ 0.23  &  42.67 $\pm$ 0.03  &  42.73  $\pm$ 0.03  &  31.16 $\pm$ 0.03  &  1.12 $\pm$ 0.31  &  0.28         &  372          &  0.31                &  G     \\
HCG 10               &  0.19 $\pm$ 0.07  &  0.00 $\pm$ 0.01  &  41.70 $\pm$ 0.14  &  41.82  $\pm$ 0.14  &  30.69 $\pm$ 0.14  &  1.79 $\pm$ 1.51  &  0.08         &  95           &  0.50                &  G     \\
NGC 720              &  0.52 $\pm$ 0.03  &  0.18 $\pm$ 0.02  &  41.20 $\pm$ 0.02  &  41.43  $\pm$ 0.02  &  30.65 $\pm$ 0.02  &  0.90 $\pm$ 0.80  &  0.16         &  65           &  0.15                &  G     \\
NGC 741              &  1.21 $\pm$ 0.09  &  2.00 $\pm$ 0.67  &  42.44 $\pm$ 0.06  &  42.50  $\pm$ 0.06  &  31.03 $\pm$ 0.06  &  1.07 $\pm$ 0.28  &  0.28         &  386          &  0.44                &  G     \\
HCG 15               &  0.93 $\pm$ 0.13  &  0.05 $\pm$ 0.03  &  42.12 $\pm$ 0.05  &  42.25  $\pm$ 0.05  &  31.26 $\pm$ 0.05  &  1.10 $\pm$ 0.68  &  0.10         &  166          &  0.32                &  G     \\
HCG 16               &  0.32 $\pm$ 0.07  &  0.13 $\pm$ 0.13  &  41.30 $\pm$ 0.11  &  41.43  $\pm$ 0.11  &  30.35 $\pm$ 0.11  &  0.12 $\pm$ 0.08  &  0.12         &  119          &  0.22                &  G     \\
NGC 1052             &  0.41 $\pm$ 0.15  &  0.00 $\pm$ 0.02  &  40.08 $\pm$ 0.15  &  40.53  $\pm$ 0.15  &  29.70 $\pm$ 0.15  &  0.13 $\pm$ 0.11  &  0.07         &  25           &  0.31                &  H     \\
HCG 22               &  0.26 $\pm$ 0.04  &  2.00 $\pm$ 0.51  &  40.68 $\pm$ 0.13  &  41.03  $\pm$ 0.13  &  30.11 $\pm$ 0.13  &  0.01 $\pm$ 0.01  &  0.07         &  47           &  0.42                &  G*    \\
NGC 1332             &  0.56 $\pm$ 0.03  &  0.15 $\pm$ 0.03  &  40.81 $\pm$ 0.02  &  40.93  $\pm$ 0.02  &  30.27 $\pm$ 0.02  &  0.39 $\pm$ 0.19  &  0.07         &  28           &  0.22                &  H     \\
NGC 1407             &  1.02 $\pm$ 0.04  &  0.23 $\pm$ 0.05  &  41.69 $\pm$ 0.02  &  41.92  $\pm$ 0.02  &  30.64 $\pm$ 0.02  &  0.62 $\pm$ 0.20  &  0.23         &  105          &  0.54                &  G     \\
NGC 1566             &  0.70 $\pm$ 0.11  &  0.00 $\pm$ 0.02  &  40.41 $\pm$ 0.05  &  40.85  $\pm$ 0.05  &  29.14 $\pm$ 0.05  &  0.30 $\pm$ 0.16  &  0.08         &  29           &  0.13                &  H     \\
NGC 1587             &  0.96 $\pm$ 0.17  &  0.47 $\pm$ 1.24  &  41.18 $\pm$ 0.09  &  41.53  $\pm$ 0.09  &  30.11 $\pm$ 0.09  &  0.09 $\pm$ 0.05  &  0.08         &  77           &  0.68                &  G     \\
NGC 1808             &  n/a              &  n/a              &  $<$40.10          &  $<$40.59           &  $<$29.37          &  n/a              &  0.07         &  21           &  0.27                &  U     \\
NGC 2563             &  1.05 $\pm$ 0.04  &  0.64 $\pm$ 0.20  &  42.50 $\pm$ 0.03  &  42.66  $\pm$ 0.03  &  31.05 $\pm$ 0.03  &  0.88 $\pm$ 0.23  &  0.28         &  359          &  0.42                &  G     \\
HCG 40               &  n/a              &  n/a              &  $<$41.04          &  $<$41.30           &  $<$29.96          &  n/a              &  0.04         &  64           &  0.35                &  U     \\
HCG 42               &  0.75 $\pm$ 0.04  &  0.29 $\pm$ 0.10  &  41.99 $\pm$ 0.02  &  42.07  $\pm$ 0.02  &  30.66 $\pm$ 0.02  &  0.67 $\pm$ 0.21  &  0.10         &  112          &  0.48                &  G     \\
NGC 3227             &  n/a              &  n/a              &  41.23 $\pm$ 0.05  &  41.28  $\pm$ 0.05  &  30.43 $\pm$ 0.05  &  n/a              &  0.12         &  56           &  0.21                &  H     \\
HCG 48               &  n/a              &  n/a              &  41.09 $\pm$ 0.04  &  41.30  $\pm$ 0.04  &  29.65 $\pm$ 0.08  &  n/a              &  0.06         &  43           &  0.51                &  G     \\
NGC 3396             &  0.74 $\pm$ 0.14  &  0.15 $\pm$ 0.10  &  40.53 $\pm$ 0.08  &  40.99  $\pm$ 0.08  &  30.70 $\pm$ 0.04  &  0.10 $\pm$ 0.04  &  0.05         &  27           &  0.20                &  H     \\
NGC 3557             &  0.24 $\pm$ 0.02  &  0.00 $\pm$ 0.01  &  42.04 $\pm$ 0.04  &  42.11  $\pm$ 0.04  &  30.93 $\pm$ 0.04  &  2.40 $\pm$ 0.98  &  0.14         &  95           &  0.74                &  G     \\
NGC 3607             &  0.35 $\pm$ 0.04  &  0.23 $\pm$ 0.10  &  41.05 $\pm$ 0.05  &  41.50  $\pm$ 0.05  &  30.02 $\pm$ 0.05  &  1.40 $\pm$ 0.60  &  0.15         &  62           &  0.15                &  G     \\
NGC 3640             &  n/a              &  n/a              &  $<$40.37          &  $<$40.74           &  $<$29.54          &  n/a              &  0.11         &  55           &  0.43                &  U     \\
NGC 3665             &  0.47 $\pm$ 0.10  &  0.17 $\pm$ 0.14  &  41.11 $\pm$ 0.08  &  41.32  $\pm$ 0.08  &  30.30 $\pm$ 0.08  &  0.10 $\pm$ 0.09  &  0.11         &  71           &  0.21                &  G     \\
NGC 3783$^{\dagger}$ &  n/a              &  n/a              &  40.76 $\pm$ 0.11  &  40.94  $\pm$ 0.11  &  30.46 $\pm$ 0.11  &  n/a              &  0.11         &  69           &  0.85                &  G     \\
HCG 58               &  n/a              &  n/a              &  $<$41.33          &  $<$41.50           &  $<$30.11          &  n/a              &  0.07         &  120          &  0.32                &  U     \\
NGC 3923             &  0.52 $\pm$ 0.03  &  0.18 $\pm$ 0.05  &  40.98 $\pm$ 0.02  &  41.07  $\pm$ 0.02  &  30.18 $\pm$ 0.02  &  0.69 $\pm$ 0.38  &  0.09         &  34           &  0.62                &  H     \\
NGC 4065             &  1.22 $\pm$ 0.08  &  0.97 $\pm$ 0.48  &  42.64 $\pm$ 0.05  &  42.78  $\pm$ 0.05  &  31.08 $\pm$ 0.05  &  1.04 $\pm$ 0.44  &  0.23         &  425          &  0.24                &  G     \\
NGC 4073             &  1.52 $\pm$ 0.09  &  0.70 $\pm$ 0.15  &  43.41 $\pm$ 0.02  &  43.48  $\pm$ 0.02  &  31.71 $\pm$ 0.02  &  1.32 $\pm$ 0.34  &  0.28         &  470          &  0.19                &  G     \\
NGC 4151             &  n/a              &  n/a              &  $<$40.20          &  $<$40.51           &  $<$29.58          &  n/a              &  0.10         &  40           &  0.20                &  U     \\
NGC 4193             &  n/a              &  n/a              &  40.63 $\pm$ 0.08  &  41.06  $\pm$ 0.08  &  29.70 $\pm$ 0.08  &  n/a              &  0.04         &  27           &  0.26                &  H     \\
NGC 4261             &  1.30 $\pm$ 0.07  &  1.23 $\pm$ 0.42  &  41.92 $\pm$ 0.03  &  42.30  $\pm$ 0.03  &  30.46 $\pm$ 0.03  &  0.19 $\pm$ 0.05  &  0.18         &  112          &  0.15                &  G     \\
NGC 4325             &  0.82 $\pm$ 0.02  &  0.50 $\pm$ 0.08  &  43.15 $\pm$ 0.01  &  43.18  $\pm$ 0.01  &  32.09 $\pm$ 0.01  &  1.08 $\pm$ 0.40  &  0.15         &  307          &  0.22                &  G     \\
NGC 4589             &  0.60 $\pm$ 0.07  &  0.08 $\pm$ 0.03  &  41.61 $\pm$ 0.05  &  41.84  $\pm$ 0.05  &  30.92 $\pm$ 0.05  &  0.84 $\pm$ 0.42  &  0.24         &  122          &  0.29                &  G     \\
NGC 4565$^{\dagger}$ &  0.36 $\pm$ 0.14  &  0.10 $\pm$ 0.15  &  40.44 $\pm$ 0.14  &  40.74  $\pm$ 0.14  &  29.51 $\pm$ 0.14  &  0.09 $\pm$ 0.18  &  0.10         &  46           &  0.13                &  H     \\
NGC 4636             &  0.84 $\pm$ 0.02  &  0.41 $\pm$ 0.05  &  41.49 $\pm$ 0.02  &  41.71  $\pm$ 0.02  &  31.04 $\pm$ 0.02  &  0.60 $\pm$ 0.31  &  0.30         &  68           &  0.18                &  G     \\
NGC 4697             &  0.32 $\pm$ 0.03  &  0.07 $\pm$ 0.02  &  41.01 $\pm$ 0.02  &  41.30  $\pm$ 0.02  &  30.13 $\pm$ 0.02  &  0.28 $\pm$ 0.19  &  0.15         &  53           &  0.21                &  H     \\
NGC 4725             &  0.50 $\pm$ 0.07  &  0.00 $\pm$ 0.01  &  40.63 $\pm$ 0.06  &  41.08  $\pm$ 0.06  &  29.61 $\pm$ 0.06  &  0.03 $\pm$ 0.03  &  0.06         &  26           &  0.10                &  H     \\
HCG 62               &  1.43 $\pm$ 0.08  &  2.00 $\pm$ 0.56  &  43.14 $\pm$ 0.04  &  43.20  $\pm$ 0.04  &  31.63 $\pm$ 0.04  &  0.77 $\pm$ 0.19  &  0.22         &  282          &  0.30                &  G     \\
NGC 5044             &  1.21 $\pm$ 0.02  &  0.69 $\pm$ 0.06  &  43.01 $\pm$ 0.01  &  43.09  $\pm$ 0.01  &  31.82 $\pm$ 0.01  &  0.94 $\pm$ 0.33  &  0.30         &  180          &  0.49                &  G     \\
NGC 5129             &  0.84 $\pm$ 0.06  &  0.66 $\pm$ 0.28  &  42.33 $\pm$ 0.04  &  42.60  $\pm$ 0.04  &  30.79 $\pm$ 0.04  &  0.87 $\pm$ 0.27  &  0.08         &  151          &  0.18                &  G     \\
NGC 5171             &  1.07 $\pm$ 0.09  &  1.47 $\pm$ 1.25  &  42.38 $\pm$ 0.06  &  42.45  $\pm$ 0.06  &  31.11 $\pm$ 0.06  &  1.43 $\pm$ 0.59  &  0.16         &  298          &  0.19                &  G     \\
HCG 67               &  0.68 $\pm$ 0.08  &  0.22 $\pm$ 0.13  &  42.02 $\pm$ 0.07  &  42.07  $\pm$ 0.07  &  30.70 $\pm$ 0.07  &  0.63 $\pm$ 0.31  &  0.11         &  222          &  0.25                &  G     \\
NGC 5322             &  0.23 $\pm$ 0.07  &  0.00 $\pm$ 0.02  &  40.71 $\pm$ 0.10  &  41.00  $\pm$ 0.10  &  29.82 $\pm$ 0.10  &  0.76 $\pm$ 0.62  &  0.07         &  43           &  0.18                &  H     \\
HCG 68               &  0.58 $\pm$ 0.06  &  0.09 $\pm$ 0.02  &  41.52 $\pm$ 0.04  &  41.77  $\pm$ 0.04  &  30.12 $\pm$ 0.04  &  0.40 $\pm$ 0.15  &  0.17         &  122          &  0.10                &  G     \\
NGC 5689             &  n/a              &  n/a              &  $<$40.24          &  $<$40.53           &  $<$29.76          &  n/a              &  0.06         &  40           &  0.20                &  U     \\
NGC 5846             &  0.73 $\pm$ 0.02  &  1.25 $\pm$ 0.69  &  41.90 $\pm$ 0.02  &  42.04  $\pm$ 0.02  &  30.66 $\pm$ 0.02  &  1.02 $\pm$ 0.30  &  0.18         &  94           &  0.43                &  G     \\
NGC 5907             &  n/a              &  n/a              &  39.69 $\pm$ 0.14  &  40.34  $\pm$ 0.14  &  29.27 $\pm$ 0.14  &  n/a              &  0.04         &  12           &  0.14                &  H     \\
NGC 5930             &  0.97 $\pm$ 0.27  &  0.17 $\pm$ 0.12  &  40.73 $\pm$ 0.07  &  41.19  $\pm$ 0.07  &  30.42 $\pm$ 0.07  &  0.14 $\pm$ 0.14  &  0.04         &  29           &  0.20                &  H     \\
NGC 6338             &  n/a              &  n/a              &  43.51 $\pm$ 0.02  &  43.57  $\pm$ 0.02  &  31.72 $\pm$ 0.02  &  n/a              &  0.28         &  619          &  0.26                &  G     \\
NGC 6574$^{\dagger}$ &  n/a              &  n/a              &  $<$40.81          &  $<$40.96           &  $<$30.81          &  n/a              &  0.10         &  61           &  1.08                &  U     \\
NGC 7144$^{\dagger}$ &  0.53 $\pm$ 0.20  &  0.00 $\pm$ 0.02  &  40.33 $\pm$ 0.13  &  40.71  $\pm$ 0.13  &  29.69 $\pm$ 0.13  &  0.02 $\pm$ 0.04  &  0.10         &  46           &  0.28                &  H     \\
HCG 90               &  0.46 $\pm$ 0.06  &  0.08 $\pm$ 0.03  &  41.49 $\pm$ 0.05  &  41.79  $\pm$ 0.05  &  30.62 $\pm$ 0.05  &  0.23 $\pm$ 0.09  &  0.16         &  101          &  0.16                &  G     \\
HCG 92               &  0.71 $\pm$ 0.06  &  0.20 $\pm$ 0.13  &  41.99 $\pm$ 0.04  &  42.19  $\pm$ 0.04  &  30.93 $\pm$ 0.04  &  1.92 $\pm$ 1.46  &  0.06         &  93           &  0.80                &  G     \\
IC  1459             &  0.39 $\pm$ 0.04  &  0.04 $\pm$ 0.01  &  41.28 $\pm$ 0.04  &  41.46  $\pm$ 0.04  &  30.35 $\pm$ 0.04  &  0.80 $\pm$ 0.45  &  0.27         &  121          &  0.12                &  G     \\
NGC 7714$^{\dagger}$ &  n/a              &  n/a              &  $<$40.03          &  $<$40.48           &  $<$29.73          &  n/a              &  0.03         &  20           &  0.49                &  U     \\
HCG 97               &  0.82 $\pm$ 0.06  &  0.23 $\pm$ 0.10  &  42.37 $\pm$ 0.05  &  42.43  $\pm$ 0.05  &  31.30 $\pm$ 0.05  &  1.38 $\pm$ 0.56  &  0.21         &  339          &  0.36                &  G     \\

%%%%%%%%%%%%%%%%%%%%%%%%%%%%%%%%%%%%%%%%%%%%%%%%%%%%%%%%%%%%%%%%%%%%%%%%%%%%%%%%%

\hline          
\end{tabular}   
\end{center}    

%%%%%%%%%%%%%%%%%%%%%%%%%%%%%%%%%%%%%%%%%%%%%%%%%%%%%%%%%%%%%%%%%%%%%%%%%%%%%%%%%

\caption        
{\label{tab_spectral}   
The spectral data (Section \ref{sec_spectral}).  The above parameters are derived
from an absorbed \MEKAL\ hot plasma model which we have fitted to 52 of our 60
datasets.  Luminosities shown without corresponding values of temperature or
metal abundance have been derived from fixed models, with \TX\ = 1 and \Z\ = 0.3.
Groups marked with $\dagger$ have \Ngal\ $<$ 4 before the luminosity cut and have
been excluded from the statistical analysis.  The final column indicates the
subsample to which the group belongs (Section \ref{sec_sample}), and those marked
with * have been manually altered from their default classification.}

%%%%%%%%%%%%%%%%%%%%%%%%%%%%%%%%%%%%%%%%%%%%%%%%%%%%%%%%%%%%%%%%%%%%%%%%%%%%%%%%%

\end{table*}

%%%%%%%%%%%%%%%%%%%%%%%%%%%%%%%%%%%%%%%%%%%%%%%%%%%%%%%%%%%%%%%%%%%%%%%%%%%%%%%%%

\subsection{Correlations in Properties}
\label{sec_relations}

We have derived 18 group parameters, listed in Table~\ref{tab_statistics}, which
we use for our statistical analysis.  All parameters were cross-correlated and
any significant relationships identified by examining the resulting plots and the
Kendall's rank correlation coefficient (\K), which corresponds to a correlation
significance in units of Gaussian sigma.  Trends were parameterised by taking the
bisector between two orthogonal least squares regression fits, as calculated by
the SLOPES software \citep{feigelson92}. We prefer to use an unweighted
orthogonal regression, since the scatter observed in the properties of galaxy
groups is primarily non-statistical \citep{helsdon00b}. It is therefore
inappropriate to weight points by their statistical errors when fitting
regression lines.

The regression parameters are summarised in Table~\ref{tab_statistics} and
relationships listed in Table~\ref{tab_relations}.  These results are presented
and discussed in the following sections.  In all figures, filled squares
represent X-ray groups (G-sample), open circles X-ray galactic-halos (H-sample),
and crosses X-ray non-detections (U-sample).

\begin{table}
\begin{center}
\scriptsize
\begin{tabular}{@{}llrrrr@{}}
\hline

%%%%%%%%%%%%%%%%%%%%%%%%%%%%%%%%%%%%%%%%%%%%%%%%%%%%%%%%%%%%%%%%%%%%%%%%%%%%%%%%%

\multicolumn{2}{l}{Parameter}     &  Mean    &  Min.   &  Max.   &  N   \\

\hline                                     

\sigmav    &  (\kmps)             &  261     &  25     &  651    &  54  \\
\D         &  (\Mpc)              &  53      &  10     &  127    &  54  \\
\rfh       &  (\Mpc)              &  0.46    &  0.23   &  0.88   &  54  \\
\rfhsigma  &  (\Mpc)              &  0.45    &  0.10   &  1.07   &  38  \\
\dengal    &  (\pMpccu)           &  30      &  6      &  131    &  54  \\
\LB        &  (log \Lsol)         &  11.16   &  10.32  &  11.80  &  54  \\
\fsp       &                      &  0.50    &  0.00   &  1.00   &  54  \\
\LBGG      &  (log \Lsol)         &  10.67   &  9.44   &  11.19  &  54  \\
\dom       &                      &  3.64    &  1.06   &  23.55  &  54  \\

\rcore     &  (\kpc)              &  6.47    &  0.07   &  81.26  &  34  \\
\betafit   &                      &  0.47    &  0.36   &  0.58   &  34  \\
\e         &                      &  1.52    &  1.04   &  3.16   &  34  \\

\TX        &  (\kev)              &  0.75    &  0.19   &  1.52   &  43  \\
\Z         &  (\Zsol)             &  0.46    &  0.00   &  2.00   &  43  \\
\LX        &  (log \ergps)        &  42.47   &  39.69  &  43.51  &  48  \\
\LXrfh     &  (log \ergps)        &  42.55   &  40.34  &  43.57  &  48  \\

\betaspec  &                      &  0.75    &  0.01   &  2.40   &  43  \\
\LXpLB     &  (log \ergpspLsol)   &  31.07   &  29.14  &  32.09  &  48  \\

%%%%%%%%%%%%%%%%%%%%%%%%%%%%%%%%%%%%%%%%%%%%%%%%%%%%%%%%%%%%%%%%%%%%%%%%%%%%%%%%%

\hline
\end{tabular}
\end{center}

%%%%%%%%%%%%%%%%%%%%%%%%%%%%%%%%%%%%%%%%%%%%%%%%%%%%%%%%%%%%%%%%%%%%%%%%%%%%%%%%%

\caption
{\label{tab_statistics}
A summary of the parameters investigated in the statistical analysis.  N is the
number of datapoints used in each calculation.}

%%%%%%%%%%%%%%%%%%%%%%%%%%%%%%%%%%%%%%%%%%%%%%%%%%%%%%%%%%%%%%%%%%%%%%%%%%%%%%%%%

\end{table}

\begin{table*}
\begin{center}
\scriptsize
\begin{tabular}{@{}r@{\hspace{0.1cm}}l@{\hspace{0.3cm}}r@{\hspace{0.1cm}}l@{\hspace{0.3cm}}c@{}r@{\hspace{0.3cm}}r@{\hspace{0.3cm}}r@{}c@{\hspace{0.3cm}}r@{\hspace{0.3cm}}r@{\hspace{0.3cm}}r@{}c@{\hspace{0.3cm}}r@{\hspace{0.3cm}}r@{\hspace{0.3cm}}c@{}r@{\hspace{0.3cm}}r@{}}

\hline

%%%%%%%%%%%%%%%%%%%%%%%%%%%%%%%%%%%%%%%%%%%%%%%%%%%%%%%%%%%%%%%%%%%%%%%%%%%%%%%%%

     &  y           &       &  x          & &  \multicolumn{3}{c}{\bf{G-Sample}}                                            & &  \multicolumn{3}{c}{H-Sample}                       & &  \multicolumn{3}{c}{All}                           & &  Figure               \\
\cline{6-8}  \cline{10-12} \cline{14-16}                                                                                                                                                                          
     &              &       &             & &  \bf{Slope}                  &  \bf{Intercept}                &  \bf{K}       & &  Slope             &  Intercept           &  K      & &  Slope             &  Intercept          &  K      & &                       \\

\hline                                                                                                                                                                                                             

log  &  \LX        &  log  &  \TX       & &  \bf{ 2.75} $\pm$ \bf{0.49}  &  \bf{ 42.38} $\pm$ \bf{0.10}   &  \bf{ 4.37}   & &  -1.05 $\pm$ 0.22  &   40.40 $\pm$ 0.11   &  -0.09  & &   3.22 $\pm$ 0.51  &   42.25 $\pm$ 0.11  &  4.537  & &  \ref{fig_LX_TX}        \\
log  &  \LX        &  log  &  \sigmav   & &  \bf{ 2.56} $\pm$ \bf{0.66}  &  \bf{ 35.73} $\pm$ \bf{1.68}   &  \bf{ 4.94}   & &   1.90 $\pm$ 0.62  &   36.61 $\pm$ 1.37   &   1.46  & &   3.10 $\pm$ 0.43  &   34.27 $\pm$ 1.05  &  6.854  & &  \ref{fig_LX_sigma}     \\
     &  \betaspec  &  log  &  \LX       & &  \bf{ 0.84} $\pm$ \bf{0.13}  &  \bf{-34.33} $\pm$ \bf{5.34}   &  \bf{ 2.73}   & &   0.86 $\pm$ 0.21  &  -34.80 $\pm$ 8.37   &   0.98  & &   0.69 $\pm$ 0.11  &  -28.14 $\pm$ 4.59  &  4.323  & &  \ref{fig_betaspec_LX}  \\
log  &  \LX        &  log  &  \LB       & &  \bf{ 2.05} $\pm$ \bf{0.21}  &  \bf{ 19.13} $\pm$ \bf{2.41}   &  \bf{ 4.78}   & &   1.28 $\pm$ 0.34  &   26.82 $\pm$ 3.66   &   0.12  & &   2.47 $\pm$ 0.19  &   14.25 $\pm$ 2.15  &  5.978  & &  \ref{fig_LX_LB}        \\
log  &  \LB        &  log  &  \TX       & &  \bf{ 1.28} $\pm$ \bf{0.20}  &  \bf{ 11.33} $\pm$ \bf{0.04}   &  \bf{ 3.95}   & &  -1.12 $\pm$ 0.31  &   10.48 $\pm$ 0.16   &  -0.27  & &   1.37 $\pm$ 0.17  &   11.30 $\pm$ 0.04  &  4.055  & &  \ref{fig_LB_TX}        \\
     &  \fsp       &  log  &  \TX       & &  \bf{-0.93} $\pm$ \bf{0.11}  &  \bf{ 0.26}  $\pm$ \bf{0.04}   &  \bf{-2.64}   & &   1.32 $\pm$ 0.19  &   1.01  $\pm$ 0.11   &   0.76  & &  -1.03 $\pm$ 0.09  &   0.27  $\pm$ 0.04  & -2.554  & &  \ref{fig_fsp_TX}       \\
     &  \fsp       &  log  &  \sigmav   & &  \bf{-0.88} $\pm$ \bf{1.88}  &  \bf{ 2.56}  $\pm$ \bf{4.63}   &  \bf{-1.78}   & &  -1.28 $\pm$ 0.21  &   3.43  $\pm$ 0.45   &  -1.67  & &  -0.93 $\pm$ 0.10  &   2.68  $\pm$ 0.24  & -3.918  & &  \ref{fig_fsp_sigma}    \\
     &  \fsp       &  log  &  \LX       & &  \bf{-0.55} $\pm$ \bf{0.22}  &  \bf{ 23.47} $\pm$ \bf{9.32}   &  \bf{-1.03}   & &  -0.84 $\pm$ 0.21  &   34.92 $\pm$ 8.63   &  -0.64  & &  -0.34 $\pm$ 0.04  &   14.68 $\pm$ 1.87  & -3.503  & &  \ref{fig_fsp_LX}       \\
     &  \fsp       &  log  &  \Z        & &  \bf{-0.53} $\pm$ \bf{0.08}  &  \bf{ 0.16}  $\pm$ \bf{0.06}   &  \bf{-3.17}   & &   0.36 $\pm$ 0.19  &   1.07  $\pm$ 0.17   &  -0.31  & &  -0.54 $\pm$ 0.10  &   0.13  $\pm$ 0.06  & -3.877  & &  \ref{fig_fsp_Z}        \\
     &  \fsp       &  log  &  \dengal   & &  \bf{ 0.78} $\pm$ \bf{0.09}  &  \bf{-0.68}  $\pm$ \bf{0.13}   &  \bf{ 2.53}   & &  -1.00 $\pm$ 0.10  &   2.02  $\pm$ 0.16   &  -0.52  & &   0.98 $\pm$ 0.04  &  -0.85  $\pm$ 0.06  &  1.943  & &  \ref{fig_fsp_dengal}   \\
log  &  \LBGG      &  log  &  \LB       & &  \bf{ 0.96} $\pm$ \bf{0.15}  &  \bf{-0.08}  $\pm$ \bf{1.73}   &  \bf{ 4.86}   & &   1.22 $\pm$ 0.17  &  -2.70  $\pm$ 1.82   &   2.68  & &   0.99 $\pm$ 0.11  &  -0.38  $\pm$ 1.19  &  6.236  & &  \ref{fig_LBGG_LB}      \\
log  &  \LBGG      &  log  &  \sigmav   & &  \bf{ 1.09} $\pm$ \bf{0.19}  &  \bf{ 7.95}  $\pm$ \bf{0.52}   &  \bf{ 1.92}   & &  -1.09 $\pm$ 0.38  &   12.77 $\pm$ 0.82   &  -0.37  & &   1.14 $\pm$ 0.14  &   7.89  $\pm$ 0.36  &  2.772  & &  \ref{fig_LBGG_sigma}   \\

%%%%%%%%%%%%%%%%%%%%%%%%%%%%%%%%%%%%%%%%%%%%%%%%%%%%%%%%%%%%%%%%%%%%%%%%%%%%%%%%%

\hline
\end{tabular}
\end{center}

%%%%%%%%%%%%%%%%%%%%%%%%%%%%%%%%%%%%%%%%%%%%%%%%%%%%%%%%%%%%%%%%%%%%%%%%%%%%%%%%%

\caption
{\label{tab_relations}
A summary of the relationships discussed in the following sections.  Values of
slope and intercept describe an unweighted orthogonal regression fit to all data,
and the G-sample (bold) and H-sample seperately.  K is Kendall's rank correlation
co-efficient, corresponding to a significance in units of Gaussian sigma.}

%%%%%%%%%%%%%%%%%%%%%%%%%%%%%%%%%%%%%%%%%%%%%%%%%%%%%%%%%%%%%%%%%%%%%%%%%%%%%%%%%

\end{table*}

%%%%%%%%%%%%%%%%%%%%%%%%%%%%%%%%%%%%%%%%%%%%%%%%%%%%%%%%%%%%%%%%%%%%%%%%%%%%%%%%%
%%%%%%%%%%%%%%%%%%%%%%%%%%%%%%%%%%%%%%%%%%%%%%%%%%%%%%%%%%%%%%%%%%%%%%%%%%%%%%%%%

\section{Comparison With Previous Work}
\label{sec_comp}

\begin{figure*}
  \begin{minipage}{241pt} 

    \includegraphics[height=\linewidth,angle=270]{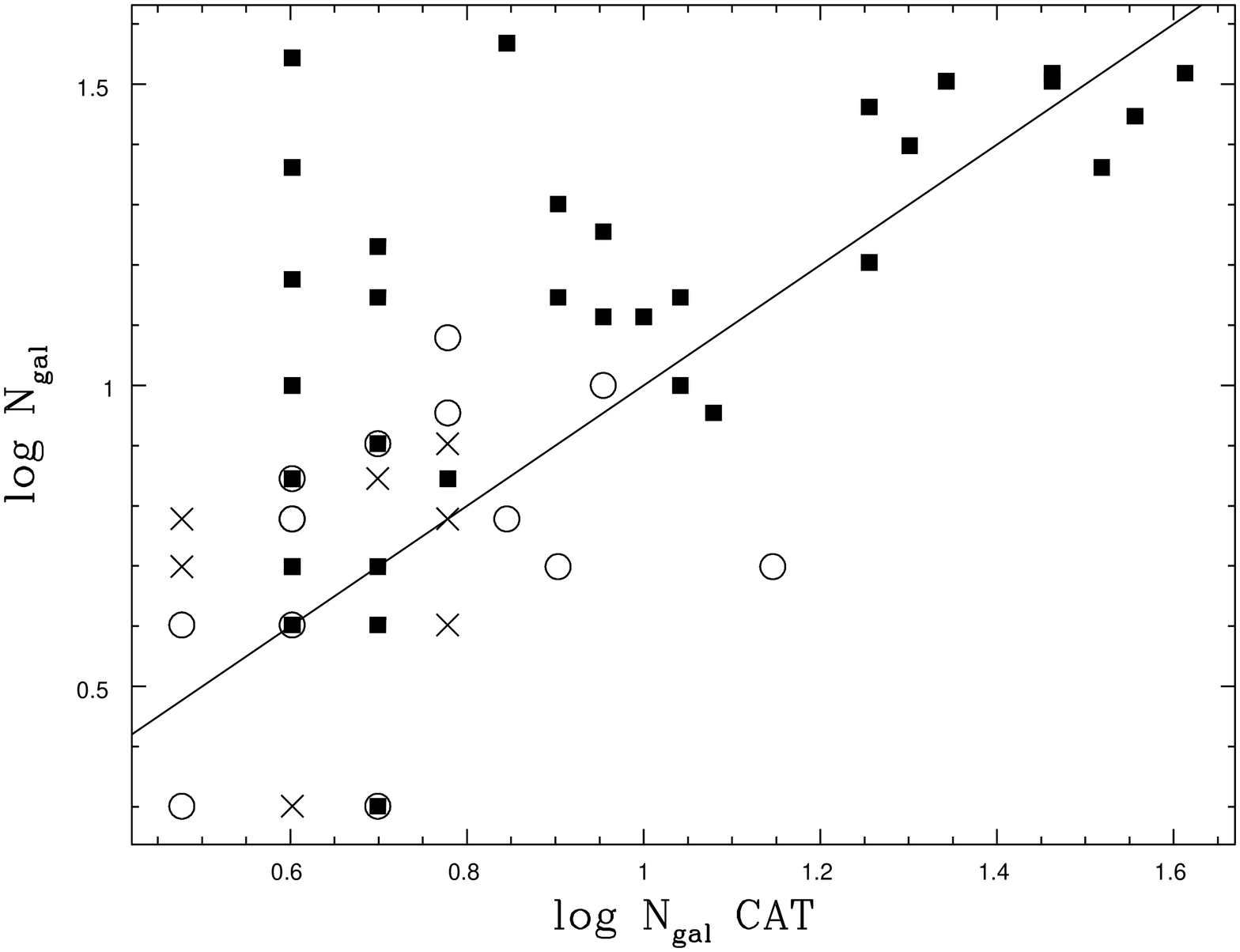}
    \caption{A comparison between our values of \Ngal, and those taken from the
             source catalogues (Table~\ref{tab_sample}).  Filled squares
             represent the G-sample, open circles the H-sample and crosses
             non-detections.  The solid line represents equality.}
    \label{fig_Ngal_Ngalcat}

  \end{minipage}\hspace{18pt}
  \begin{minipage}{241pt}

    \includegraphics[height=\linewidth,angle=270]{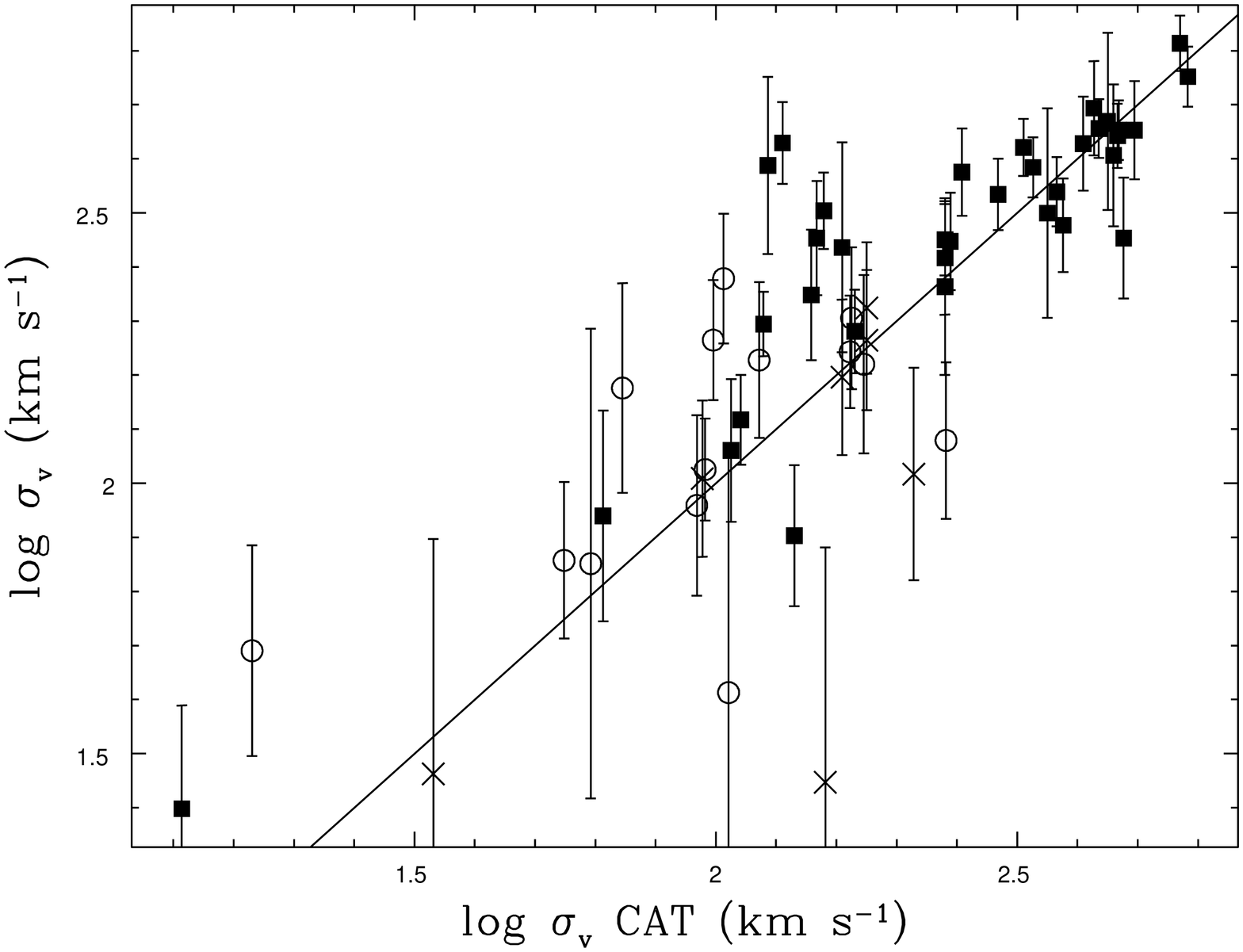}
    \caption{A comparison between our values of \sigmav, and those taken from the
             source catalogues (Table~\ref{tab_sample}).  Filled squares
             represent the G-sample, open circles the H-sample and crosses
             non-detections.  The solid line represents equality.}
    \label{fig_sigma_sigmacat}

  \end{minipage}\\
  \begin{minipage}{241pt} 

    \includegraphics[height=\linewidth,angle=270]{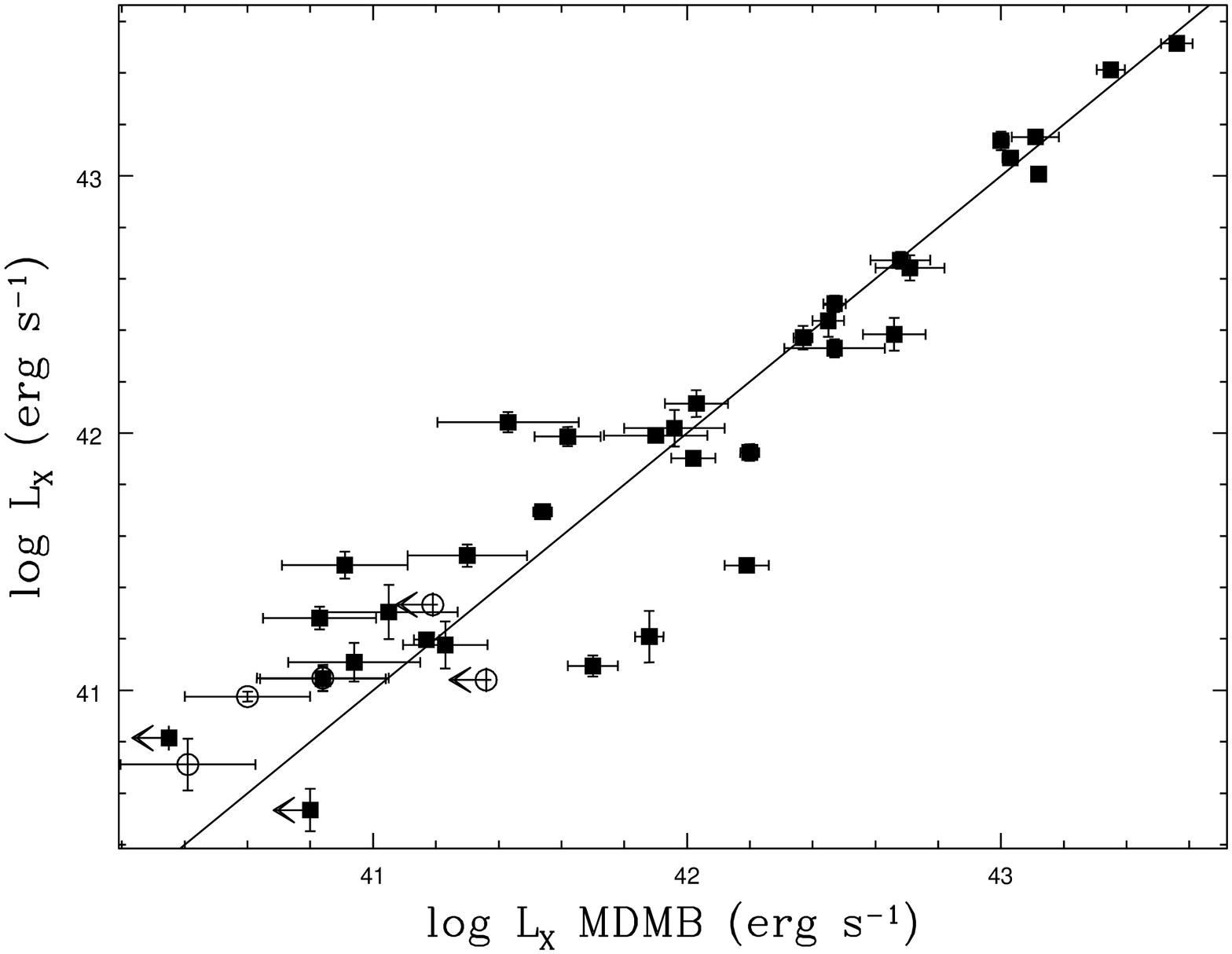}
    \caption{A comparison between our values of \LX, and those taken from
             \citet{mulchaey03}.  Filled squares represent the G-sample, open
             circles the H-sample and arrows represent upper-limits from
             non-detections.  The solid line represents equality.}
    \label{fig_LX_LXMDMB}

  \end{minipage}\hspace{18pt}
  \begin{minipage}{241pt}

    \includegraphics[height=\linewidth,angle=270]{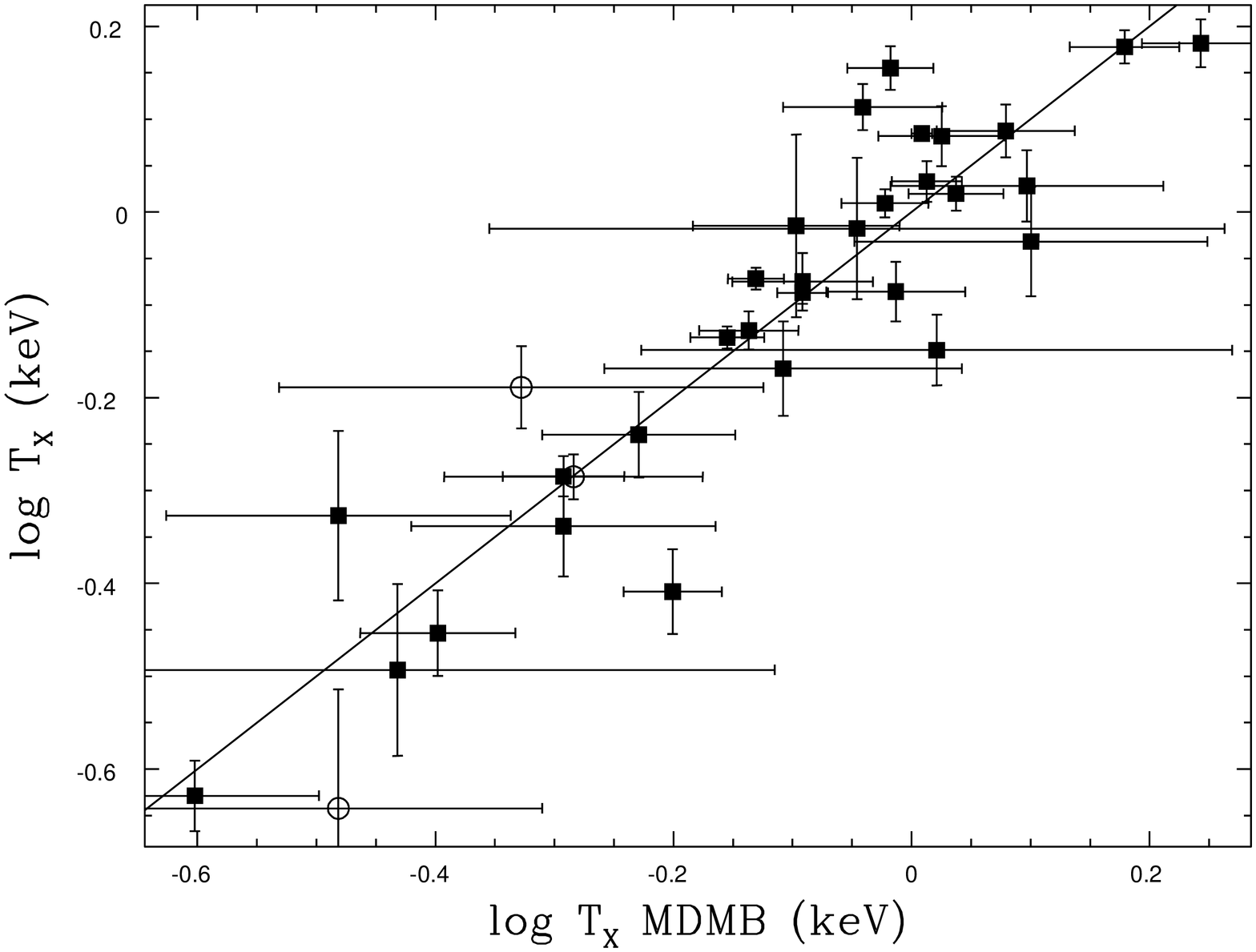}
    \caption{A comparison between our values of \TX, and those taken from
             \citet{mulchaey03}.  Filled squares represent the G-sample and open
             circles the H-sample.  The solid line represents equality.}
    \label{fig_TX_TXMDMB}

  \end{minipage}
\end{figure*}

We compare our derived values of \Ngal\ and \sigmav\ to those given in the group
catalogues from which our groups are drawn.  Figure~\ref{fig_Ngal_Ngalcat} shows
a reasonable agreement between values of \Ngal\ in all but the compact groups,
where we typically find many more galaxy members, since the original compact
group search included only a compact core of galaxies in what is generally a much
larger group \citep[e.g][]{zabludoff98}.  We also find a reasonable match between
our values of \sigmav, and those taken from the source catalogues
(Figure~\ref{fig_sigma_sigmacat}).

The recently published atlas by \citet[][hereafter referred to as MDMB]{mulchaey03},
includes 109 \ROSAT-observed groups, larger than present sample, with X-ray
emission detected from 61.  X-ray fluxes in this study have not been corrected to
\rfh, and optical properties have been drawn from a variety of group catalogues,
rather than re-extracted in a uniform manner as in our sample.  However,
\citeauthor{mulchaey03} subject all their groups to a uniform X-ray data analysis
similar in many ways to ours, so that comparisons with our results provide a
valuable check.  In particular they adopt the same procedure for choosing \rcut, and
quoted luminosities are bolometric.  In the spectral fitting the netural hydrogen
column density is fixed at the galactic value and unconstrained metallicities are
fixed at 0.3 solar.

There is an overlap of 43 groups between the two samples.  Plotting \LX\ against
\LXMDMB\ (Figure~\ref{fig_LX_LXMDMB}) shows a good agreement for X-ray bright
groups, and reasonable agreement amongst groups with lower luminosity, but with
a good deal of scatter in the latter. The reason for this scatter is not clear.
It does not, in general,  appear to be related to the radius out to which
emission has been integrated in the two studies, nor (see below) to systematic
differences in the spectral properties derived. We explored the comparison in
more detail for three of the groups for which the disagreement with MDMB was
strongest.  NGC\,315 has a value of \LX\ which is a factor of $\approx$ 5 less
than the MDMB value of \LX\ = 10$^{41.88}$ \ergps.  However this difference is
accounted for by the removal of the central AGN in our analysis
(Section~\ref{sec_reduction}).  Our \LX\ for HCG\,48 is a factor of $\approx$ 2
less than the MDMB value of 10$^{41.70}$ \ergps.  This value has been extracted
from a circular region with a value of \rcut\ equivalent to 60\% of the MDMB
radius of 72 kpc.  We have used a smaller radius in order to minimise X-ray
contamination from the nearby cluster Abell\,1060 and it is this difference which
accounts for the deficit in \LX.  NGC\,4636 has \LX\ a factor of $\approx$ 5 less
than the MDMB value of 10$^{42.19}$ \ergps, and is extracted from a similar size
region.  Furthermore the diffuse emission is so extensive that replacing the
central AGN only changes the overall \LX\ by a small proportion.  The value of
\LX\ derived by \citet{helsdon00a} for the same system is 10$^{42.18}$ \ergps, in
good agreement with the MDMB value.  However both of these studies take the group
velocity from the source catalogue, and applying our iterative membership
calculation decreases this catalogued value (and hence the distance inferred from
it) by a factor of $\sim$ 2, accounting for the majority of the discrepancy in
\LX.  Our value of distance is also in much better agreement with that of the
BGG, NGC\,4636 (\D\ = 10 \Mpc).

We find a good agreement between our values of \TX\ and those taken from
MDMB (Figure~\ref{fig_TX_TXMDMB}), though the latter have not been corrected for
cool cores. Since the MDMB study is based on the same \ROSAT\ data that we are
using, this comparison does not address the issue of the whether \ROSAT\ spectra
yield reliable temperatures.  The results of \citet{hwang99} suggest that for hot
plasmas with \TX\ $>$ 1.5 \kev, \ROSAT\ \PSPC\ temperatures are biased low (by
about 30\%) relative to those derived using the superior spectral capabilities of
\ASCA, whilst for cooler systems temperatures from the two instruments are in
reasonable agreement. Since the hottest system in our sample has \TX\ = 1.52 \kev,
any such bias should have only minor effects on our study.

%%%%%%%%%%%%%%%%%%%%%%%%%%%%%%%%%%%%%%%%%%%%%%%%%%%%%%%%%%%%%%%%%%%%%%%%%%%%%%%%%
%%%%%%%%%%%%%%%%%%%%%%%%%%%%%%%%%%%%%%%%%%%%%%%%%%%%%%%%%%%%%%%%%%%%%%%%%%%%%%%%%

\section{The Radii of Galaxy Groups}
\label{sec_radius}

As discussed in Section~\ref{sec_optical}, we have sought in this study to
extract group properties within a consistently defined overdensity radius,
corresponding to 500 times the critical density of the Universe at the current
epoch.  The best way to define such a radius for each group would be to derive
total mass profiles, to directly measure the radius within which the desired mean
density is obtained.  This could be done (under the assumption of hydrostatic
equilibrium) if gas density and temperature profiles could be extracted
from our data.  Unfortunately, the quality of the data does not permit reliable
gas temperature profiles to be extracted for most of these groups, and in no case
could such a profile be extended to \rfh.  Failing this, three methods for
estimating \rfh\ were considered, as described in Section~\ref{sec_optical},
based on the use of X-ray temperature, galaxy velocity dispersion or total
optical (blue) luminosity.  The principle behind all these, is that a system
virialising at $z$ = 0 should have a given mean density within its virial radius,
and hence all overdensity radii should scale as the third power of system mass.
Total optical luminosity (\LB) can be used to estimate system mass under the
assumption that star formation efficiency, and mean mass-to-light ratio of
galaxies, are independent of other group properties.

Unfortunately, these assumptions are debatable. Semi-analytical models of galaxy
formation predict a correlation between \MpL\ and halo size.  For example,
\citet{benson00} find that \MpL\ drops by a factor $\sim$ 3 in their models,
between halos of mass 10$^{14}$ \Msol\ and 10$^{12}$ \Msol, i.e. in the group and
galaxy regime.  Observational evidence on this issue is mixed.  Most X-ray
studies, such as those by \citet{hradecky00} and \citet{sanderson03b}, have found
the mass-to-light ratio and star formation efficiency in groups and clusters to
be essentially independent of temperature.  However, the study of
\citet{hoekstra01}, based on the weak lensing signal from a set of stacked
groups, found \MpL\ lower than that in clusters, and a compilation of a variety
of measurements on group to cluster scales, led \citet{bahcall02} to conclude
that \MpL\ rises gently, as \TX$^{0.3}$, across the temperature range from 1 \kev\
to 12 \kev.

Alternatively, for a system in virial equilibrium, the characteristic velocity
dispersion of the galaxies, and the gas temperature, should be related to system
mass via the virial theorem.  This leads to

\begin{equation}
\TX ~ \propto ~ \sigmav^2 ~ \propto ~ M/R ~ \propto ~ M^{2/3},
\label{eqn_tx}
\end{equation}

\noindent where the final step involves the assumption of constant mean density
for newly-virialised systems. Results from cosmological simulations suggest that
a scaling relation $M$ $\propto$ \TX$^{1.5}$ can give a robust and reliable
measure of mass.  \citet{evrard96}, in an analysis of an ensemble of simulated
clusters (including some incorporating feedback), found that mass estimates using
a \TX$^{1.5}$ formula with an appropriate normalisation, scattered about the true
masses, with a standard deviation of only 15\%.  On the other hand, a number of
studies \citep*[e.g.][]{finoguenov01b} find that the $M$-\TX\ relation has a slope
steeper than 1.5, and  \citet{sanderson03a} find observational evidence, by
comparing X-ray derived masses with the results obtained from simple scaling
formulae, that a \TX$^{0.5}$ scaling can overestimate virial radii, especially in
cool systems, by up to 40\%, leading to a corresponding overestimate in virial
mass.

It is known from previous studies (c.f. Section~\ref{subsec_sigT}) that the
energy per unit mass in gas tends to be higher than that in galaxies (i.e.
\betaspec\ $<$ 1) for poor clusters and groups, and that in groups there appears
to be a great deal of scatter in \betaspec.  This implies that either \TX\ or
\sigmav\ (or both) is an unreliable indicator of system mass. {\it A priori} one
could think of reasons to suspect either parameter: \sigmav\ is usually
statistically poorly determined in groups, due to the low number of galaxy
redshifts available, and might also be affected by a variety of biases and
physical effects, whilst \TX\ could be vulnerable to the effects which are
believed to have raised the entropy of the gas in groups relative to that
expected on the basis of what is seen in clusters \citep{ponman99}.

\begin{figure}

  \includegraphics[height=\linewidth,angle=270]{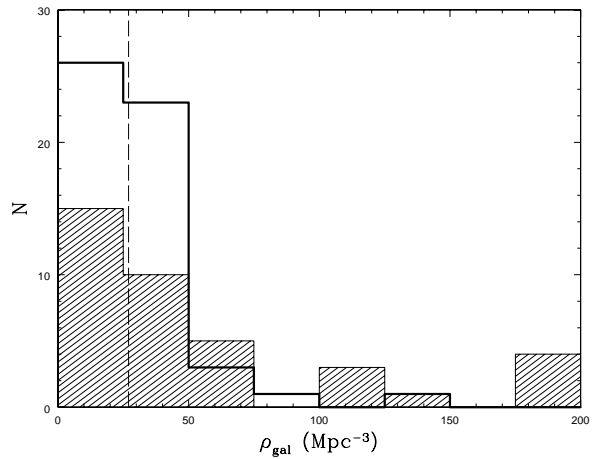}
  \caption{Distribution of mean galaxy densities within \rfh\ for the \GEMS\ sample
           when the \rfh\ radii are evaluated using (a) group velocity dispersion
           (shaded histogram) and (b) X-ray temperature (bold line).  The upper
           most bin of the \sigmav-based histogram contains 2 groups with much
           larger densities than indicated (\dengal\ = 284 \& 653).  The expected
           mean density of galaxies down to our luminosity cut is shown as a
           vertical dashed line.}
  \label{fig_dengal_hist}

\end{figure}

\begin{figure}

  \includegraphics[height=\linewidth,angle=270]{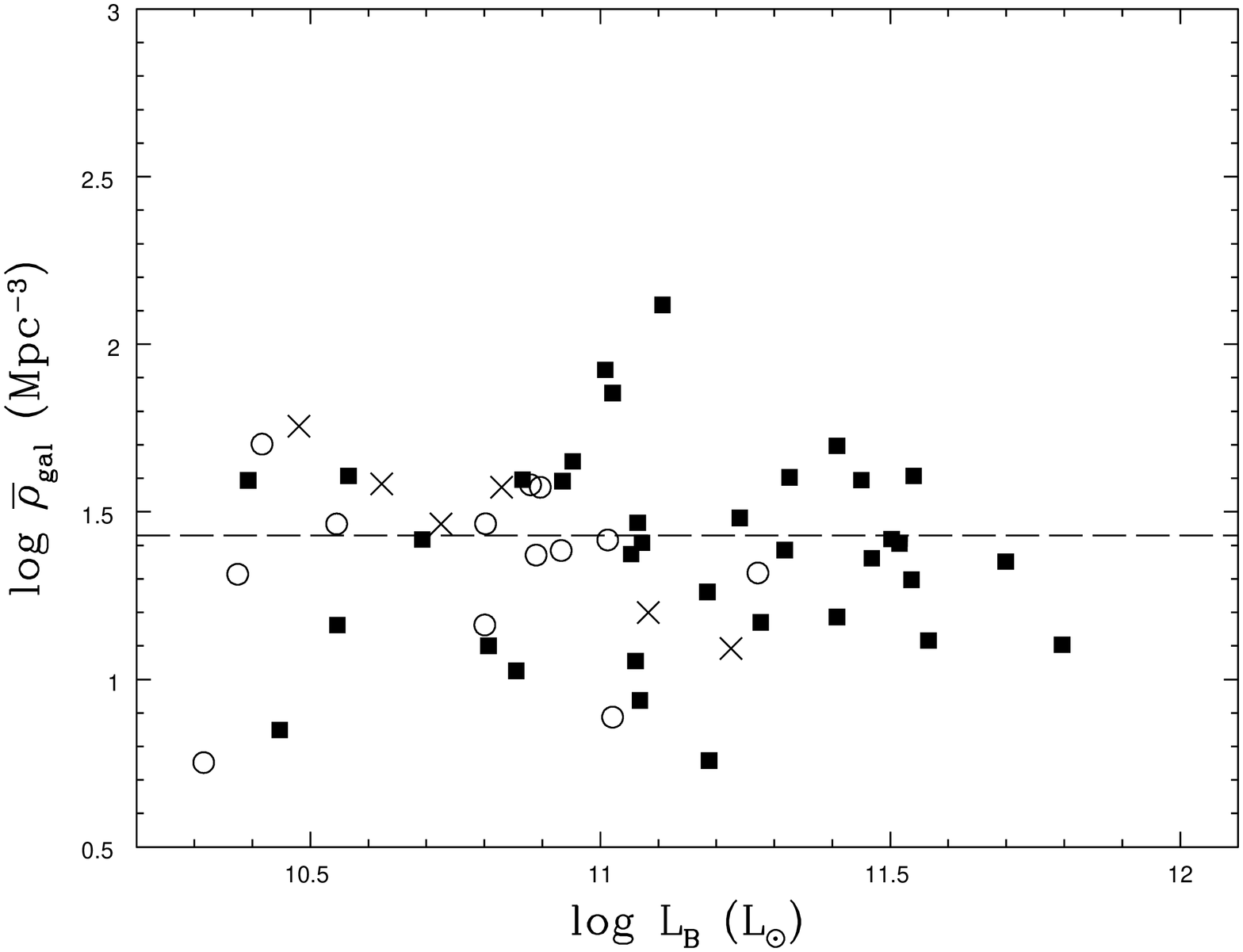}
  \includegraphics[height=\linewidth,angle=270]{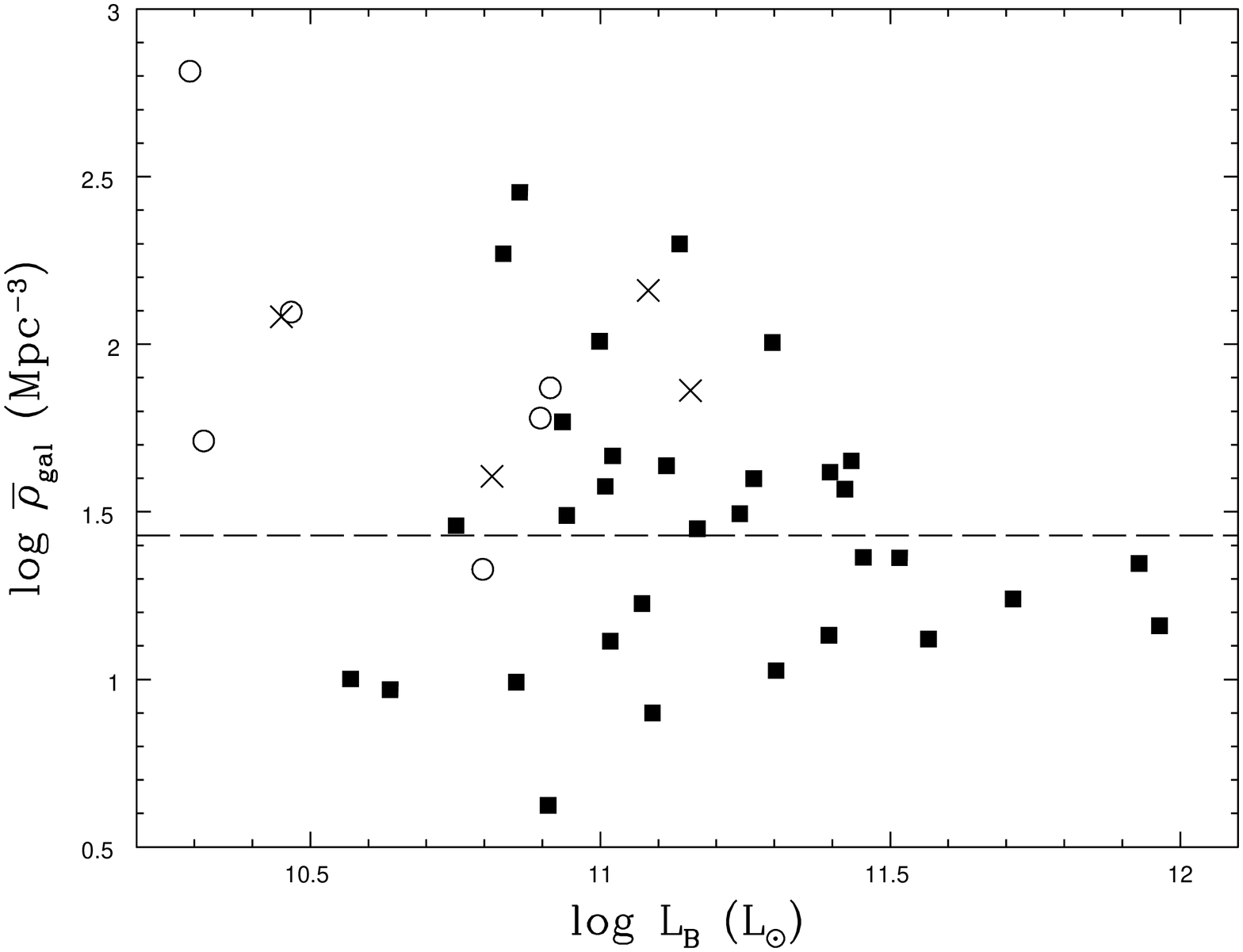}
  \caption{The relationship between \dengal\ and \LB\ using (top) gas temperature,
           and (bottom) galaxy velocity dispersion, to calculate \rfh\ and derive
           group membership.  Filled squares represent the G-sample, open circles
           the H-sample and crosses non-detections.  The expected mean density of
           galaxies down to our luminosity cut is shown as a horizontal dashed
           line.}
  \label{fig_dengal_LB}

\end{figure}

To explore this further, we tried both methods (Equations~\ref{eqn_r500_TX} and
\ref{eqn_r500_sig}) for the evaluation of \rfh, and extracted  the group members
for each of the two resulting definitions.  It is instructive to consider the
mean density of galaxies,

\begin{equation}
\dengal ~ = ~ \frac{\Ngal}{\Vfh} ~ = ~ \frac{\Ngal}{\frac{4}{3}\pi\rfh^{3}} ~ \pMpccu,
\label{eqn_dengal}
\end{equation} 

\noindent for the \GEMS\ sample, computed by each method.  Histograms showing the
distribution of \dengal\ values obtained are shown in
Figure~\ref{fig_dengal_hist}.  For comparison, we calculated the expected mean
galaxy density within \rfh\ from the average Sloan Digital Sky Survey (SDSS)
luminosity function of \citet{blanton03}, by integrating their Schechter function
down to our luminosity cut, giving a predicted mean galaxy density

\begin{equation}
\dengal(\mathrm{pred}) ~ = ~ \frac{500}{\omegam} \int^{\infty}_{\Lcut} \phi(L) dL ~ = ~ 27 ~ \pMpccu,
\label{eqn_dengalpred}
\end{equation} 

\noindent where \omegam\ is the density of ordinary matter, as a fraction of the
critical density, and is assumed to be 0.3.  The predicted mean galaxy density is
marked in Figures~\ref{fig_dengal_hist} and \ref{fig_dengal_LB}.   

It can be seen that using the \TX-based estimate of \rfh, the expected density is
close to the median of our derived values (\dengal\ = 25), whilst the
\sigmav-based estimates lead to a much wider scatter in \dengal, with some values
(mostly for very poor groups) over an order of magnitude higher than expected.
The standard deviations in the \dengal\ distributions for the \TX\ and
\sigmav-based estimates are 21 \pMpccu\ and 115 \pMpccu, respectively.

In Figure~\ref{fig_dengal_LB}, we compare the derived densities for the two
methods, plotted against total optical luminosity of the groups.  It can be seen
that not only does the \TX-based analysis give a smaller scatter in \dengal, but
that the inferred densities show no discernable  trend with \LB. It seems that
any effects of non-self-similar entropy scaling are not acting to systematically
raise \TX\ in lower mass systems, otherwise we would see a trend towards lower
apparent \dengal\ in the poorest systems.

The good agreement between our observed and expected galaxy densities, appears to
conflict with the conclusions of the \citet{sanderson03a} analysis, discussed
above, since a 40\% overestimate in \rfh\ would lead to our densities being
underestimated by a factor of 2.7, which does not seem consistent with the
results shown in Figure~\ref{fig_dengal_LB}.  Moreover a number of recent studies
(e.g. \citealt*{nevalainen00a}; \citealt{sato00,finoguenov01b}), have indicated
that the $M$-\TX\ relation for clusters and groups is significantly steeper than
the self-similar ($M$ $\propto$ \TX$^{1.5}$) relation, although \citet*{allen01d}
find a relation consistent with self-similarity from a high-quality \Chandra\
study of a small sample of rich, relaxed clusters, with a 2500 overdensity
radius.  If the $M$-\TX\ relation {\it does} have a slope steeper than 1.5, then
it follows that the \TX$^{0.5}$ scaling for \rfh\ is too flat, and will
presumably tend to overestimate the radius in the group regime.

It should be noted that the good agreement between our median value of galaxy
density and the expected value, assumes that galaxies are not biased relative to
mass on group scales. Recent results from the 2dF galaxy redshift survey
\citep{verde02} suggest that light is essentially unbiased relative to mass on
scales larger than 5 \Mpc. However, as we discussed earlier in this section,
there is some evidence from both observations and simulations
\citep[e.g.][]{bahcall02,benson00} that light may be biased on smaller scales. A
recent study of the K-band mass-to-light ratio \citep*{lin03}, based on 2MASS
luminosities which provide a measure of the stellar mass relatively unaffected
by recent star-formation history, coupled with mass estimates based on X-ray
temperatures, found that $M/L_K$ dropped by a factor $\sim$ 2 over the mass range
$M$(\rfh) $\sim$ 10$^{15}$ \Msol\ to $\sim$ 10$^{14}$ \Msol.  It may therefore be
that the apparent good agreement between our derived galaxy densities and the
prediction from the universal mean is fortuitous, and that our application of
Equation~\ref{eqn_r500_TX} leads to an overestimate of \rfh, and hence an
underestimate of \dengal, which cancels the factor of $\sim$ 2-3 by which these
densities are biased upward relative to the Universal mean. Derivation of
reliable X-ray masses with \XMM\ may eventually resolve this issue.

Our conclusion is that the use of Equation~\ref{eqn_r500_TX} appears to provide
a more stable estimate of \rfh\ than the use of a \sigmav-based scaling relation,
although there is some danger that all our radii may be somewhat overestimated by
the \TX$^{0.5}$ formula.  Where no value of \TX\ is available, we adopt an
estimate based on the scaling of  scaling of mass with \LB, using
Equation~\ref{eqn_r500_LB}.  In the latter case, an iterative process is
involved, since \LB\ depends upon the group membership within \rfh, whilst \rfh,
in turn, depends on \LB.

%%%%%%%%%%%%%%%%%%%%%%%%%%%%%%%%%%%%%%%%%%%%%%%%%%%%%%%%%%%%%%%%%%%%%%%%%%%%%%%%%
%%%%%%%%%%%%%%%%%%%%%%%%%%%%%%%%%%%%%%%%%%%%%%%%%%%%%%%%%%%%%%%%%%%%%%%%%%%%%%%%%

\section{Global Scaling Relations}
\label{sec_scale}

Scaling relations between the major global parameters of galaxy systems -- \LX,
\TX, \sigmav\ and \LB\ -- are of great interest in studying the extent to which
groups are related to clusters through simple similarity scalings. Previous work
\citep{helsdon00a,helsdon00b,xue00,mulchaey00} has shown that even where scaling
relations follow self-similar forms for rich clusters, this behaviour does not
usually extend to the group regime. In terms of X-ray properties, we are
interested here in the scaling properties of the hot IGM, so we concentrate
primarily on those systems designated as having ``group'' emission. In order to
compare with cluster properties, we make use of the sample of \citet{horner01},
based on a homogeneous analysis of data from the \ASCA\ observatory.  We remove
cool (\TX\ $<$ 2 \kev), low luminosity (\LX\ $<$ 2$\times$10$^{43}$ \ergps)
groups from the systems studied by Horner, to give a sample of 230 clusters, 105
of which have velocity dispersions available.  The X-ray luminosities for these
systems were corrected to \rth\ by Horner, assuming a standard $\beta$-model with
$\beta$ = 0.67, and core radius which scales as \LX$^{0.28}$, following the
empirical result of \citet{bohringer00}.  We have used the same model to correct
each of these cluster luminosities instead to \rfh, for comparison with our group
values.  Temperatures for these clusters have been derived by Horner from \MEKAL\
model fits to integrated \ASCA\ spectra from within some extraction radius.
Although no attempt was made to remove any emission from a central cool core,
we have seen that this has had only a small effect on our own temperatures, so
that the two samples can reasonably be compared.  Velocity dispersions for a
subset of his clusters were collected by Horner from the literature. These will
therefore be derived in a heterogeneous fashion. However, for all but three of
the clusters, these velocity dispersions are based on more than 10 redshifts.

%%%%%%%%%%%%%%%%%%%%%%%%%%%%%%%%%%%%%%%%%%%%%%%%%%%%%%%%%%%%%%%%%%%%%%%%%%%%%%%%%

\subsection{The \LX-\TX\ Relation}
\label{subsec_LT}

Strong correlations exist between X-ray luminosity and both gas temperature and
velocity dispersion, reflecting the fact that deeper potential wells generally
contain more hot gas.  It has been clear for many years that the \LX-\TX\
relation for clusters does not follow the \LX\ $\propto$ \TX$^2$ law expected for
self-similar systems radiating bremsstrahlung X-rays.  Most authors
(e.g. \citealt*{white97b}; \citealt{arnaud99}) have found logarithmic slopes
close to 3 in the cluster regime, though attempts to remove the effects of
central cooling flows \citep{allen98,markevitch98} have produced rather flatter
relations.  Studies of the relation for galaxy groups have mostly found
considerably steeper slopes.  \citet{helsdon00a,helsdon00b} obtained a slope of
4.9$\pm$0.8 for a sample of X-ray bright loose groups, and 4.3$\pm$0.5 for a
larger sample (36 systems) including both loose and compact groups.
\citet{xue00}, found a slope of 5.6$\pm$1.8 from data for 38 groups drawn from
the literature.

\begin{figure*}
  \begin{minipage}{241pt} 

    \includegraphics[height=\linewidth,angle=270]{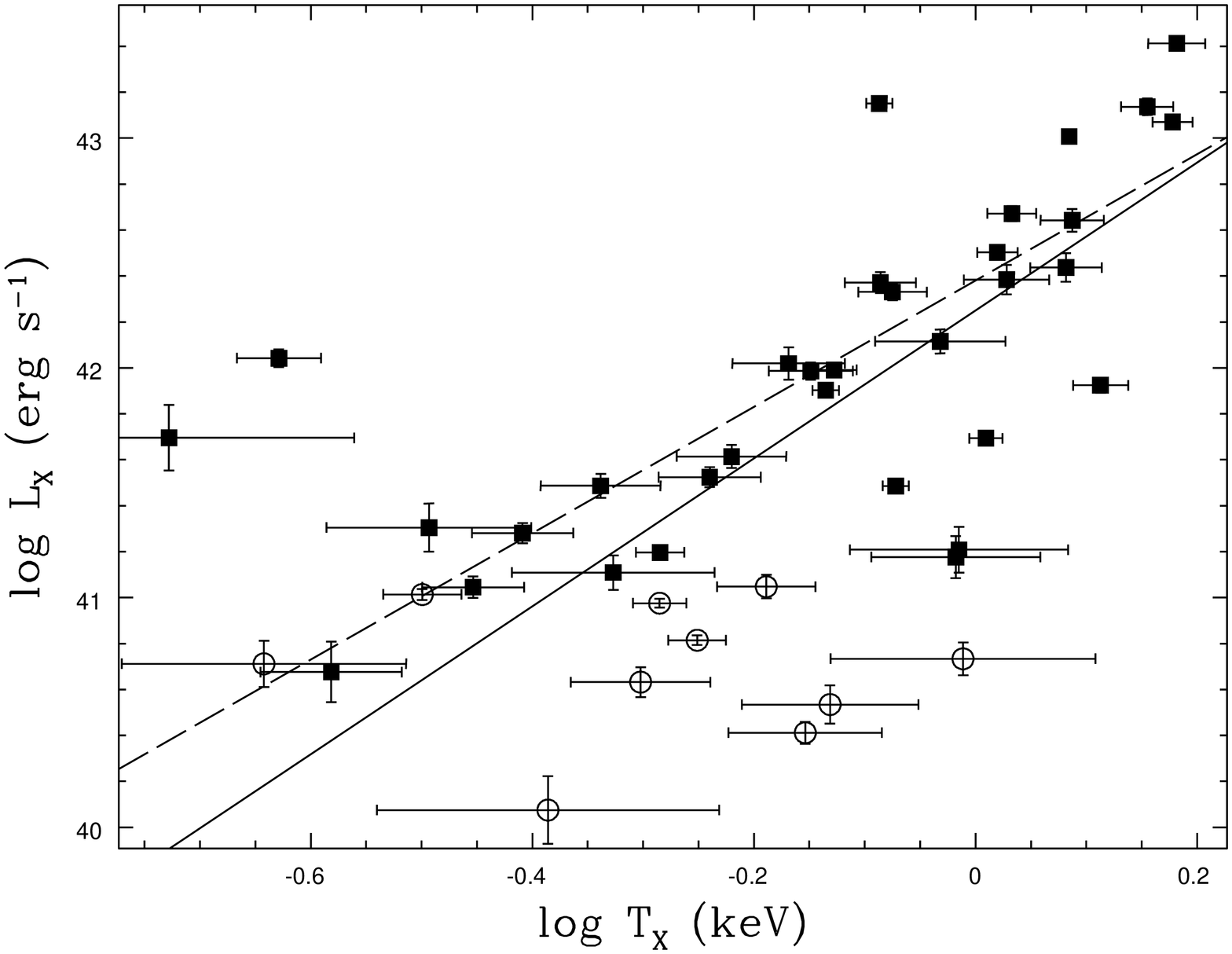}
    \caption{The relationship between \LX\ and \TX\ for the \GEMS\ groups.  Filled
             squares represent the G-sample and open circles the H-sample.  The
             solid line represents an unweighted orthogonal regression fit to all
             points, and the dashed line to the G-sample only.}
    \label{fig_LX_TX}

  \end{minipage}\hspace{18pt}
  \begin{minipage}{241pt}

    \includegraphics[height=\linewidth,angle=270]{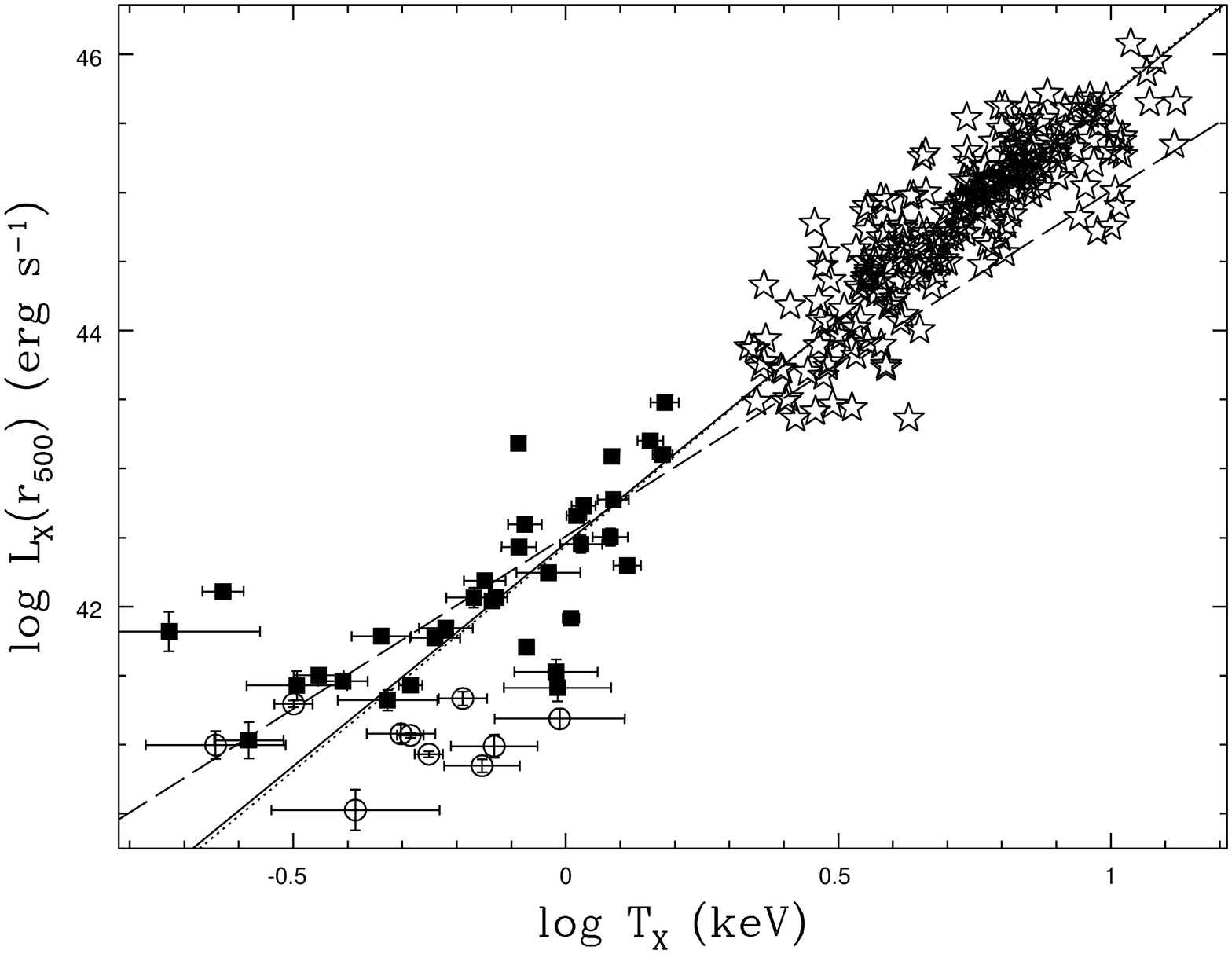}
    \caption{The relationship between \LXrfh\ and \TX\ for the \GEMS\ groups
             (squares and circles) and the Horner clusters (stars). The dashed
             line represents a fit to the G-sample, dotted line to the
             clusters, and solid line to clusters plus the G-sample.  The
             H-sample is excluded from the fitting, but is plotted for
             comparison.}
    \label{fig_LXrfh_TX_comp}

  \end{minipage}\\
  \begin{minipage}{241pt} 

    \includegraphics[height=\linewidth,angle=270]{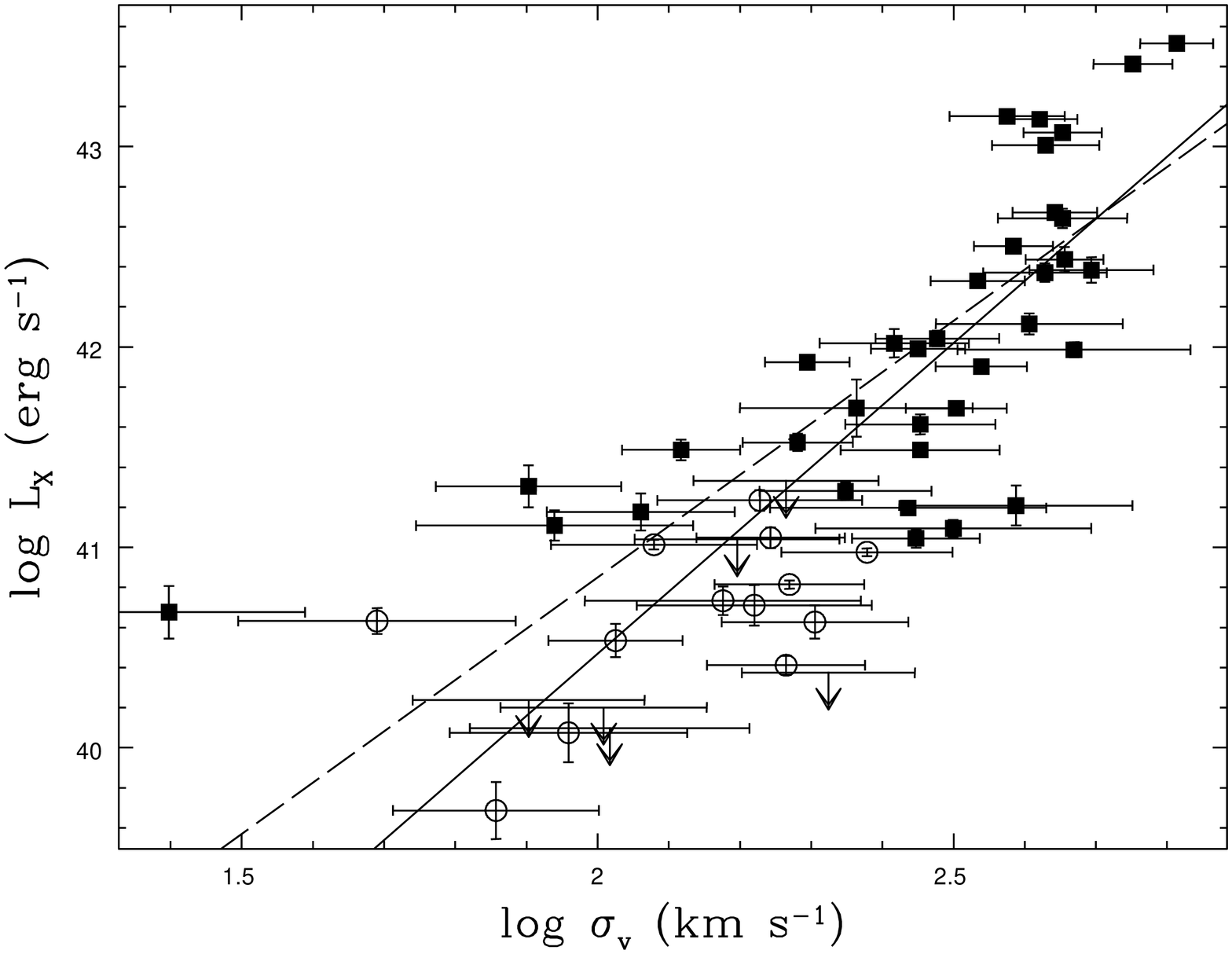}
    \caption{The relationship between \LX\ and \sigmav\ for the \GEMS\ groups.
             Filled squares represent the G-sample, open circles the H-sample
             and arrows represent upper-limits from non-detections.  The solid
             line represents an unweighted orthogonal regression fit to all
             points, and the dashed line to the G-sample only.} 
    \label{fig_LX_sigma}

  \end{minipage}\hspace{18pt}
  \begin{minipage}{241pt}

    \includegraphics[height=\linewidth,angle=270]{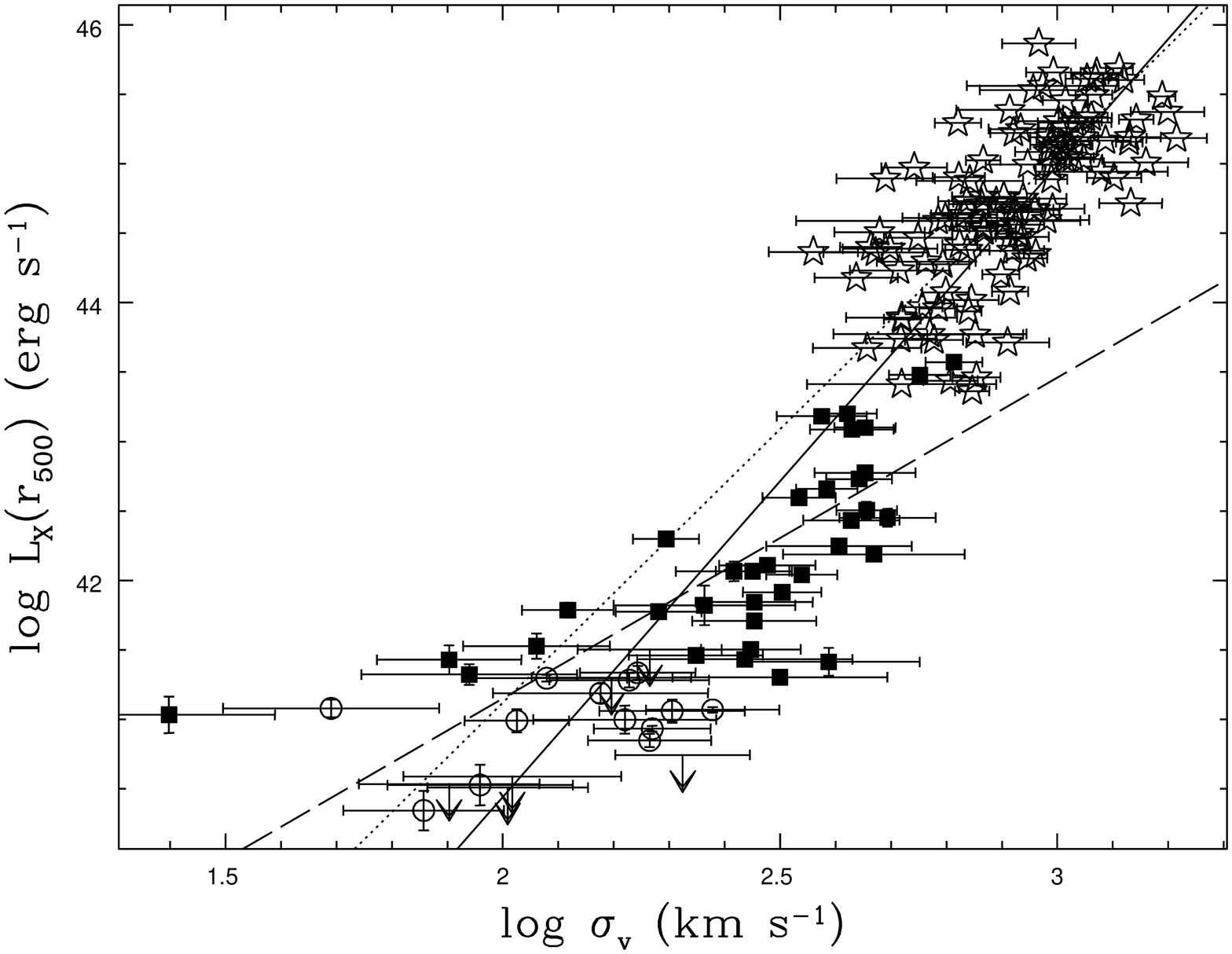}
    \caption{The relationship between \LXrfh\ and \sigmav\ for the \GEMS\ groups
             (squares, circles and arrows) and the Horner clusters (stars). The
             dashed line represents a fit to the G-sample, dotted line to the
             clusters, and solid line to clusters plus the G-sample. The H-sample
             and non-detections are excluded from the fitting but are plotted for
             comparison.}
    \label{fig_LXrfh_sigma_comp}

  \end{minipage}
\end{figure*}

Our result from the \GEMS\ sample, shown in Table~\ref{tab_relations} and Figure
\ref{fig_LX_TX}, for the subsample of 45 groups with fitted temperatures, is
significantly flatter than the above group results, and appears close to the
slope seen in clusters.  This is especially striking if we restrict our
attention to G-sample systems (slope = 2.75$\pm$0.46) and flattens still further
(2.50$\pm$0.42) if we use \LX\ values extrapolated to \rfh.  In
Figure~\ref{fig_LXrfh_TX_comp}, we plot the G-sample systems, with extrapolated
luminosities, alongside the Horner cluster sample.  The parameters for the three
trend lines are given in Table~\ref{tab_comparison}, and that for the G-sample
groups is actually somewhat {\it flatter} than the cluster relation.  Can we
conclude from this that the earlier results of a steeper \LX-\TX\ relation in
groups were incorrect?

\begin{table*}
\begin{center}
\scriptsize
\begin{tabular}{@{}ll@{}r@{\hspace{0.5cm}}rr@{}r@{\hspace{0.5cm}}rr@{}r@{\hspace{0.5cm}}rr@{}rr@{}}
\hline

%%%%%%%%%%%%%%%%%%%%%%%%%%%%%%%%%%%%%%%%%%%%%%%%%%%%%%%%%%%%%%%%%%%%%%%%%%%%%%%%%

\multicolumn{2}{@{}l}{Relation}  & &  \multicolumn{2}{c}{Groups}           & &  \multicolumn{2}{c}{Clusters}          & &  \multicolumn{2}{c}{All}               & &  Figure                      \\

\cline{4-5} \cline{7-8} \cline{10-11}                                                           

             &                   & &  Slope            &  Intercept         & &  Slope            & Intercept          & &  Slope            &  Intercept         & &                             \\

\hline

log \LXrfh\  & log \TX           & &  2.50 $\pm$ 0.42  &  42.51 $\pm$ 0.09  & &  3.26 $\pm$ 0.12  &  42.44 $\pm$ 0.10  & &  3.23 $\pm$ 0.10  &  42.46 $\pm$ 0.07  & &  \ref{fig_LXrfh_TX_comp}    \\
log \LXrfh\  & log \sigmav       & &  2.31 $\pm$ 0.61  &  36.53 $\pm$ 1.54  & &  3.94 $\pm$ 0.33  &  33.24 $\pm$ 0.97  & &  4.55 $\pm$ 0.25  &  31.34 $\pm$ 0.72  & &  \ref{fig_LXrfh_sigma_comp} \\
log \sigmav\ & log \TX           & &  1.15 $\pm$ 0.26  &  2.60  $\pm$ 0.03  & &  0.78 $\pm$ 0.05  &  2.36  $\pm$ 0.04  & &  0.71 $\pm$ 0.05  &  2.43  $\pm$ 0.03  & &  \ref{fig_sigma_TX_comp}    \\

%%%%%%%%%%%%%%%%%%%%%%%%%%%%%%%%%%%%%%%%%%%%%%%%%%%%%%%%%%%%%%%%%%%%%%%%%%%%%%%%%

\hline
\end{tabular}
\end{center}

%%%%%%%%%%%%%%%%%%%%%%%%%%%%%%%%%%%%%%%%%%%%%%%%%%%%%%%%%%%%%%%%%%%%%%%%%%%%%%%%%

\caption
{\label{tab_comparison}
A comparison of scaling relations betweeen groups and clusters.  Group relations
are derived from the G-sample and cluster relations from the sample of
\citet{horner01}.}

%%%%%%%%%%%%%%%%%%%%%%%%%%%%%%%%%%%%%%%%%%%%%%%%%%%%%%%%%%%%%%%%%%%%%%%%%%%%%%%%%

\end{table*}

To explore the origin of the differences from our own earlier results, we
examined the subset of 16 of our \GEMS\ groups which overlap with the sample of 24
groups studied by \citet{helsdon00a}.  Our regression line through these 16
systems has a slope of 4.3$\pm$0.9, close to the result of \citet{helsdon00a}.
The use of luminosities extrapolated to \rfh\ flattens this regression line only
slightly, to a slope of 3.7$\pm$1.0.  These tests strongly suggest that the
flatter slope from our G-sample systems is primarily related to differences in
the group sample used here, rather than in the analysis techniques employed.  The
systems studied by \citet{helsdon00a} were selected on the basis of significant
X-ray flux, and therefore constitute a sample of X-ray bright groups, whereas the
\GEMS\ sample was deliberately designed to cover a wider spectrum of X-ray
properties, as discussed in Section~\ref{sec_sample}.  As a result, our present
sample includes a much larger number of cool groups (\TX\ $<$ 0.7 \kev) than that
of \citet{helsdon00a}, and most other previous studies. Two groups in particular,
HCG\,10 and NGC\,3557, have \TX\ $<$ 0.25 \kev, and yet have moderately high X-ray
luminosities.

In the present study, we are also pushing closer to the statistical limits of
what can be achieved with \ROSAT\ data. It is well-established that there is
considerably larger real scatter in the scaling relations for galaxy groups than
is seen in clusters.  This scatter introduces three sources of bias into our
regression process.  Firstly, the result of truncating this scattered trend at
low \LX\ (since we will either reject systems with very low \LX\ as galaxy halo
sources, or fail to detect them altogether) will be to flatten the fitted
relation. Secondly there is a `logarithmic bias' whereby the scatter in log \TX\
(which dominates the statistical scatter about the trend line) will be asymmetric
(if the scatter in \TX\ is fairly symmetric) with larger scatter towards low log
\TX.  Since the statistical errors are largest in systems with lowest \LX, this
will also tend to flatten the regression line.  Thirdly, there is an additional
bias which couples with the scatter in \TX.  At temperatures towards the bottom
of the \ROSAT\ bandpass, and close to the absorption cut-off due to interstellar
gas and dust, the unabsorbed bolometric flux corresponding to a given \PSPC\
count rate rises quite sharply as \TX\ falls (e.g. for an absorbing column of
4$\times$10$^{20}$ \pcmsq\ it rises by 67\% as \TX\ falls from 0.6 \kev\ to 0.3
\kev).  Thus it follows that points in the \LX-\TX\ plane which scatter down in
temperature, will also scatter up in \LX, which further magnifies the flattening
effect discussed above.  Since all three of these biases are related to the large
scatter in the data, we investigated the effect of clipping the outliers.
Iteratively discarding G-sample points which lie more than 2 sigma from the
regression line does indeed steepen the fitted slope, from 2.5 to 3.0 where 8 of
the 33 points are clipped in this analysis.  Note, however, that this steeper
slope is still fully consistent with the cluster \LX-\TX\ relation.

One final source of bias which becomes important when pushing the sample down to
very poor groups, is the impact of contamination from point sources which are
unresolved by \ROSAT.  Since the fractional contribution of such sources will
tend to be larger in the lowest luminosity groups, it will tend to flatten the
\LX-\TX\ relation.  \citet*{helsdon04} find, from a comparison of \Chandra\ and
\ROSAT\ results for two very cool groups, that the level of unresolved point
source contribution to the diffuse flux derived from the \PSPC\ analysis is
30-40\%.  So this effect is smaller than the correction (a factor 2-3 for such
poor systems) arising from extrapolation to \rfh, and works in the opposite
direction.

A further difference in our present analysis, compared to earlier studies, is
that we separate off systems in which the X-ray emission appears to be related to
a central galaxy, rather than to the group as a whole.  In studying the
properties of groups, for comparison with clusters, this seems the appropriate
thing to do.  As can be seen from Figure~\ref{fig_LX_TX} and
Table~\ref{tab_relations}, these halo sources do fall at the bottom of the
\LX-\TX\ plot, and their inclusion steepens the fitted \LX-\TX\ relation.
However, this will have had little impact on the earlier work of
\citet{helsdon00a}, since their X-ray bright sample contained very few objects
which might be classified as galaxy halo sources.

In summary, we conclude that although the \LX-\TX\ relation obtained from our
G-sample groups is close to continuous with the cluster relation, albeit with
increased scatter, this may be a misleading result, since we have identified a
number of biases, all of which work towards flattening the fitted \LX-\TX\
relation in the low \TX, low \LX\ regime.  Extension of the data towards lower
\LX\ would be necessary to reduce these biases and establish definitively whether
the \LX-\TX\ relation does steepen in groups. This may ultimately be possible
with \XMM, but will not be  straightforward, in a regime where luminosities are
comparable to those of individual halos around early-type group member galaxies.
What is clear from our results, is that groups show a considerably larger real
scatter about the mean \LX-\TX\ trend than is seen in clusters -- spanning at
least a factor of 30 in \LX, at a given value of \TX, or a factor 3-4 in \TX\ at
given \LX.

%%%%%%%%%%%%%%%%%%%%%%%%%%%%%%%%%%%%%%%%%%%%%%%%%%%%%%%%%%%%%%%%%%%%%%%%%%%%%%%%%

\subsection{\LX-\sigmav\ relation}
\label{subsec_Lsig}

Our relationship between \LX\ and \sigmav\ (Figure~\ref{fig_LX_sigma}) has a
slope of 2.56$\pm$0.56 for the G-sample, which is flatter than the value of
4.5$\pm$1.1 found by \citet{helsdon00a}.  Although there is general agreement
that the \LX-\sigmav\ relation does not {\it steepen} in groups, unlike the
\LX-\TX\ relation, there is disagreement between studies
\citep[e.g][]{ponman96,mulchaey98,helsdon00a,mahdavi01} which find that groups
are consistent with the cluster-relation slope of $\approx$ 4, and those
\citep{mahdavi97,xue00,helsdon00b,mahdavi00} which find significantly flatter
relations in groups.  The result from the \GEMS\ sample is clearly
(Figure~\ref{fig_LXrfh_sigma_comp}) substantially flatter than the cluster trend,
and has a slope in good agreement with that of \citet{helsdon00b} (2.4$\pm$0.4)
and \citet{xue00} (2.35$\pm$0.21).  Extrapolation of the luminosity to \rfh\
results, as expected, in a slightly lower slope, of 2.31$\pm$0.62
(Table~\ref{tab_comparison}).

As with the \LX-\TX\ relation, biases are at work which will tend to lead to
some spurious flattening of our regression results.  Since the groups with lowest
\LX\ tend to be the poorest, and hence to have the largest fractional errors in
\sigmav, three of the four sources of bias discussed in the last section (the
truncation and logarithmic biases, and point source contamination) will also
apply to the \LX-\sigmav\ relation.  Nonetheless, there are some groups in our
sample with remarkably low velocity dispersion, which appear to show group-scale
emission.  This is hard to understand, since such X-ray emission presumably
implies that a group is collapsed, if not virialised, and as \citet{mamon94} has
argued, the requirement that collapsed systems should have some minimum
overdensity sets a lower bound (at a given mass or radius) to the velocity
dispersion which they can have.  This bound is $\sim$ 100-200 \kmps\ for poor
groups.  A study with \Chandra\ \citep{helsdon04}, of two of the low \sigmav\
systems in our sample, NGC\,1587 and NGC\,3665, has confirmed that the diffuse
X-ray emission identified by \ROSAT\ is not grossly misleading, although point
source contamination and inaccurate spectral characterisation can lead to
overestimation of \LX\ by $\sim$ 30-40\%.

The fact that groups with such low values of \sigmav\ can contain a significant
IGM, with properties which accord reasonably with those of other groups, strongly
suggests that the observed values of \sigmav\ are not reflecting the depth of the
potential well in the way one expects.  We will return to this in the following
section.

%%%%%%%%%%%%%%%%%%%%%%%%%%%%%%%%%%%%%%%%%%%%%%%%%%%%%%%%%%%%%%%%%%%%%%%%%%%%%%%%%

\subsection{\sigmav-\TX\ and \betaspec}
\label{subsec_sigT}

\begin{figure}

  \includegraphics[height=\linewidth,angle=270]{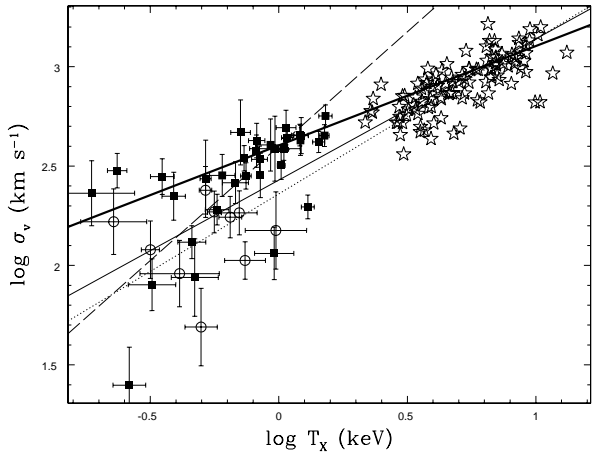}
  \caption{The relationship between \sigmav\ and \TX\ for the \GEMS\ groups
           (squares and circles) and the Horner clusters (stars). The dashed line
           represents a fit to the G-sample, dotted line to the clusters, and
           solid line to clusters plus the G-sample. The H-sample is excluded
           from the fitting but is plotted for comparison and the bold line
           represents \betaspec\ = 1.}
  \label{fig_sigma_TX_comp}

\end{figure}

A number of previous studies have found that the relationship between velocity
dispersion and gas temperature departs slightly from the virial theorem
expectation (\sigmav\ $\propto$ \TX$^{0.5}$) in clusters
(\citealt*{bird95}; \citealt{girardi98}; \citealt*{wu99}).  The evidence for
groups is more controversial, with some authors \citep[e.g.][]{xue00,mulchaey00}
finding that groups fall on the cluster trend, and others
\citep{helsdon00a,helsdon00b} finding that the relation steepens in groups with
\TX\ $<$ 1 \kev, to a slope $\ge$ 1.  Our result, presented in
Figure~\ref{fig_sigma_TX_comp}, shows that there is a great deal of
non-statistical scatter in the groups, in addition to the large statistical
errors in both \TX, and especially \sigmav, in the poorest systems.  This appears
to be the origin of the controversy over whether the relation does or does not
steepen in the group regime. Formally, we find that it does, with a best fit
slope to the \GEMS\ G-sample systems of 1.15$\pm$0.29.  However it is clear that
the best fit line to the combined group+cluster sample (with a slope of
0.71$\pm$0.05) passes through the centre of the scatter of group points, and also
represents the trend in the cluster regime quite adequately.  This slope is
somewhat steeper than that found in most previous studies. 

Comparison with the line \betaspec\ = 1 in Figure~\ref{fig_sigma_TX_comp}, shows
that many of the G-sample groups are actually consistent with this energy
equipartition line (as are many clusters), but that there is a significant
subsample of points which scatter well below it.  These systems have velocity
dispersions typically a factor 3 below what would be expected for their X-ray
temperatures. All of the groups with \sigmav\ \ltsim\ 100 \kmps\ fall into this
category.

\begin{figure}

  \includegraphics[height=\linewidth,angle=270]{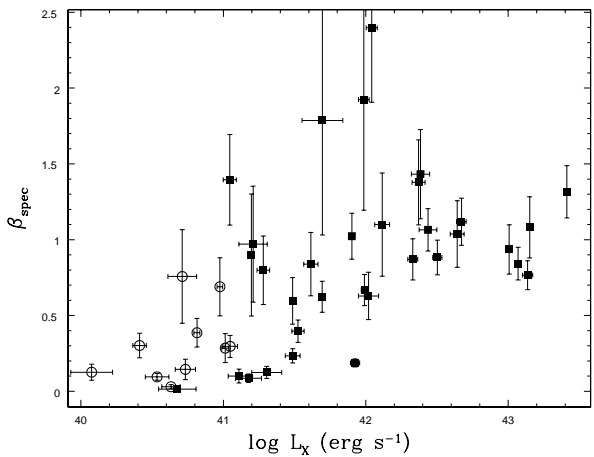}
  \caption{The relationship between \betaspec\ and \LX.  Filled squares
           represent the G-sample and open circles the H-sample.}
  \label{fig_betaspec_LX}

\end{figure}

Figure~\ref{fig_betaspec_LX} shows clearly that high X-ray luminosity (\LX\ $>$
10$^{42}$ \ergps) groups have \betaspec\ $\sim$ 1, whilst groups with \LX\
\ltsim\ 10$^{41.5}$ \ergps\ scatter widely in \betaspec, spanning the range
\betaspec\ = 0.1-1.0. The evidence discussed in Section~\ref{sec_radius} above,
led us to the conclusion that \TX\ gives a much more reliable measure of system
mass and radius than does \sigmav.  It follows from this that the ``problem'' in
some of the poorest systems, which leads to their remarkably low values of
\betaspec, lies not with \TX, but with \sigmav. Some of these groups, from both G
and H-samples, have extremely low velocity dispersions, giving them low
\betaspec, and also small radii (and hence high \dengal) when these are
calculated on the basis of \sigmav.

As can be seen from Table~\ref{tab_optical}, most of these low \sigmav\ systems
have $\le$ 6 galaxy redshifts available for the computation of \sigmav.  Under
these circumstances, a number of biases may affect the derived value of \sigmav\
\citep{helsdon04b}.  We have already corrected for a statistical bias which
results if one uses the normal unbiased estimator for $\sigmav^2$, and then takes
the square root to obtain \sigmav\ (which is then {\it not} unbiased).  This is
the origin of the term $3/2$ (rather than 1) in the denominator of
Equation~\ref{eqn_sigmav}.  However, another downward bias may arise if (as is
normally the case for X-ray bright groups) one of the galaxies is at rest at the
bottom of the group potential well.  This galaxy will not contribute to the sum
of squared deviations from the mean velocity, but will be included in the
denominator.  For a group with only four members, this reduces \sigmav\ by over
20\%.  However, such statistical biases cannot provide anything approaching the
large factors ($\sim$ 2-3) by which we infer that \sigmav\ has been reduced in
some of these groups.

Physical effects which might lead to such low values of \sigmav, are discussed in
greater detail in \citet{helsdon04}.  Observed galaxy velocity dispersions might
be reduced if (a) galaxy orbits decay due to dynamical friction, (b) orbital
energy is converted into internal energy of the galaxies via tidal heating, or
(c) orientation effects result in most of the galaxy velocity vectors for some
systems lying close to the plane of the sky.  The last of these has, perhaps, the
greatest potential to achieve really substantial reductions in observed
line-of-sight velocity dispersions.

%%%%%%%%%%%%%%%%%%%%%%%%%%%%%%%%%%%%%%%%%%%%%%%%%%%%%%%%%%%%%%%%%%%%%%%%%%%%%%%%%

\subsection{Scaling between \LB\ and X-ray properties}

\begin{figure}

  \includegraphics[height=\linewidth,angle=270]{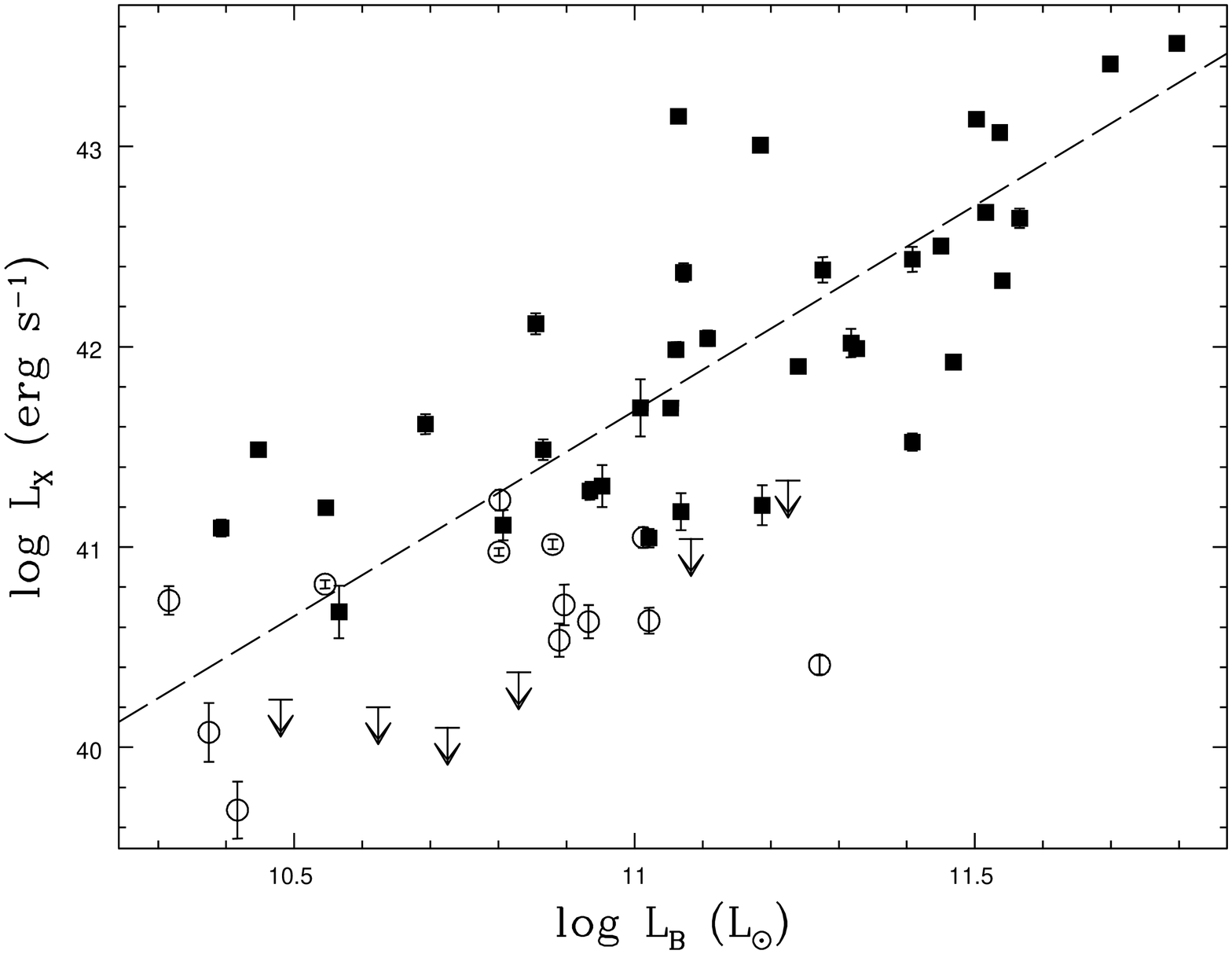}
  \caption{The relationship between \LX\ and \LB.  Filled squares represent the
           G-sample, open circles the H-sample and arrows represent upper-limits
           from non-detections.  The dashed line represents a fit to the
           G-sample.}
   \label{fig_LX_LB}

  \includegraphics[height=\linewidth,angle=270]{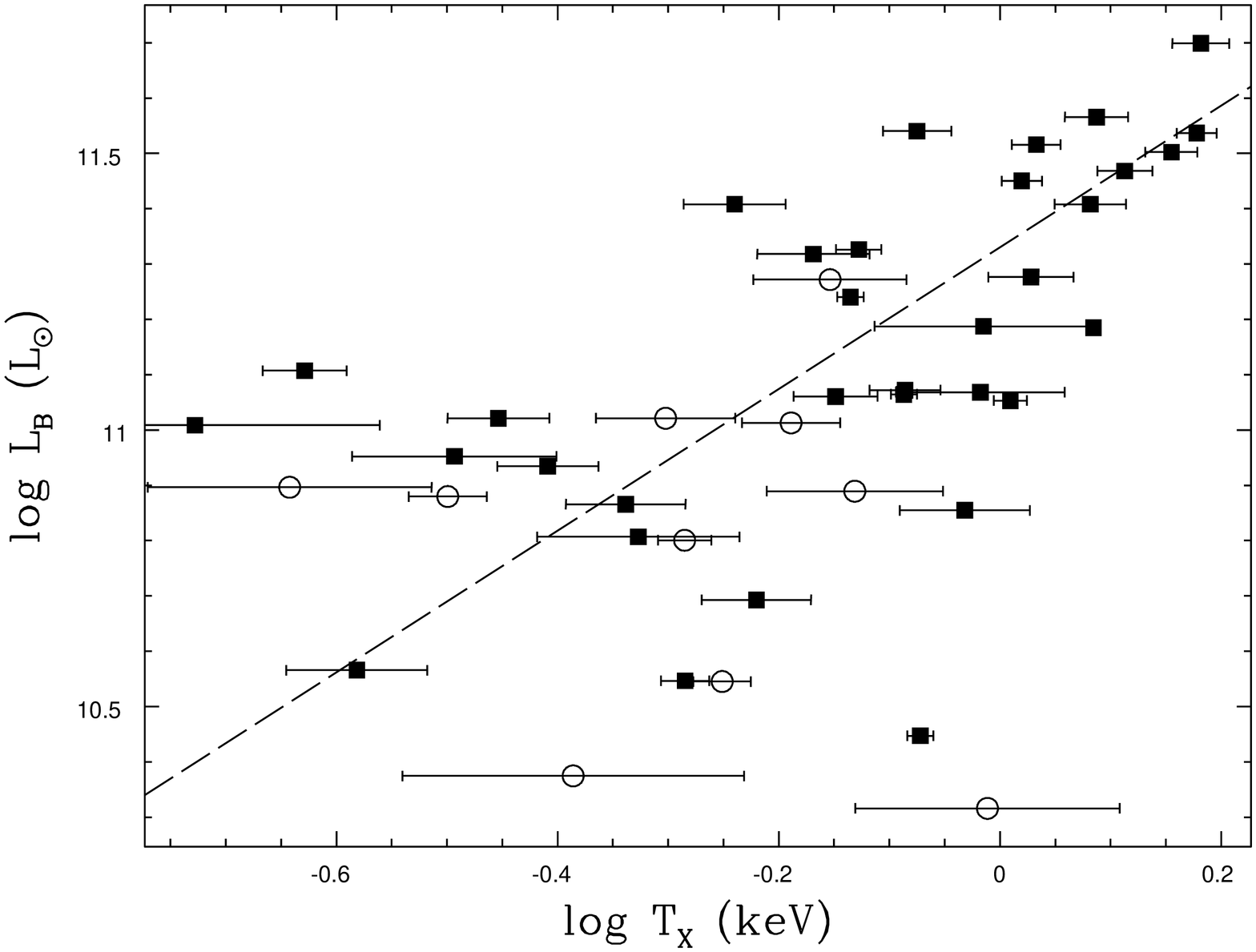}
  \caption{The relationship between \LB\ and \TX.  Filled squares represent the
           G-sample and open circles the H-sample.  The dashed line represents a
           fit to the G-sample.}
  \label{fig_LB_TX}

\end{figure}

In an ensemble of self-similar groups with constant mean density, the X-ray
luminosity would scale linearly with galaxy mass, and hence with optical
luminosity.  \citet{helsdon03b} found a much steeper relation, \LX\ $\propto$
\LB$^{2.69\pm0.29}$. Our fitted relation for our full sample of X-ray detected
systems is consistent with this (Table~\ref{tab_relations}).  The trend for the
G-sample systems (Figure~\ref{fig_LX_LB}) is somewhat flatter (2.05$\pm$0.21),
but still much steeper than the self-similar expectation.  This could be
explained in any of three ways:\\

\noindent (a) \fgas\ rises with system mass,\\
(b) gas density is higher in more massive systems,\\
(c) star formation is more efficient in poorer groups.\\

\noindent There are indications that all three of these factors may contribute,
but the results of \citet{ponman99} and \citet{sanderson03a} suggest that the
dominant effect is probably a reduction in gas density in the inner regions of
poor systems, relative to richer ones.

If \TX\ gives a reliable measure of group size ($R$ $\propto$ \TX$^{0.5}$) and
hence mass, as we are assuming here (since we use this relation to define our
value of \rfh), then \LB\ $\propto$ \TX$^{1.5}$ would result if star formation
efficiency were independent of system mass.  Our observed relation
(Figure~\ref{fig_LB_TX}) has a slope 1.28$\pm$0.19, somewhat flatter than the
slope of 1.64$\pm$0.23 found by \citet{helsdon03b}, but still consistent with the
value of 1.5 expected from self-similarity and constant SFE.  If, however, the
$M$-\TX\ relation has a slope steeper than the value of 1.5 (i.e. $M$ $\propto$
\TX$^{1.5}$) as has been found by a number of previous studies -- e.g. 
\citet{nevalainen00a}, \citet{finoguenov01b} and \citet{sanderson03a} suggest a
slope $\approx$ 1.8 -- then our \LB-\TX\ slope of 1.28 suggests a star formation
efficiency which is significantly higher in lower mass systems, as discussed in
Section~\ref{sec_radius}.

%%%%%%%%%%%%%%%%%%%%%%%%%%%%%%%%%%%%%%%%%%%%%%%%%%%%%%%%%%%%%%%%%%%%%%%%%%%%%%%%%
%%%%%%%%%%%%%%%%%%%%%%%%%%%%%%%%%%%%%%%%%%%%%%%%%%%%%%%%%%%%%%%%%%%%%%%%%%%%%%%%%

\section{$\beta$ and the Role of Heating}
\label{sec_beta}

\citet{mulchaey98} and \citet{helsdon00a} found that most X-ray bright groups
with data of sufficiently high quality, require two-component $\beta$-models to
adequately represent their surface brightness distributions.  The central
component is identified either with a halo associated with the brightest cluster
galaxy (BCG) \citep{mulchaey98}, or a central group cooling region
\citep{helsdon03b}, whilst the outer component clearly arises from a group-scale
intergalactic medium.  \citet{helsdon00a} investigated the effect of fitting a
single component $\beta$-model to a group where a two-component model is more
appropriate, and found that the value of \betafit\ obtained with a
single-component model typically scattered by $\sim$ 0.1 (but {\it in extremis}
by up to 0.3), either upward or downward from the value for the group component
given by a two-component fit. We therefore place more credibility in the
two-component results from our present fits.

The {\it largest} value of \betafit\ obtained for any of the \GEMS\ groups is 0.58,
and the median value from our G-sample fits is 0.45, in good agreement with the
value of 0.46 derived by \citet{helsdon00a}.  This is clearly well below the
typical value, \betafit\ $\approx$ 0.67, found in rich clusters
(\citealt{arnaud99}; \citealt*{mohr99}).  A trend towards lower values of
\betafit\ in poorer clusters has been  reported by a number of authors. However
\citet{mohr99} found that this trend disappeared when the systems in their study
(which all had \TX\ \gtsim\ 2 \kev)  were fitted with two-component models, and
hence concluded that the effect was spurious.  This explanation cannot account
for the low values of \betafit\ obtained here, nor in the studies of
\citet{helsdon00a} or \citet{mulchaey03}, since two-component fits are employed
wherever possible, and found to give \betafit\ $\sim$ 0.4-0.5. 

However, some uncertainty arises due to the fact that X-ray emission in groups
can typically be traced only to a modest fraction of the radii to which they
should be virialised \citep{mulchaey00}.  It has been argued on the basis of
simulations \citep*{navarro95}, and analytical models \citep{wu02}, that gas
density profiles steepen progressively, and hence that low values of \betafit\
naturally arise when profiles are fitted only within r $\sim$ 0.3\rth.  However
there is some evidence that this is unlikely to explain the low values of
\betafit\ we observe.  \citet{sanderson03a} examined the effects of truncating
the data for galaxy clusters when fitting $\beta$-models, and found little
evidence for any significant drop in \betafit. Moreover observed radial surface
brightness profiles are generally found to be modelled remarkably well by simple
power laws, outside the central core.  \citealt*{vikhlinin99c} studied a sample
of rich clusters out to r $\sim$ \rth\ with the \ROSAT\ \PSPC, and found only a
slight steepening (by $\Delta\beta$ $\approx$ 0.05) in the outer regions of
typical clusters.  \citet{rasmussen04} were able to trace the emission from two
rich groups (\TX\ $\sim$ 2 \kev) out to \rfh, and found no evidence for
steepening of the profiles beyond the simple $\beta$-model fits.

\begin{figure}

  \includegraphics[height=\linewidth,angle=270]{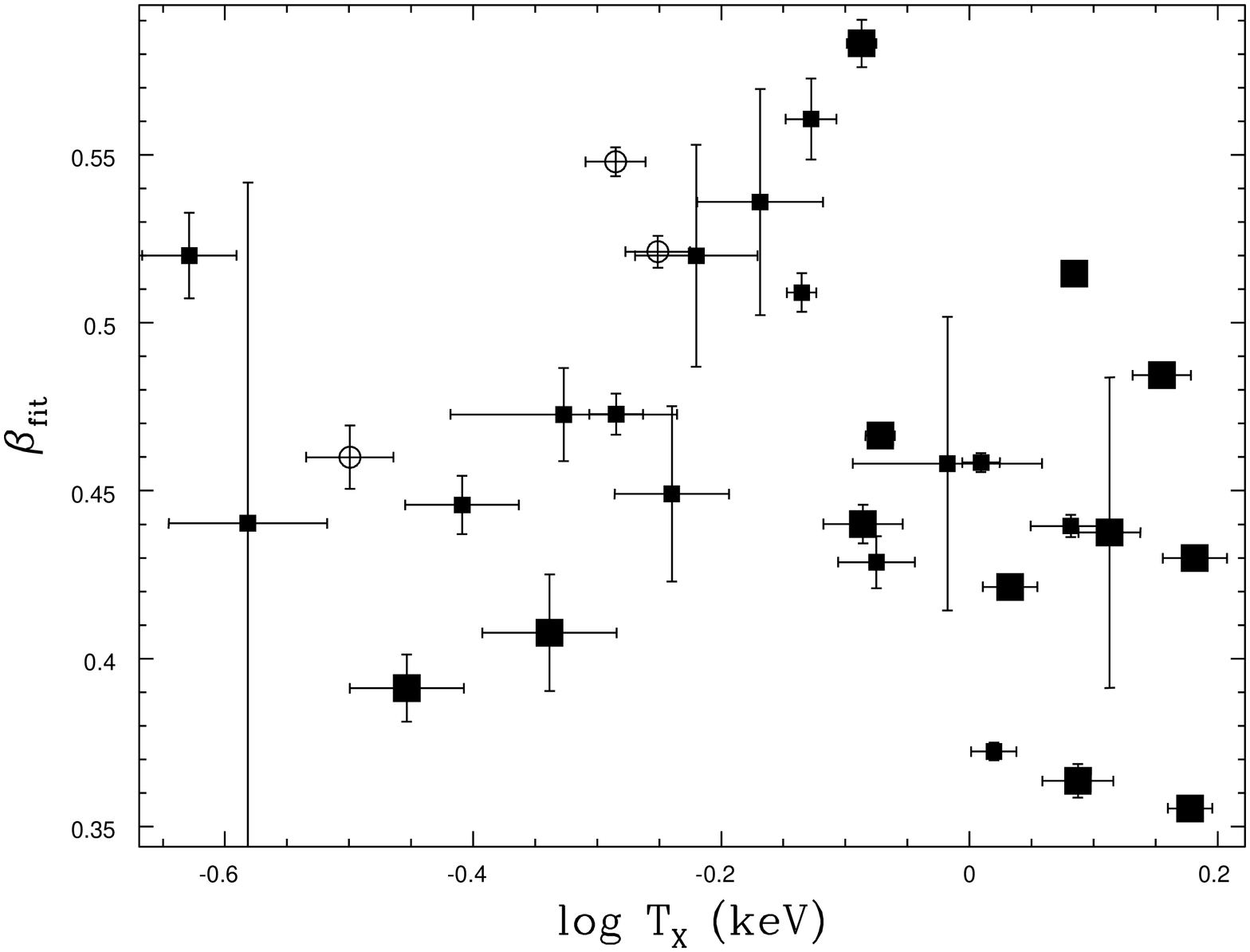}
  \caption{The relationship between \betafit\ and \TX.  Filled squares
           represent the G-sample and open circles the H-sample.  Larger points
           denote systems in which a two-component $\beta$-model was fitted.}
   \label{fig_betafit_TX}

  \includegraphics[height=\linewidth,angle=270]{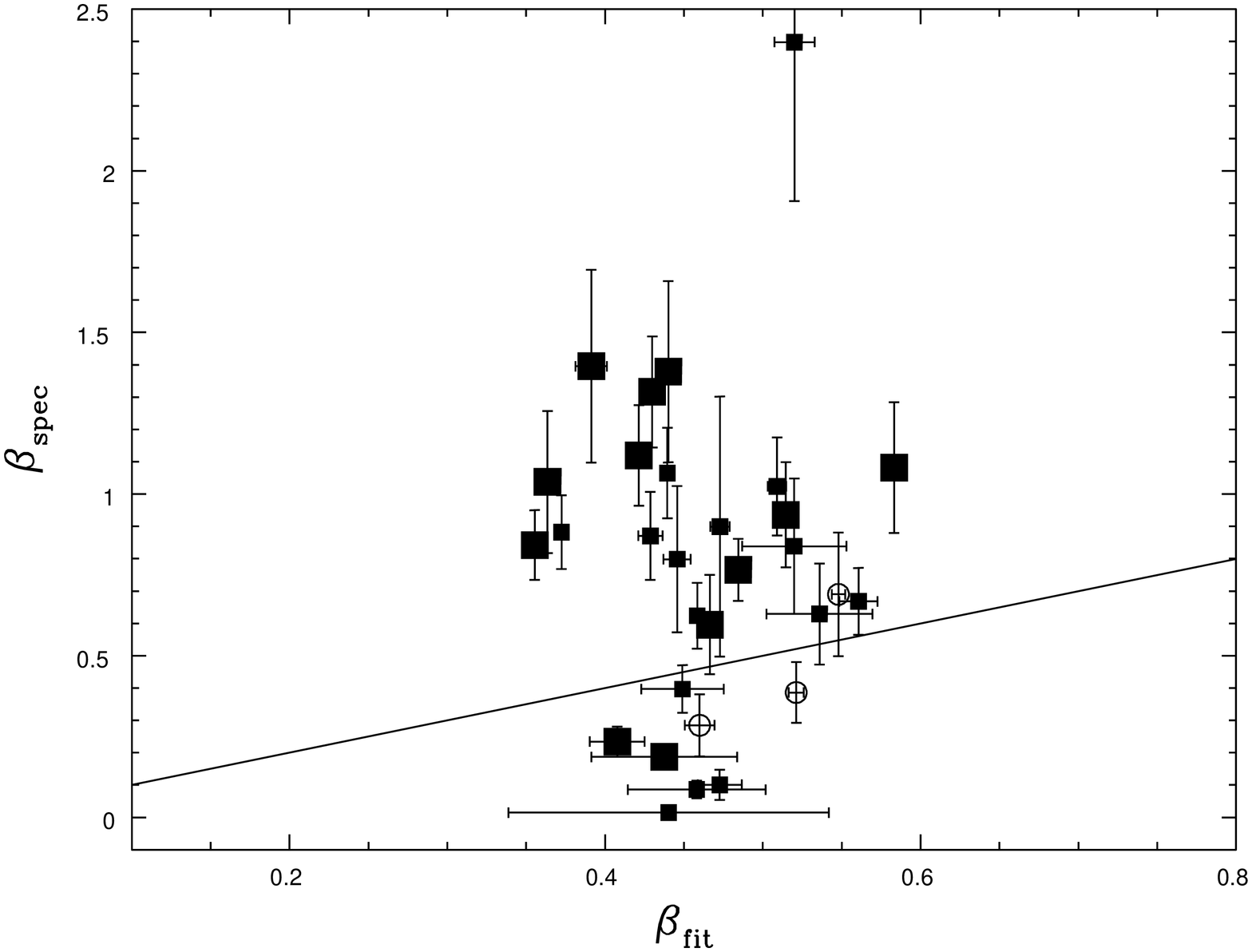}
  \caption{The relationship between \betaspec\ and \betafit.  Filled squares
           represent the G-sample and open circles the H-sample.  Larger points
           denote systems in which a two-component $\beta$-model was fitted, and
           the solid line represents equality.}
  \label{fig_betaspec_betafit}

\end{figure}

Whilst the \betafit\ values of groups appear to be lower than those of clusters,
no previous studies have detected any significant correlation between \betafit\
and \TX\ within the group regime.  Figure~\ref{fig_betafit_TX} shows the
relationship between these two variables for the \GEMS\ groups. Results from
two-component fits are marked as larger symbols. No correlation is apparent for
the G-sample systems; in fact a weak (1.4$\sigma$) anti-correlation is present in
these data. The results for groups with two-component fits (i.e. the most
reliable values), show no significant trend at all.

It is interesting to compare the values of \betafit\ and \betaspec\ for the \GEMS\
groups, and this is plotted in Figure~\ref{fig_betaspec_betafit}.  In the simple
case of an isothermal hot gas, and a set of galaxies with isotropic velocity
dispersion, both in equilibrium in the gravitational potential of a mass
distribution having the $\beta$-model form, one would expect to find \betafit\ =
\betaspec.  Even though the assumptions underlying this model are restrictive and
unrealistic, one might in general expect to find a correlation between the two
$\beta$s if heating of the IGM is affecting the gas density profiles, as is often
assumed \citep*{cavaliere99,balogh99}, since energy input into the gas
is expected to directly reduce \betaspec, and also to flatten the gas profile,
lowering \betafit\ \citep[c.f.][]{muanwong02}.  In practice, no such relationship
is seen. \betaspec\ covers a much wider range than \betafit, with many systems
having \betaspec\ $\sim$ 1, and some having extremely low values \betaspec\
$\sim$ 0.1.  However, the latter do not have remarkably low values of \betafit.

How can we understand the lack of significant correlations involving \betafit? If
group mass, coupled to a simple universal preheating model, were responsible for
the flatter profiles in groups compared to clusters, then strong correlations
would be seen.  Hence, one plausible suggestion is that additional properties
peculiar to individual groups, such as star formation efficiency or merger
history, play a substantial role, and introduce a large amount of non-statistical
scatter.  Flat X-ray surface brightness profiles are associated with lower
central gas densities and consequently with higher entropy compared to
self-similar expectations.  Recent results (\citealt*{ponman03};
\citealt{pratt03,sun03,mushotsky03}) suggest that the earlier hypothesis
\citep{ponman99} of a universal ``floor'' value of entropy was incorrect. It
appears \citep{ponman03} that entropy scales in a non-self-similar way with
system mass, and that there is also significant scatter \citep{sun03,mushotsky03}
between the magnitude of the entropy in groups of a given mean temperature.  This
would lead to scatter in the value of \betafit, making it hard to detect any
trend with temperature without a large and accurately modelled sample. Since
biases are present where only one-component fits are performed, we actually have
only 15 reliable values of \betafit\ in our sample.

The lack of correlation between \betafit\ and \betaspec, is not obviously
explained by this explanation of individual scatter.  What is required is some
way of breaking the link between the two $\beta$ values.  The obvious way to do
this is through effects on \sigmav, since the expectation that the two $\beta$s
should be related, is based on the assumption that \sigmav\ provides a measure of
the gravitational potential.  We have already seen, from the discussion in
Sections \ref{sec_radius} and \ref{subsec_sigT}, that there is good reason to
doubt this assumption.

%%%%%%%%%%%%%%%%%%%%%%%%%%%%%%%%%%%%%%%%%%%%%%%%%%%%%%%%%%%%%%%%%%%%%%%%%%%%%%%%%
%%%%%%%%%%%%%%%%%%%%%%%%%%%%%%%%%%%%%%%%%%%%%%%%%%%%%%%%%%%%%%%%%%%%%%%%%%%%%%%%%

\section{Galaxy morphology}
\label{sec_morph}

\subsection{Spiral fraction in groups}
\label{subsec_fsp}

\begin{figure*}
  \begin{minipage}{241pt}
  \centering
  
    \includegraphics[height=\linewidth,angle=270]{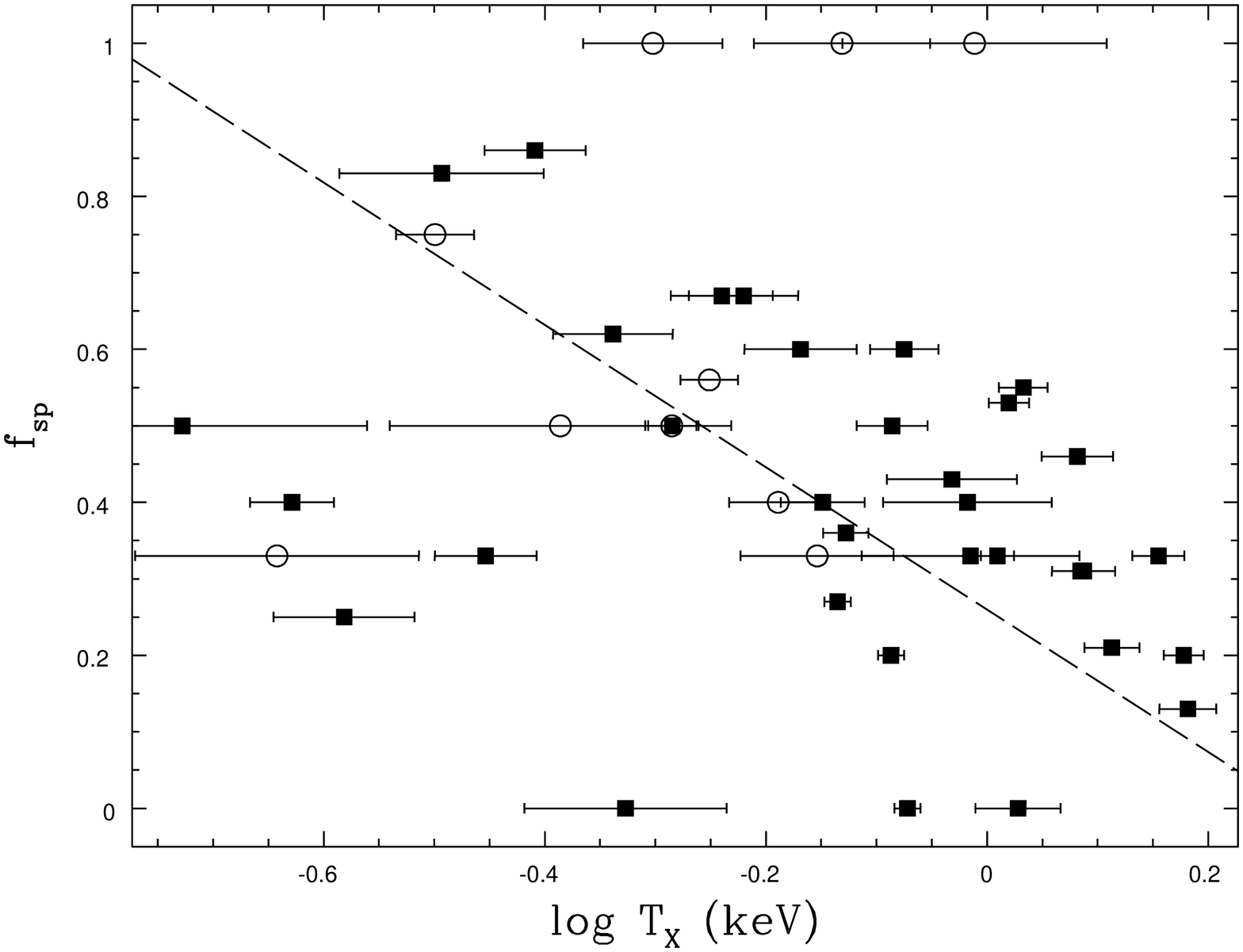}
    \caption{The relationship between \fsp\ and \TX.  Filled squares
             represent the G-sample and open circles the H-sample.  The dashed
             line represents a fit to the G-sample.}
     \label{fig_fsp_TX}

  \end{minipage}\hspace{18pt}
  \begin{minipage}{241pt}
  \centering

    \includegraphics[height=\linewidth,angle=270]{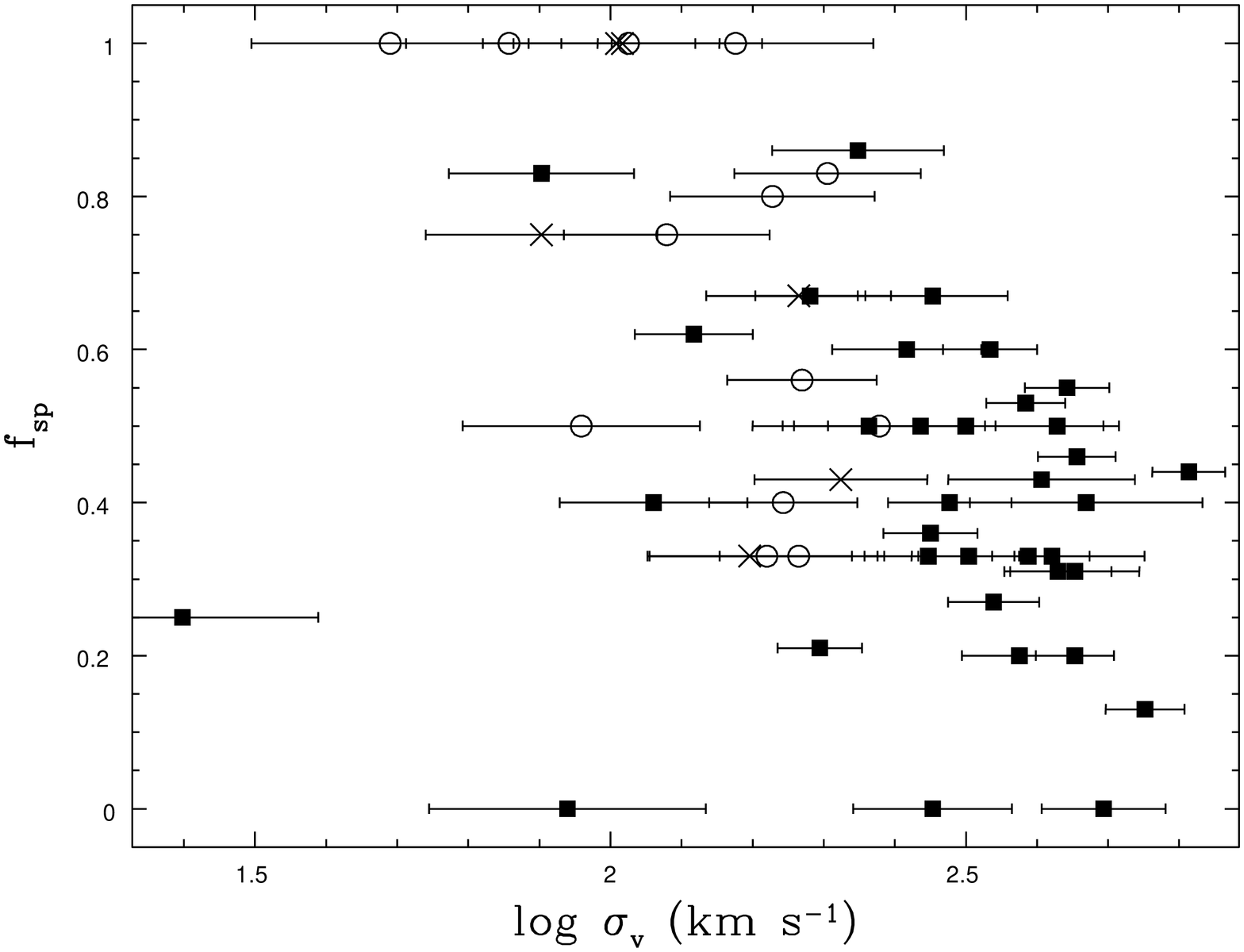}
    \caption{The relationship between \fsp\ and \sigmav.  Filled squares
             represent the G-sample, open circles the H-sample and crosses
             non-detections.}
    \label{fig_fsp_sigma}

  \end{minipage}\\
  \begin{minipage}{241pt}
  \centering

    \includegraphics[height=\linewidth,angle=270]{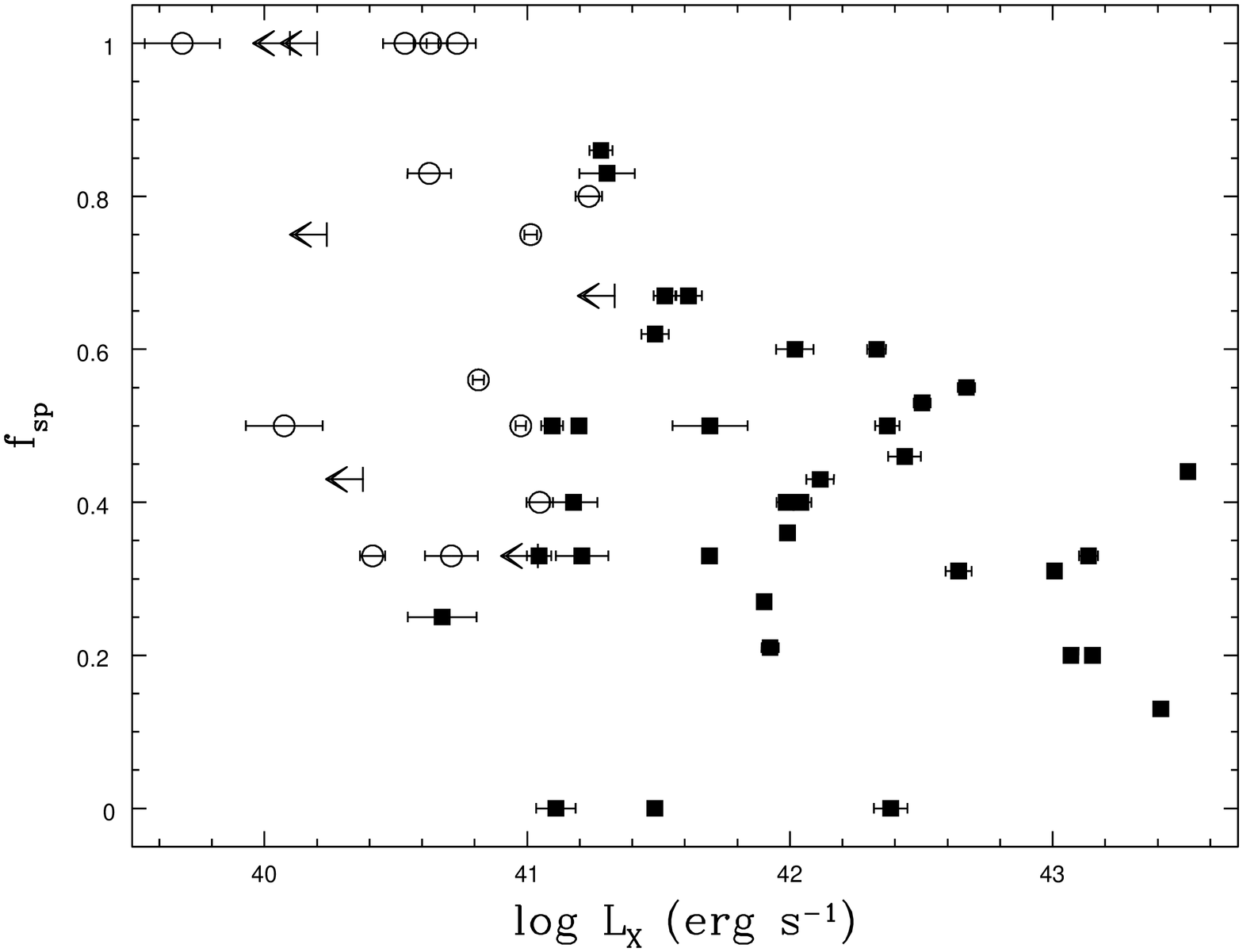}
    \caption{The relationship between \fsp\ and \LX.  Filled squares
             represent the G-sample, open circles the H-sample and arrows
             represent upper-limits from non-detections.}
    \label{fig_fsp_LX}

  \end{minipage}\hspace{18pt}
  \begin{minipage}{241pt}
  \centering

    \includegraphics[height=\linewidth,angle=270]{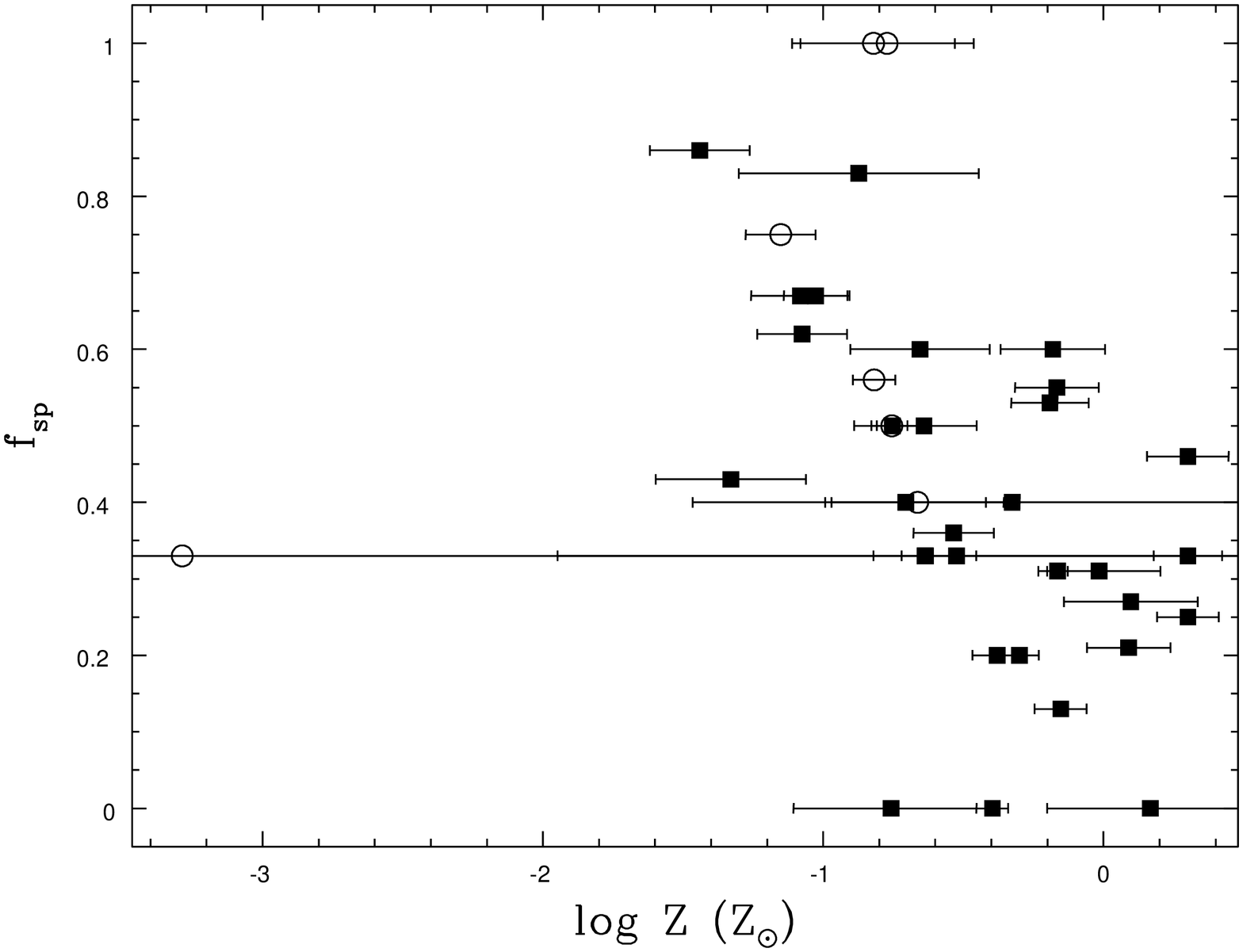}
    \caption{The relationship between \fsp\ and \Z.  Filled squares represent
             the G-sample and open circles the H-sample.}
    \label{fig_fsp_Z}

  \end{minipage}\\
  \begin{minipage}{241pt}

    \includegraphics[height=\linewidth,angle=270]{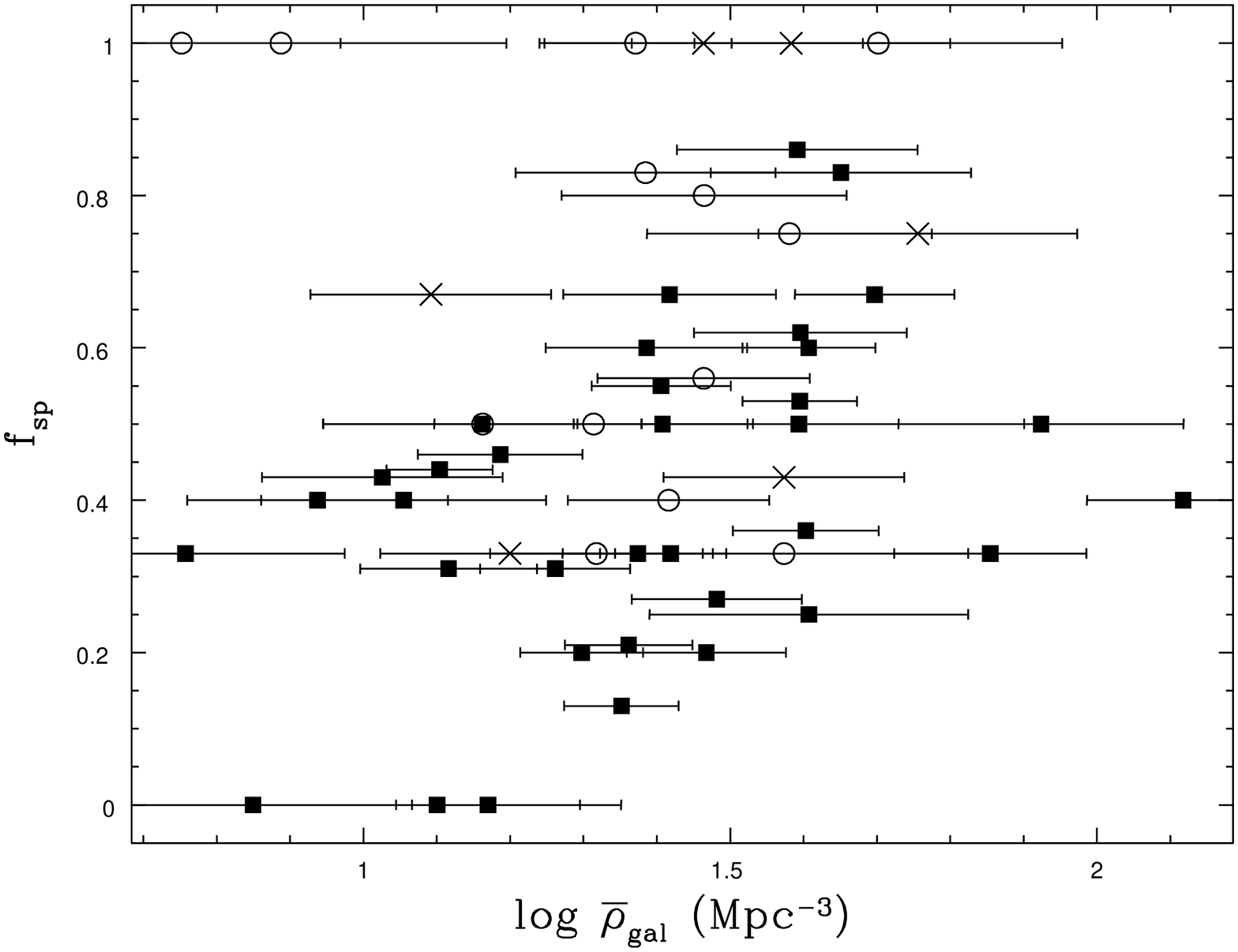}
    \caption{The relationship between \fsp\ and \dengal.  Filled squares
             represent the G-sample, open circles the H-sample and crosses
             non-detections.}
    \label{fig_fsp_dengal}

  \end{minipage}\hspace{18pt}
  \begin{minipage}{241pt}

  \end{minipage}
\end{figure*}

Strong correlations between \fsp\ and both \LX\ and \TX\ are seen in clusters
\citep{edge91b}. In groups, such trends have proved to be surprisingly weak
\citep{ponman96,mulchaey00,helsdon03b}, despite the strong correlation between
X-ray emission and the morphology of the central group galaxy (discussed below).
For the G-sample systems, we also find only a weak (1$\sigma$) tendency for \fsp\
to be higher in low \LX\ groups, but the correlations of \fsp\ with \TX\ and
\sigmav\ (see Table~\ref{tab_relations}) are stronger, though they still show a
large amount of scatter (Figures~\ref{fig_fsp_TX} and \ref{fig_fsp_sigma}).  This
is consistent with what \citet{ponman96} found in their study of Hickson compact
groups.  Since \TX\ and \sigmav\ are primarily determined by the depth of the
potential well, whilst \LX\ is related to the mass and density of hot gas, the
relative strengths of these correlations, now found for both loose and compact
groups, is that galaxy morphology is related to the depth of the potential rather
than its gas content.

Comparing the \fsp-\TX\ correlation with that seen in clusters \citep{edge91c},
there is a large offset between the two relations. The cluster relation shows
\fsp\ rising from $\sim$ 0.1 in hot clusters like Coma, to $\sim$ 0.5 in systems
like Virgo, with \TX\ = 2-3 \kev. In contrast, a best fit trend through the
G-sample \fsp-\TX\ data, is fitted by $\fsp = 0.26-0.93\ \mathrm{log}\ \TX$,
corresponding to a typically low value of \fsp\ for groups with \TX\ $>$ 1 \kev.
Proper comparison of the group and cluster samples requires the application of
consistent absolute magnitude limits to all systems, and lies beyond the scope
of the present paper, but would be very worthwhile.

Although the relationship between \fsp\ and \LX\ is not strong within the
G-sample, when the full sample is considered, it becomes a great deal stronger
(3.5$\sigma$), as shown in Figure~\ref{fig_fsp_LX}.  It is clear that galaxy
halo sources, and X-ray undetected sources (which of course have lower \LX\ than
most G-sample groups) tend to have systematically higher \fsp\ than do X-ray
bright groups.  It seems that the {\it presence} of detectable hot
intergalactic gas is related to galaxy morphology much more strongly that its
luminosity.

In general, metallicities derived from \ROSAT\ \PSPC\ spectra must be regarded
with considerable caution, since the spectral resolution of the instrument is
very limited, and it is well known that serious biases can result if
multi-temperature gas is fitted with a single-temperature hot plasma model
\citep{buote98}. Nonetheless, the 3$\sigma$ anticorrelation between metallicity
and spiral fraction (Figure~\ref{fig_fsp_Z}) is sufficiently strong to be
worthy of comment.  Taken at face value, this would seem to imply that most of
the intergalactic metals have their origin in early-type galaxies.  Such a result
can be understood if the formation of early-type galaxies, either by galaxy
merging or through some primordial collapse (or more likely early-epoch multiple
merging) picture, results in the ejection of much of the enriched interstellar
medium of the galaxy due to energy input from type II supernovae
\citep{matteucci87,kauffmann98}.  However, the role of biases in \ROSAT\
metallicity estimates is potentially sufficiently serious that this result
requires careful checking with CCD quality spectra.  This is underway at present.

Finally, in Figure~\ref{fig_fsp_dengal}, we show the relationship between
spiral fraction and mean galaxy density.  It was shown by \citet{helsdon03a} that
groups display a morphology-density relation rather similar to clusters, in the
sense that galaxies in regions of high surface density are more likely to be of
early type.  Here we see the {\it opposite} effect, in terms of the mean density
of groups as a whole: there is a tendency, with a great deal of scatter, but
significant at the 2.5$\sigma$ level (Table~\ref{tab_relations}), for groups with
the highest mean density to have {\it higher} \fsp. This is not necessarily in
conflict with the \citet{helsdon03a} result, which was a {\it local} effect.  If
all these groups had virialised recently, we would actually expect them all to
show the same mean density (as discussed in Section~\ref{sec_radius} above), so
that no correlation at all would be expected between \fsp\ and \dengal, even if a
local morphology-density relation were present. The fact that we see a systematic
trend, suggests that the range of values of \dengal\ that we observe (which span
a factor of $\sim$ 30) is not simply due to statistical scatter, but is subject
to some systematic effects, which are related to the morphological mix in groups.
For example, most of the groups with mean densities significantly higher than the
expected value (marked in the Figure) have reasonably high values of \fsp. This
could result from underestimation of their \rfh\ values, which are based on gas
temperatures for most groups. One explanation could be that these groups are not
fully virialised, so that their gas temperature has yet to rise to its final
value.

%%%%%%%%%%%%%%%%%%%%%%%%%%%%%%%%%%%%%%%%%%%%%%%%%%%%%%%%%%%%%%%%%%%%%%%%%%%%%%%%%

\subsection{Brightest group galaxies}
\label{subsec_bgg}

\begin{figure}

  \includegraphics[height=\linewidth,angle=270]{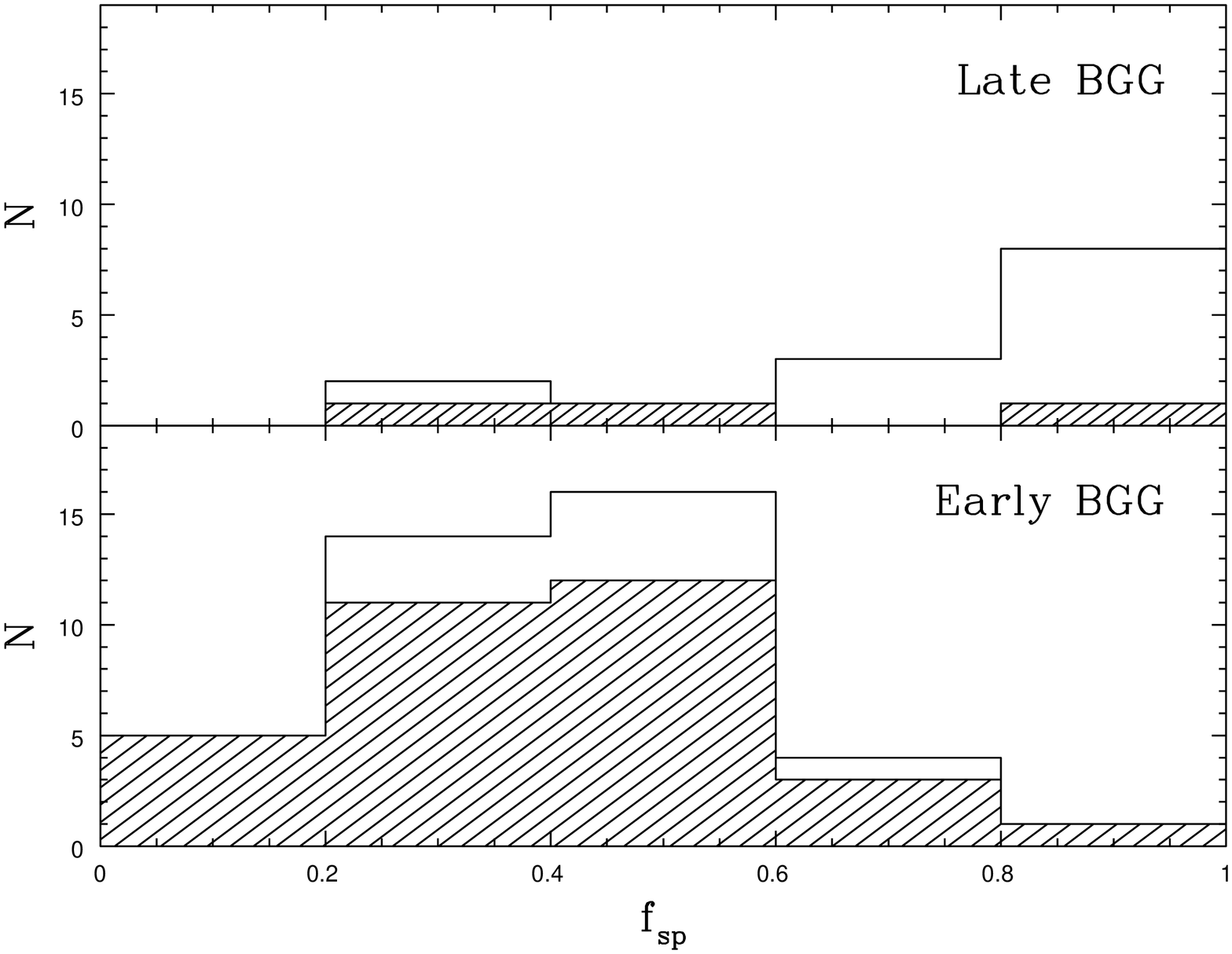}
  \caption{A histogram showing the distribution of \fsp\ for systems dominated
           by a late type galaxy, and those dominated by an early type galaxy.
           The shaded region represents the G-sample.}
  \label{fig_fsp_hist}

  \includegraphics[height=\linewidth,angle=270]{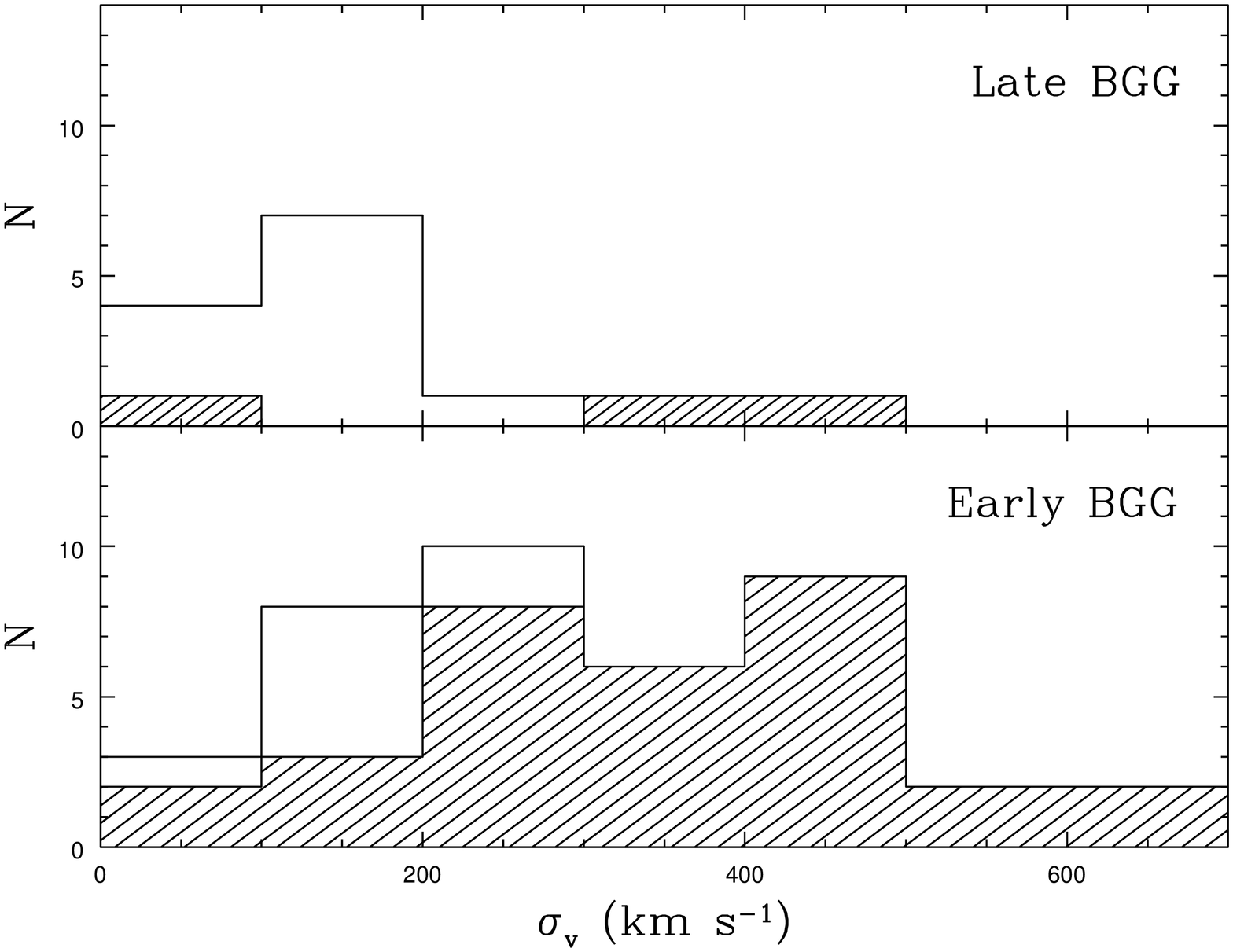}
  \caption{A histogram showing the distribution of \sigmav\ for systems dominated
           by a late type galaxy, and those dominated by an early type galaxy.
           The shaded region represents the G-sample.}
  \label{fig_sigma_hist}

\end{figure}

The most striking correlation between the X-ray properties of groups, and their
galaxy contents, which has been noted since early \ROSAT\ studies of groups
\citep*{ebeling94}, is the presence of a bright early-type galaxy at the
centre of the X-ray emission in virtually every X-ray bright group.  This
phenomenon is clearly apparent in the \GEMS\ sample, as can be seen from
Figure~\ref{fig_fsp_hist}.  Of our 35 G-sample systems, 32 have an early-type
BGG, and in all but four of these, this galaxy lies at the centre of the group's
X-ray halo.  As described in Section~\ref{sec_optical}, we have defined our
``brightest group galaxy'' (BGG) to be the most optically luminous within
0.25\rfh\ of the group centre, in order to avoid picking up outlying galaxies
which may have recently fallen into a group.  With this definition, we find that
all but three of the systems with group-scale emission (hatched regions) have
early-type BGGs.  These three exceptions are HCG\,16, HCG\,48 and HCG\,92.
HCG\,16 is well known as an unusual group dominated by spirals which shows
significant intergalactic hot gas \citep{ponman96,dossantos99}, and probably
corresponds to a system which is collapsed but not yet virialised
\citep{belsole03}.  HCG\,92 is the well studied system {\it Stephan's Quintet},
which contains a number of strongly interacting galaxies.  None of the member
galaxies lie at the centre of the X-ray emission, and high resolution
observations \citep{pietsch97,sulentic01} have shown sharp features in the X-ray
emission, interpreted as intergalactic shocks.  Like HCG\,16, this appears to be
a newly collapsed group.  Finally, in HCG\,48 the late-type BGG does appear to
lie at the centre of the diffuse X-ray emission, although a bright elliptical is
seen 71 kpc away in projection.  The unusual properties of this group may be
related to the fact that it appears to be falling into the nearby cluster
Abell\,1060, and its X-ray morphology is distorted into a coma-like shape.

The distribution of spiral fraction in detected and undetected groups can be
seen from Figure~\ref{fig_fsp_hist} to differ primarily in that there is a
set of groups with high \fsp\ and late-type BGGs, which almost invariably show no
group-scale X-ray emission.  It can be seen from Figure~\ref{fig_sigma_hist} that
these systems (and indeed all groups without detectable group-scale hot gas) also
have low velocity dispersions.  It seems very likely that these systems are not
yet fully collapsed and virialised, so that their IGM has not been heated and
compressed to a point where it radiates detectable X-rays.  Note however (e.g.
from Figure~\ref{fig_dengal_LB}) that these H and U-sample systems do {\it not}
in general have low inferred mean densities.  This is not entirely surprising,
since they have been selected from group catalogues which are compiled using
techniques which rely on a significant overdensity in the inferred 3-dimensional
density, typically corresponding, for the catalogues we draw from, to a threshold
factor of $\sim$ 20-80, relative to the mean density of the Universe -- although
all our systems appear to lie well above this threshold.

\begin{figure}

  \includegraphics[height=\linewidth,angle=270]{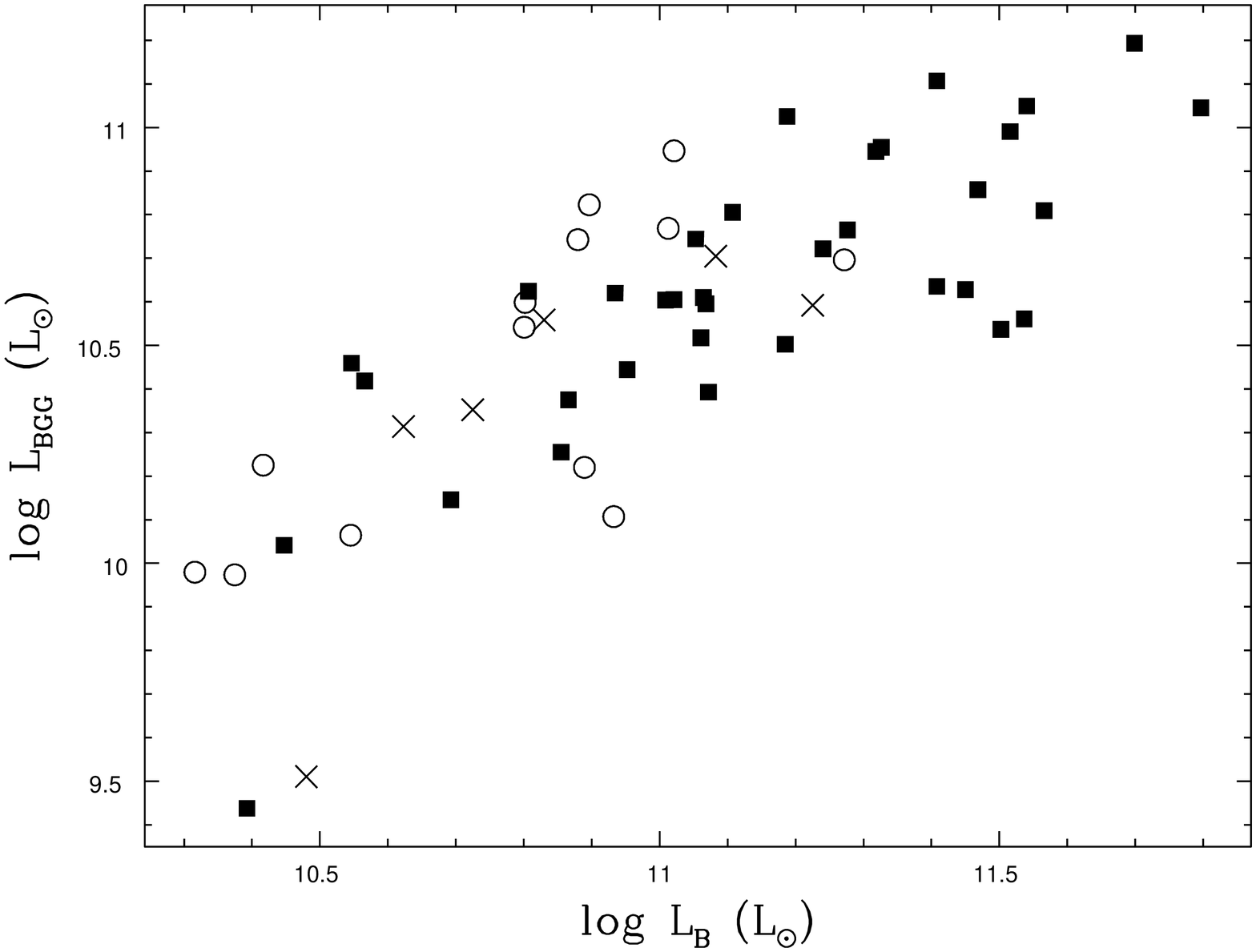}
  \caption{The relationship between \LBGG\ and \LB.  Filled squares
           represent the G-sample, open circles the H-sample and crosses
           non-detections.}
  \label{fig_LBGG_LB}

  \includegraphics[height=\linewidth,angle=270]{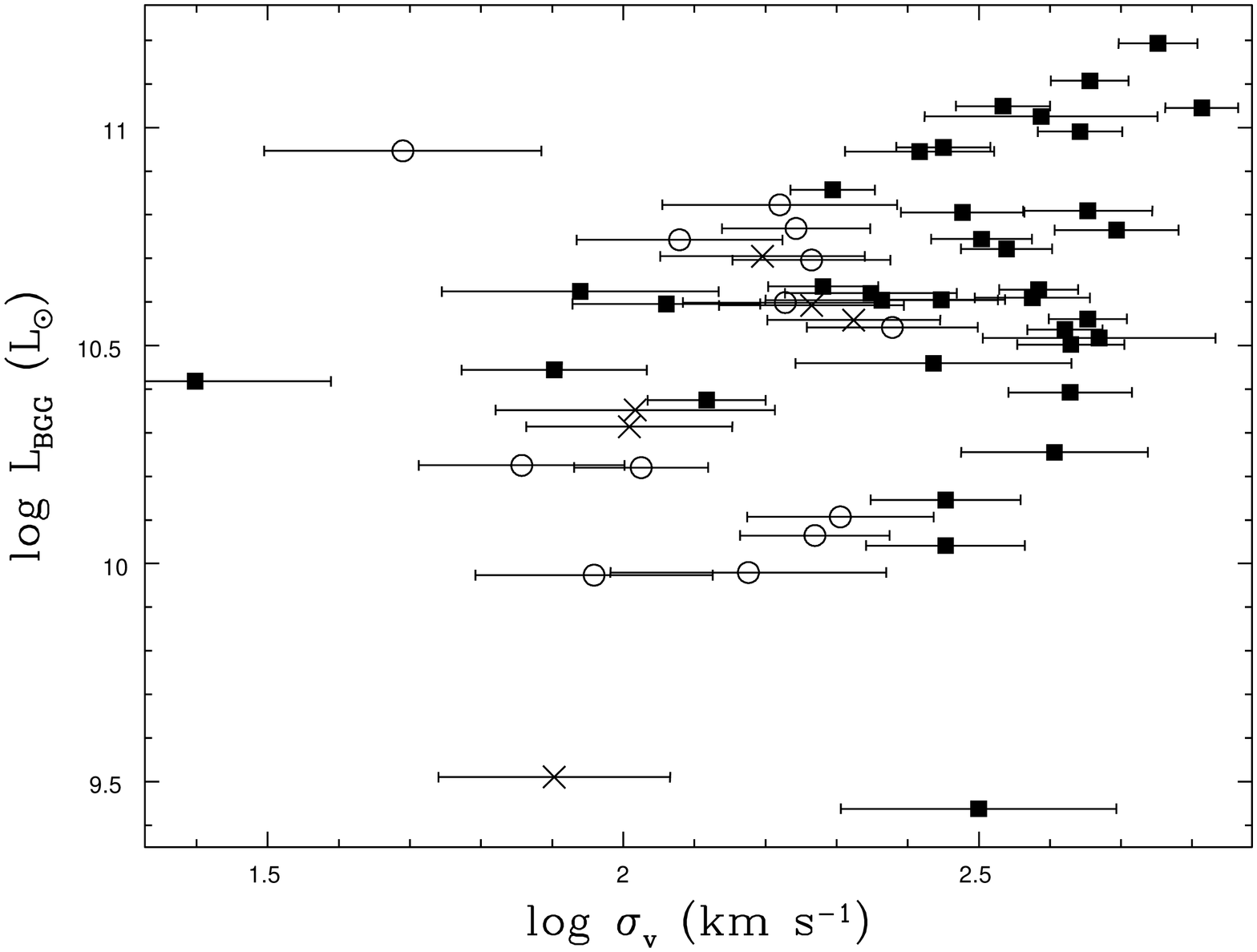}
  \caption{The relationship between \LBGG\ and \sigmav.  Filled squares represent
           the G-sample, open circles the H-sample and crosses non-detections.}
  \label{fig_LBGG_sigma}

\end{figure}

Figure~\ref{fig_LBGG_LB} shows a highly significant (6$\sigma$, including all
points in the plot) tendency for brighter BGGs to be found in richer groups (i.e.
with higher total \LB).  A similar, but less significant (2.5$\sigma$),
correlation exists between \LBGG\ and \LX.  These relationships are similar in
character to those seen in the BCGs, found in richer systems, and represent an
extension towards poorer systems of the trends with X-ray luminosity and
temperature reported by \citet{edge91a}.  Studies of the K-band luminosities of
BCGs (\citealt*{burke00}; \citealt{brough02}) in clusters spanning a range of
redshifts and X-ray luminosities, show a correlation between \LX\ and \LBGG,
although \citet{brough02} argue that this relation is weak in low redshift ($z$
$<$ 0.1) clusters.  Ours is a low redshift sample, and yet clearly does show such
a correlation.  This is in accord with the results of \citet{burke00}, who found
BCGs in the most X-ray luminous clusters to be standard candles, whilst for poor
clusters (\LX\ $<$ 10$^{44}$ \ergps\ in our cosmology) they find that lower
luminosity clusters tend to contain less bright BCGs. Our results show that this
trend continues through the group regime.

It seems very likely \citep{dubinski98} that BCGs and BGGs are primarily
assembled through galaxy merging.  This idea is also supported, from the present
study, by the fact that early-type BGGs tend to be more luminous (on average by
a factor 1.8) than late-type BGGs.  Since (other things being equal) the merger
rate should be higher in systems with low velocity dispersion, it is of some
interest to examine the relationship between \LBGG\ and \sigmav.  This is shown
in Figure~\ref{fig_LBGG_sigma}, and shows a tendency (1.9$\sigma$) towards
brighter BGGs in {\it higher} velocity dispersion systems.  There is a strong
clue here, that much of the merging involving the galaxies which we now see in
the more luminous groups, must have happened at earlier epochs.  We will return
to this in Section~\ref{sec_conc}.

Finally we examine the issue of the dominance of the group BGGs, and how this
relates to other group properties.  There is a class of groups, constituting
$\sim$ 10\% of X-ray bright systems \citep{jones03}, in which the galaxy contents
are dominated by a single luminous elliptical at least two magnitudes brighter
than the second ranked galaxy.  These have been dubbed ``fossil groups''
\citep{ponman94}, or ``overluminous elliptical galaxies'' \citep{vikhlinin99c}.
These are found to have unusually high ratios of X-ray to optical luminosity, and
are probably old groups in which the orbits of the major galaxies have had time
to decay, leading to their merger into a central remnant \citep{jones03}.  It has
been suggested that the high \LXpLB\ ratios may result from high gas density (and
hence high \LX) resulting from an early formation epoch \citep{jones03}, or from
an unusually low star formation efficiency, leading to low \LB\
\citep{vikhlinin99c}.  One might hope to derive some insights into the origin and
properties of fossil groups from a study of the optical luminosity ratio between
the first and second ranked galaxies (\dom) in the \GEMS\ sample.

\begin{figure}

  \includegraphics[height=\linewidth,angle=270]{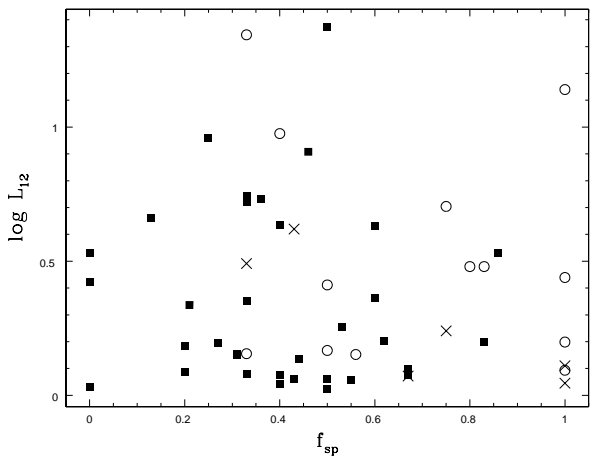}
  \caption{The relationship between \dom\ and \fsp.  Filled squares represent
           the G-sample, open circles the H-sample and crosses non-detections.}
  \label{fig_dom_fsp}

\end{figure}

Only one of the \GEMS\ systems qualifies as a fossil group, according to the
definition of \citet{jones03}, which requires \LX\ $>$ 10$^{42}$ \hfifty$^{-2}$
\ergps\ (\hfifty\ = \Hzero/50) and \dom\ $>$ 6.3 (2 magnitudes).  This system is
NGC\,741, which safely passes both thresholds.  However, it does {\it not}
display the feature of high \LXpLB, which is a feature of previously studied
fossil groups, nor does it display a high \LX\ for its temperature, which
\citet{jones03} suggest may be another characteristic property of fossil systems.
A study of the relationship between \dom\ and other statistical parameters,
across the \GEMS\ sample, reveals nothing very striking. For example, given the
hypothesised link to merging, one might have expected to see a relationship with
the spiral fraction or velocity dispersion of groups.  However, neither of these
shows any significant correlation (the relationship with \fsp\ is shown in
Figure~\ref{fig_dom_fsp}), nor is there any very substantial difference between
the distribution of \dom\ values for systems with early-type and late-type BGGs.

On the other hand, a parallel optical study of the luminosity function
\citep{miles04} has shown the presence of a distinctive dip in the luminosity
function at \MB\ $\sim$ -18 (which can be seen in Figure~\ref{fig_lumfun}),
especially in low \LX\ groups.  A study of the structure of \GEMS\ galaxies
\citep{khosroshahi04a}, provides further motivation for the suggestion that this
dip might be generated by the effects of galaxy merging, which will generate
brighter galaxies, whilst leaving the dwarfs relatively unaffected (due to their
small merger cross-sections).  If this is the case, then it appears that the
opening up of a dip at intermediate magnitudes is not reflected strongly in the
difference between first and second ranked galaxies, until the fossil stage is
reached.  This is understandable if merging affects all the brighter galaxies in
a group, rather than simply the central BGG.

%%%%%%%%%%%%%%%%%%%%%%%%%%%%%%%%%%%%%%%%%%%%%%%%%%%%%%%%%%%%%%%%%%%%%%%%%%%%%%%%%
%%%%%%%%%%%%%%%%%%%%%%%%%%%%%%%%%%%%%%%%%%%%%%%%%%%%%%%%%%%%%%%%%%%%%%%%%%%%%%%%%

\section{Concluding Discussion}
\label{sec_conc}

Although the \GEMS\ groups cannot be regarded as a proper statistically-selected
sample, they do span a wider range in group properties than any previous sample
subjected to a thorough X-ray analysis and statistical investigation.  This has
resulted in a number of new results, and some challenges to findings from earlier
work.  Here we summarise and discuss some of the most important of these.

%%%%%%%%%%%%%%%%%%%%%%%%%%%%%%%%%%%%%%%%%%%%%%%%%%%%%%%%%%%%%%%%%%%%%%%%%%%%%%%%%

\subsection{\LX-\TX}

It has become widely accepted that the \LX-\TX\ relation steepens in the group
regime, following in part the results of Ponman and co-workers, and much
theoretical effort has been expended in trying to reproduce this steepening.  It
therefore comes as something of an embarrassment that the \GEMS\ \LX-\TX\ relation
is consistent with an extrapolation of the \LX\ $\propto$ \TX$^3$ trend seen in
clusters.  As we discussed in Section~\ref{subsec_LT}, this flatter \LX-\TX\
slope cannot be regarded at present as a secure result, since a number of biases
resulting from the large scatter in the properties of poor groups, coupled with
the observational selection effect against detecting groups with very low \LX,
conspire to systematically flatten the fitted relation.  However, earlier results
\citep[e.g.][]{helsdon00b} based on X-ray-selected group samples, showing a slope
of 4-5, must now be regarded as questionable.  The origin of the flatter relation
found here, appears to be primarily the inclusion of more groups with very low
temperatures, some of which have surprisingly high X-ray luminosities.

Recent developments from studies of the entropy of groups and clusters tend to
support the idea of a greater continuity in properties between the IGM in groups
and clusters.  The idea of a universal ``entropy floor'' \citep{ponman99}, which
could have led to an isentropic IGM in the poorest systems \citep{babul02}, and
hence to an \LX-\TX\ slope of $\sim$ 5, has now been shown to be incorrect in two
ways.  Firstly, individual groups do not show isentropic cores, as would be
expected in such a simple preheating model 
(\citealt{pratt03,ponman03,sun03,mushotsky03,rasmussen04};
\citealt*{khosroshahi04b}), and secondly the scaling properties of entropy in
clusters and groups appears to take the form of a ``ramp'', rather than a
``floor'' \citep{ponman03} - following the non-self-similar power law form $S
\propto \TX^{0.65}$, rather than breaking from the self-similar slope of unity at
a well-defined temperature \TX\ $\sim$ 1-2 \kev, as was previously suggested
\citep*{lloyddavies00}.  Given this continuity of properties, a
(non-self-similar) power law form for the \LX-\TX\ relation is what one would
expect.

%%%%%%%%%%%%%%%%%%%%%%%%%%%%%%%%%%%%%%%%%%%%%%%%%%%%%%%%%%%%%%%%%%%%%%%%%%%%%%%%%

\subsection{\sigmav}

We find several indications that the velocity dispersions of some poor groups are
anomalously low.  There are groups in our sample, such as HCG\,16, HCG\,22 and
NGC\,3665, which appear to have diffuse X-ray emission from a hot IGM (confirmed
in the case of NGC\,3665 and NGC\,1587 by the \Chandra\ observations of
\citet{helsdon04}), and yet have velocity dispersions so  low (\ltsim\ 100
\kmps) that it is hard to understand how they can be virialised. These groups
also have very low values of \betaspec (\ltsim\ 0.3), and if \sigmav\ is used to
calculate their overdensity radii, then the inferred values of galaxy density tend
to be very high (Figure~\ref{fig_dengal_hist}). In Section~\ref{subsec_sigT}, we
discussed a number of statistical and physical effects which might lead to low
values of \sigmav, however, statistical {\it scatter} cannot account for the
lack of poor systems with high values of \betaspec, and statistical {\it biases}
seem too weak to account for the very low values we observe.  A more detailed
investigation is required to determine whether alignment effects or tidal
interactions can produce velocity dispersions as low as those observed.

%%%%%%%%%%%%%%%%%%%%%%%%%%%%%%%%%%%%%%%%%%%%%%%%%%%%%%%%%%%%%%%%%%%%%%%%%%%%%%%%%

\subsection{Group radii}

In order to make meaningful comparisons between the properties of groups and
clusters, it is important to derive physical values within some well-defined
overdensity radius.  However, it has become apparent in the present study just
how difficult it is to define a reliable value of \rfh\ (\rth\ is even more
difficult) in galaxy groups.  Their low surface brightness prohibits the mapping
of gas pressure (which would allow the mass to be inferred) out to large radii,
even in X-ray bright groups.  Virial analyses are  unreliable due to the sparse
galaxy populations, and lensing studies unfeasible on individual groups due to
their low projected mass densities.

Having investigated the use of scaling formulae based on \sigmav, \TX\ and \LB,
we have reservations about all three. \sigmav\ appears to be the worst of the
three, due to the large unexplained biases in velocity dispersion in some poor
groups, discussed above.  \TX\ gives a much more stable and satisfactory estimate
of group sizes, but there is some evidence that it may overestimate \rfh\ by up
to 40\% throughout the group regime.  \LB\ also appears to give fairly stable
results, but since there are both theoretical and observational reasons to
believe that the mass-to-light ratio may drop in low mass systems, a scaling
formula based on constant star formation efficiency seems unwise.  More work
employing deep \XMM\ observations, to clarify the relationships between group
masses and the value for \TX\ and \LB\ would be of great value for future
scaling studies.

%%%%%%%%%%%%%%%%%%%%%%%%%%%%%%%%%%%%%%%%%%%%%%%%%%%%%%%%%%%%%%%%%%%%%%%%%%%%%%%%%

\subsection{\betafit}

As was discussed in Section~\ref{sec_beta}, it seems curious that the lower value
of \betafit\ compared to clusters, which we believe to be a robust result, is not
reflected in any significant correlation between \betafit\ and \TX\ within the
\GEMS\ sample.  This appears to be another nail in the coffin of simple preheating
models, which would lead to systematically flatter gas density profiles in the
shallower potential wells of cool groups.  We suggest that in our sample, such a
weak trend may be obscured by fluctuations in the values of gas entropy between
groups.  This in turn may be driven by differences in the merger and star
formation histories.

%%%%%%%%%%%%%%%%%%%%%%%%%%%%%%%%%%%%%%%%%%%%%%%%%%%%%%%%%%%%%%%%%%%%%%%%%%%%%%%%%

\subsection{\fsp}

\GEMS\ groups typically contain a much higher fraction of early-type galaxies than
would be expected if one extrapolates the trends between \fsp\ and \TX\ or \LX\
established for clusters.  This result is almost certainly related to the results
of \citet{helsdon03b}, who found that X-ray bright groups have lower \fsp\ than
clusters, at a given projected density, and even at a given inferred
3-dimensional density.  They interpreted this result as evidence that galaxy
morphology may be related to the effects of galaxy interactions and mergers,
which are enhanced in low velocity dispersion systems.

It seems clear (Figures~\ref{fig_fsp_hist} and \ref{fig_sigma_hist}) that
there is a class of groups with low \sigmav, and fairly high spiral fraction
(including, in most cases, a late-type BGG) which show no group-scale X-ray
emission.  Since a hot, X-ray emitting IGM requires a collapsed group, both to
heat the gas, and to raise its density to the point where its emission becomes
detectable, it seems very probable that these undetected groups are not yet
virialised, and represent late-forming groups. In a study of galaxy photometry
for a subsample of 25 of the \GEMS\ groups, \citet{miles04} find that groups with
\LX\ $<$ 5$\times$10$^{41}$ \ergps\ (including some which are undetected) show a
pronounced dip in their galaxy luminosity functions, which may result from the 
effects of recent galaxy merging activity taking place in these low \sigmav\
virialising systems.

%%%%%%%%%%%%%%%%%%%%%%%%%%%%%%%%%%%%%%%%%%%%%%%%%%%%%%%%%%%%%%%%%%%%%%%%%%%%%%%%%

\subsection{BGGs}

We confirm the very strong relationship between the presence of intergalactic
X-ray emission, and the presence of a centrally located luminous early-type
galaxy, which has been noted by many previous authors.  This points strongly to
the effects of galaxy merging having played an important role during the
evolution of these systems.  We do, however, find four groups from our G-sample,
for which the early-type BGG does not lie at the centre of the X-ray emission,
and a further three systems in which the BGG is of late-type, as discussed in
Section~\ref{subsec_bgg}, above.  The presence of a small fraction of systems in
which the BGG is offset is not surprising, given that groups are expected to
experience mergers, during which their gravitational potentials will be
perturbed.  The three G-sample systems with late-type BGGs are all groups with
compact cores.  Two of them appear to be systems in the process of virialisation,
in which the gas has been recently heated, but a luminous early-type BGG has yet
to form. The status of HCG\,48 is still problematical, and warrants more detailed
study with better quality X-ray data.

Our richer groups tend to have more luminous BGGs (Figures~\ref{fig_LBGG_LB} and 
\ref{fig_LBGG_sigma}), continuing a trend which has been noted in moderately rich
clusters.  Given the likelihood \citep{dubinski98} that these dominant early-type
galaxies have been formed through multiple mergers, coupled with the strong
inverse dependence of the merger cross-section on \sigmav\ \citep{makino97}, the
positive correlation seen in Figure~\ref{fig_LBGG_sigma} strongly suggests that
much of this merging has taken place at earlier epochs, in substructures which
have since merged to form the groups we observe today, since at the present epoch
merging should be more effective in groups with {\it low} values of \sigmav.

%%%%%%%%%%%%%%%%%%%%%%%%%%%%%%%%%%%%%%%%%%%%%%%%%%%%%%%%%%%%%%%%%%%%%%%%%%%%%%%%%
%%%%%%%%%%%%%%%%%%%%%%%%%%%%%%%%%%%%%%%%%%%%%%%%%%%%%%%%%%%%%%%%%%%%%%%%%%%%%%%%%

\section{Acknowledgements}

We would like to thank the other members of the \GEMS\ project, especially Somak
Raychaudhury, Duncan Forbes, Stephen Helsdon and Frazer Pearce, for their
contributions to this work. We are also grateful to Alastair Sanderson and Don
Horner for their help in making comparisons with cluster properties, and to
the referee for pointing out a number of areas where clarification
was required. This
research has made use of the \LEDAS\ database at the University of Leicester, the
\NASA-\IPAC\ Extragalactic Database, and optical images from the STScI Digitised
Sky Survey.  The authors also acknowledge the support of a studentship (JPFO) and
a Senior Fellowship (TJP) from the Particle Physics and Astronomy Research
Council.

%%%%%%%%%%%%%%%%%%%%%%%%%%%%%%%%%%%%%%%%%%%%%%%%%%%%%%%%%%%%%%%%%%%%%%%%%%%%%%%%%
%%%%%%%%%%%%%%%%%%%%%%%%%%%%%%%%%%%%%%%%%%%%%%%%%%%%%%%%%%%%%%%%%%%%%%%%%%%%%%%%%

%%%%%%%%%%%%%%%%%%%%%%%%%%%%%%%%%%%%%%%%%%%%%%%%%%%%%%%%%%%%%%%%%%%%%%%%%%%%%%%%%
%%%%%%%%%%%%%%%%%%%%%%%%%%%%%%%%%%%%%%%%%%%%%%%%%%%%%%%%%%%%%%%%%%%%%%%%%%%%%%%%%

\label{lastpage}

\begin{thebibliography}{}

\bibitem[\protect\citeauthoryear{Allen \& Fabian}{Allen \&
  Fabian}{1998}]{allen98}
Allen S.~W.,  Fabian A.~C.,  1998, MNRAS, 297, 57

\bibitem[\protect\citeauthoryear{Allen, Schmidt \& Fabian}{Allen
  et~al.}{2001}]{allen01d}
Allen S.~W.,  Schmidt R.~W.,    Fabian A.~C.,  2001, MNRAS, 328, L37

\bibitem[\protect\citeauthoryear{Arnaud \& Evrard}{Arnaud \&
  Evrard}{1999}]{arnaud99}
Arnaud M.,  Evrard A.~E.,  1999, MNRAS, 305, 631

\bibitem[\protect\citeauthoryear{Babul, Balogh, Lewis \& Poole}{Babul
  et~al.}{2002}]{babul02}
Babul A.,  Balogh M.~L.,  Lewis G.~F.,    Poole G.~B.,  2002, MNRAS, 330, 329

\bibitem[\protect\citeauthoryear{Bahcall \& Comerford}{Bahcall \&
  Comerford}{2002}]{bahcall02}
Bahcall N.~A.,  Comerford J.~M.,  2002, ApJ, 565, L5

\bibitem[\protect\citeauthoryear{Balogh, Babul \& Patton}{Balogh
  et~al.}{1999}]{balogh99}
Balogh M.~L.,  Babul A.,    Patton D.~R.,  1999, MNRAS, 307, 463

\bibitem[\protect\citeauthoryear{Barton, Geller, Ramella, Marzke \& da
  Costa}{Barton et~al.}{1996}]{barton96}
Barton E.,  Geller M.~J.,  Ramella M.,  Marzke R.~O.,    da Costa L.~N.,  1996,
  AJ, 112, 871

\bibitem[\protect\citeauthoryear{Belsole, Sauvageot, Ponman \& Bourdin}{Belsole
  et~al.}{2003}]{belsole03}
Belsole E.,  Sauvageot J.~L.,  Ponman T.~J.,    Bourdin H.,  2003, A\&A, 398, 1

\bibitem[\protect\citeauthoryear{Benson, Cole, Frenk, Baugh \& Lacey}{Benson
  et~al.}{2000}]{benson00}
Benson A.~J.,  Cole S.,  Frenk C.~S.,  Baugh C.~M.,    Lacey C.~G.,  2000,
  MNRAS, 311, 793

\bibitem[\protect\citeauthoryear{Bird, Mushotzky \& Metzler}{Bird
  et~al.}{1995}]{bird95}
Bird C.~M.,  Mushotzky R.~F.,    Metzler C.~A.,  1995, ApJ, 453, 40

\bibitem[\protect\citeauthoryear{Blanton, Hogg, Bahcall, Brinkmann, Britton,
  Connolly, Csabai, Fukugita, Loveday, Meiksin, Munn, Nichol, Okamura, Quinn,
  Schneider, Shimasaku, Strauss, Tegmark, Vogeley \& Weinberg}{Blanton
  et~al.}{2003}]{blanton03}
Blanton M.~R.,  Hogg D.~W.,  Bahcall N.~A.,  Brinkmann J.,  Britton M.,
  Connolly A.~J.,  Csabai I.,  Fukugita M.,  Loveday J.,  Meiksin A.,  Munn
  J.~A.,  Nichol R.~C.,  Okamura S.,  Quinn T.,  Schneider D.~P.,  Shimasaku
  K.,  Strauss M.~A.,  Tegmark M.,  Vogeley M.~S.,    Weinberg D.~H.,  2003,
  ApJ, 592, 819

\bibitem[\protect\citeauthoryear{B{\" o}hringer, Voges, Huchra, McLean,
  Giacconi, Rosati, Burg, Mader, Schuecker, Simi{\c c}, Komossa, Reiprich,
  Retzlaff \& Tr{\" u}mper}{B{\" o}hringer et~al.}{2000}]{bohringer00}
B{\" o}hringer H.,  Voges W.,  Huchra J.~P.,  McLean B.,  Giacconi R.,  Rosati
  P.,  Burg R.,  Mader J.,  Schuecker P.,  Simi{\c c} D.,  Komossa S.,
  Reiprich T.~H.,  Retzlaff J.,    Tr{\" u}mper J.,  2000, ApJS, 129, 435

\bibitem[\protect\citeauthoryear{Brough, Collins, Burke, Mann \& Lynam}{Brough
  et~al.}{2002}]{brough02}
Brough S.,  Collins C.~A.,  Burke D.~J.,  Mann R.~G.,    Lynam P.~D.,  2002,
  MNRAS, 329, L53

\bibitem[\protect\citeauthoryear{Buote \& Fabian}{Buote \&
  Fabian}{1998}]{buote98}
Buote D.~A.,  Fabian A.~C.,  1998, MNRAS, 296, 977

\bibitem[\protect\citeauthoryear{Burke, Collins \& Mann}{Burke
  et~al.}{2000}]{burke00}
Burke D.~J.,  Collins C.~A.,    Mann R.~G.,  2000, ApJ, 532, L105

\bibitem[\protect\citeauthoryear{Cavaliere, Menci \& Tozzi}{Cavaliere
  et~al.}{1999}]{cavaliere99}
Cavaliere A.,  Menci N.,    Tozzi P.,  1999, MNRAS, 308, 599

\bibitem[\protect\citeauthoryear{Dickey \& Lockman}{Dickey \&
  Lockman}{1990}]{dickey90}
Dickey J.~M.,  Lockman F.~J.,  1990, ARA\&A, 28, 215

\bibitem[\protect\citeauthoryear{Dos~Santos \& Mamon}{Dos~Santos \&
  Mamon}{1999}]{dossantos99}
Dos~Santos S.,  Mamon G.~A.,  1999, A\&A, 352, 1

\bibitem[\protect\citeauthoryear{Dubinski}{Dubinski}{1998}]{dubinski98}
Dubinski J.,  1998, ApJ, 502, 141

\bibitem[\protect\citeauthoryear{Ebeling, Voges \& Boehringer}{Ebeling
  et~al.}{1994}]{ebeling94}
Ebeling H.,  Voges W.,    Boehringer H.,  1994, ApJ, 436, 44

\bibitem[\protect\citeauthoryear{Edge}{Edge}{1991}]{edge91a}
Edge A.~C.,  1991, MNRAS, 250, 103

\bibitem[\protect\citeauthoryear{Edge \& Stewart}{Edge \&
  Stewart}{1991a}]{edge91b}
Edge A.~C.,  Stewart G.~C.,  1991a, MNRAS, 252, 414

\bibitem[\protect\citeauthoryear{Edge \& Stewart}{Edge \&
  Stewart}{1991b}]{edge91c}
Edge A.~C.,  Stewart G.~C.,  1991b, MNRAS, 252, 428

\bibitem[\protect\citeauthoryear{Evrard, Metzler \& Navarro}{Evrard
  et~al.}{1996}]{evrard96}
Evrard A.~E.,  Metzler C.~A.,    Navarro J.~F.,  1996, ApJ, 469, 494

\bibitem[\protect\citeauthoryear{Feigelson \& Babu}{Feigelson \&
  Babu}{1992}]{feigelson92}
Feigelson E.~D.,  Babu G.~J.,  1992, ApJ, 397, 55

\bibitem[\protect\citeauthoryear{Finoguenov, Reiprich \& B{\"
  o}hringer}{Finoguenov et~al.}{2001}]{finoguenov01b}
Finoguenov A.,  Reiprich T.~H.,    B{\" o}hringer H.,  2001, A\&A, 368, 749

\bibitem[\protect\citeauthoryear{Fouque, Gourgoulhon, Chamaraux \&
  Paturel}{Fouque et~al.}{1992}]{fouque92}
Fouque P.,  Gourgoulhon E.,  Chamaraux P.,    Paturel G.,  1992, A\&AS, 93, 211

\bibitem[\protect\citeauthoryear{Frederic}{Frederic}{1995}]{frederic95a}
Frederic J.~J.,  1995, ApJS, 97, 259

\bibitem[\protect\citeauthoryear{Fukugita, Shimasaku \& Ichikawa}{Fukugita
  et~al.}{1995}]{fukugita95}
Fukugita M.,  Shimasaku K.,    Ichikawa T.,  1995, PASP, 107, 945

\bibitem[\protect\citeauthoryear{Garcia}{Garcia}{1993}]{garcia93}
Garcia A.~M.,  1993, A\&AS, 100, 47

\bibitem[\protect\citeauthoryear{Geller \& Huchra}{Geller \&
  Huchra}{1983}]{geller83}
Geller M.~J.,  Huchra J.~P.,  1983, ApJS, 52, 61

\bibitem[\protect\citeauthoryear{Girardi, S., Giuricin, Mardirossian \&
  Mezzetti}{Girardi et~al.}{1998}]{girardi98}
Girardi M.,  S. B.,  Giuricin G.,  Mardirossian F.,    Mezzetti M.,  1998, ApJ,
  506, 45

\bibitem[\protect\citeauthoryear{Giudice}{Giudice}{1999}]{giudice99}
Giudice G.,  1999, in Giuricin G.,  Mezzetti M.,   Salucci P.,  eds, ASP Conf.
  Ser. 176: Observational Cosmology: The Development of Galaxy Systems
  Vol.~176, Properties of nearby groups of galaxies.
Bolzano, p.~136

\bibitem[\protect\citeauthoryear{Helsdon \& Ponman}{Helsdon \&
  Ponman}{2000a}]{helsdon00a}
Helsdon S.~F.,  Ponman T.~J.,  2000a, MNRAS, 315, 356

\bibitem[\protect\citeauthoryear{Helsdon \& Ponman}{Helsdon \&
  Ponman}{2000b}]{helsdon00b}
Helsdon S.~F.,  Ponman T.~J.,  2000b, MNRAS, 319, 933

\bibitem[\protect\citeauthoryear{Helsdon \& Ponman}{Helsdon \&
  Ponman}{2003a}]{helsdon03a}
Helsdon S.~F.,  Ponman T.~J.,  2003a, MNRAS, 339, L29

\bibitem[\protect\citeauthoryear{Helsdon \& Ponman}{Helsdon \&
  Ponman}{2003b}]{helsdon03b}
Helsdon S.~F.,  Ponman T.~J.,  2003b, MNRAS, 340, 485

\bibitem[\protect\citeauthoryear{Helsdon \& Ponman}{Helsdon \&
  Ponman}{2004}]{helsdon04b}
Helsdon S.~F.,  Ponman T.~J.,  2004, in prep.

\bibitem[\protect\citeauthoryear{Helsdon, Ponman \& Mulchaey}{Helsdon
  et~al.}{2004}]{helsdon04}
Helsdon S.~F.,  Ponman T.~J.,    Mulchaey J.~S.,  2004, submitted to ApJ

\bibitem[\protect\citeauthoryear{Hickson}{Hickson}{1982}]{hickson82}
Hickson P.,  1982, ApJ, 255, 382

\bibitem[\protect\citeauthoryear{Hickson}{Hickson}{1997}]{hickson97}
Hickson P.,  1997, ARA\&A, 35, 357

\bibitem[\protect\citeauthoryear{Hoekstra, Franx, Kuijken, Carlberg, Yee, Lin,
  Morris, Hall, Patton, Sawicki \& Wirth}{Hoekstra et~al.}{2001}]{hoekstra01}
Hoekstra H.,  Franx M.,  Kuijken K.,  Carlberg R.~G.,  Yee H. K.~C.,  Lin H.,
  Morris S.~L.,  Hall P.~B.,  Patton D.~R.,  Sawicki M.,    Wirth G.~D.,  2001,
  ApJ, 548, L5

\bibitem[\protect\citeauthoryear{Horner}{Horner}{2001}]{horner01}
Horner D.~J.,  2001, PhD Thesis, University of Maryland

\bibitem[\protect\citeauthoryear{Hradecky, Jones, Donnelly, Djorgovski, Gal \&
  Odewahn}{Hradecky et~al.}{2000}]{hradecky00}
Hradecky V.,  Jones C.,  Donnelly R.~H.,  Djorgovski S.~G.,  Gal R.~R.,
  Odewahn S.~C.,  2000, ApJ, 543, 521

\bibitem[\protect\citeauthoryear{Huchra \& Geller}{Huchra \&
  Geller}{1982}]{huchra82}
Huchra J., Geller M.,  1982, ApJ, 257, 423

\bibitem[\protect\citeauthoryear{Hwang, Mushotzky, Burns, Fukazawa \&
  White}{Hwang et~al.}{1999}]{hwang99}
Hwang U.,  Mushotzky R.~F.,  Burns J.~O.,  Fukazawa Y.,    White R.~A.,  1999,
  ApJ, 516, 604

\bibitem[\protect\citeauthoryear{Jones, Ponman, Horton, Babul, Ebeling \&
  Burke}{Jones et~al.}{2003}]{jones03}
Jones L.~R.,  Ponman T.~J.,  Horton A.,  Babul A.,  Ebeling H.,    Burke D.~J.,
   2003, MNRAS, 343, 627

\bibitem[\protect\citeauthoryear{Kauffmann \& Charlot}{Kauffmann \&
  Charlot}{1998}]{kauffmann98}
Kauffmann G.,  Charlot S.,  1998, MNRAS, 294, 705

\bibitem[\protect\citeauthoryear{Khosroshahi, Jones \& Ponman}{Khosroshahi
  et~al.}{2004}]{khosroshahi04b}
Khosroshahi H.~G.,  Jones L.~R.,    Ponman T.~J.,  2004, submitted to MNRAS

\bibitem[\protect\citeauthoryear{Khosroshahi, Raychaudhury, Ponman, Miles \&
  Forbes}{Khosroshahi et~al.}{2004}]{khosroshahi04a}
Khosroshahi H.~G.,  Raychaudhury S.,  Ponman T.~J.,  Miles T.~M.,    Forbes
  D.~A.,  2004, submitted to MNRAS

\bibitem[\protect\citeauthoryear{Ledlow, Loken, Burns, Hill \& White}{Ledlow
  et~al.}{1996}]{ledlow96}
Ledlow M.~J.,  Loken C.,  Burns J.~O.,  Hill J.~M.,    White R.~A.,  1996, AJ,
  112, 388

\bibitem[\protect\citeauthoryear{Lin, Mohr \& Stanford}{Lin
  et~al.}{2003}]{lin03}
Lin Y.,  Mohr J.~J.,    Stanford S.~A.,  2003, ApJ, 591, 749

\bibitem[\protect\citeauthoryear{Lloyd-Davies, Ponman \& Cannon}{Lloyd-Davies
  et~al.}{2000}]{lloyddavies00}
Lloyd-Davies E.~J.,  Ponman T.~J.,    Cannon D.~B.,  2000, MNRAS, 315, 689

\bibitem[\protect\citeauthoryear{Mahdavi}{Mahdavi}{2001}]{mahdavi01}
Mahdavi A.,  2001, ApJ, 546, 812

\bibitem[\protect\citeauthoryear{Mahdavi, Boehringer, Geller \&
  Ramella}{Mahdavi et~al.}{1997}]{mahdavi97}
Mahdavi A.,  Boehringer H.,  Geller M.~J.,    Ramella M.,  1997, ApJ, 483, 68

\bibitem[\protect\citeauthoryear{Mahdavi, B{\" o}hringer, Geller \&
  Ramella}{Mahdavi et~al.}{2000}]{mahdavi00}
Mahdavi A.,  B{\" o}hringer H.,  Geller M.~J.,    Ramella M.,  2000, ApJ, 534,
  114

\bibitem[\protect\citeauthoryear{Makino \& Hut}{Makino \& Hut}{1997}]{makino97}
Makino J.,  Hut P.,  1997, ApJ, 481, 83

\bibitem[\protect\citeauthoryear{Mamon}{Mamon}{1994}]{mamon94}
Mamon G.~A.,  1994, in Durret F.,  Mazure A.,   Tran Thanh~Van J.,  eds,
  Clusters of Galaxies Gif-sur-Yvette, Frontieres, p.~291

\bibitem[\protect\citeauthoryear{Markevitch}{Markevitch}{1998}]{markevitch98}
Markevitch M.,  1998, ApJ, 504, 27

\bibitem[\protect\citeauthoryear{Matteucci \& Tornambe}{Matteucci \&
  Tornambe}{1987}]{matteucci87}
Matteucci F.,  Tornambe A.,  1987, A\&A, 185, 51

\bibitem[\protect\citeauthoryear{Mewe, Lemen \& van~den Oord}{Mewe
  et~al.}{1986}]{mewe86}
Mewe R.,  Lemen J.~R.,    van~den Oord G. H.~J.,  1986, A\&AS, 65, 511

\bibitem[\protect\citeauthoryear{Miles, Raychaudhury, Forbes \&
  Goudfrooij}{Miles et~al.}{2004}]{miles04}
Miles T.~A.,  Raychaudhury S.~R.,  Forbes D.~A.,    Goudfrooij P.,  2004,
  submitted to MNRAS

\bibitem[\protect\citeauthoryear{Mohr, Mathiesen \& Evrard}{Mohr
  et~al.}{1999}]{mohr99}
Mohr J.~J.,  Mathiesen B.,    Evrard A.~E.,  1999, ApJ, 517, 627

\bibitem[\protect\citeauthoryear{Muanwong, Thomas, Kay \& Pearce}{Muanwong
  et~al.}{2002}]{muanwong02}
Muanwong O.,  Thomas P.~A.,  Kay S.~T.,    Pearce F.~R.,  2002, MNRAS, 336, 527

\bibitem[\protect\citeauthoryear{Mulchaey}{Mulchaey}{2000}]{mulchaey00}
Mulchaey J.~S.,  2000, ARA\&A, 38, 289

\bibitem[\protect\citeauthoryear{Mulchaey, Davis, Mushotzky \&
  Burstein}{Mulchaey et~al.}{1996}]{mulchaey96}
Mulchaey J.~S.,  Davis D.~S.,  Mushotzky R.~F.,    Burstein D.,  1996, ApJ,
  456, 80

\bibitem[\protect\citeauthoryear{Mulchaey, Davis, Mushotzky \&
  Burstein}{Mulchaey et~al.}{2003}]{mulchaey03}
Mulchaey J.~S.,  Davis D.~S.,  Mushotzky R.~F.,    Burstein D.,  2003, ApJS,
  145, 39

\bibitem[\protect\citeauthoryear{Mulchaey \& Zabludoff}{Mulchaey \&
  Zabludoff}{1998}]{mulchaey98}
Mulchaey J.~S.,  Zabludoff A.~I.,  1998, ApJ, 496, 73

\bibitem[\protect\citeauthoryear{Mushotzky}{Mushotzky}{2003}]{mushotsky03}
Mushotzky R.,  2003, in IAU Symposium X-ray emission from clusters of galaxies

\bibitem[\protect\citeauthoryear{Navarro, Frenk \& White}{Navarro
  et~al.}{1995}]{navarro95}
Navarro J.~F.,  Frenk C.~S.,    White S. D.~M.,  1995, MNRAS, 275, 720

\bibitem[\protect\citeauthoryear{Nevalainen, Markevitch \& Forman}{Nevalainen
  et~al.}{2000}]{nevalainen00a}
Nevalainen J.,  Markevitch M.,    Forman W.,  2000, ApJ, 532, 694

\bibitem[\protect\citeauthoryear{Nolthenius}{Nolthenius}{1993}]{nolthenius93}
Nolthenius R.,  1993, ApJS, 85, 1

\bibitem[\protect\citeauthoryear{O'Sullivan, Ponman \& Collins}{O'Sullivan
  et~al.}{2003}]{osullivan03}
O'Sullivan E.~J., Ponman T. J., Collins R. S.,  2003, MNRAS, 340, 1375

\bibitem[\protect\citeauthoryear{O'Sullivan \& Ponman}{O'Sullivan \&
  Ponman}{2004}]{osullivan04}
O'Sullivan E.~J., Ponman T. J., 2004, submitted to MNRAS

\bibitem[\protect\citeauthoryear{Pietsch, Trinchieri, Arp \& Sulentic}{Pietsch
  et~al.}{1997}]{pietsch97}
Pietsch W.,  Trinchieri G.,  Arp H.,    Sulentic J.~W.,  1997, A\&A, 322, 89

\bibitem[\protect\citeauthoryear{Pildis, Bregman \& Evrard}{Pildis
  et~al.}{1995}]{pildis95}
Pildis R.~A.,  Bregman J.~N.,    Evrard A.~E.,  1995, ApJ, 443, 514

\bibitem[\protect\citeauthoryear{Ponman, Allan, Jones, Merrifield, McHardy,
  Lehto \& Luppino}{Ponman et~al.}{1994}]{ponman94}
Ponman T.~J.,  Allan D.~J.,  Jones L.~R.,  Merrifield M.,  McHardy I.~M.,
  Lehto H.~J.,    Luppino G.~A.,  1994, Nature, 369, 462

\bibitem[\protect\citeauthoryear{Ponman, Bourner, Ebeling \&
  B{\"{o}}hringer}{Ponman et~al.}{1996}]{ponman96}
Ponman T.~J.,  Bourner P. D.~J.,  Ebeling H.,    B{\"{o}}hringer H.,  1996,
  MNRAS, 283, 690

\bibitem[\protect\citeauthoryear{Ponman, Cannon \& Navarro}{Ponman
  et~al.}{1999}]{ponman99}
Ponman T.~J.,  Cannon D.~B.,    Navarro J.~F.,  1999, Nature, 397, 135

\bibitem[\protect\citeauthoryear{Ponman, Sanderson \& Finoguenov}{Ponman
  et~al.}{2003}]{ponman03}
Ponman T.~J.,  Sanderson A. J.~R.,    Finoguenov A.,  2003, MNRAS, 343, 331

\bibitem[\protect\citeauthoryear{Pratt \& Arnaud}{Pratt \&
  Arnaud}{2003}]{pratt03}
Pratt G.~W.,  Arnaud M.,  2003, A\&A, 408, 1

\bibitem[\protect\citeauthoryear{Ramella, Pisani \& Geller}{Ramella
  et~al.}{1997}]{ramella97}
Ramella M.,  Pisani A.,    Geller M.~J.,  1997, AJ, 113, 483

\bibitem[\protect\citeauthoryear{Rasmussen \& Ponman}{Rasmussen \&
  Ponman}{2004}]{rasmussen04}
Rasmussen J.,  Ponman T.~J.,  2004, submitted to MNRAS

\bibitem[\protect\citeauthoryear{Sanderson, Ponman, Finoguenov, Lloyd-Davies \&
  Markevitch}{Sanderson et~al.}{2003}]{sanderson03a}
Sanderson A. J.~R.,  Ponman T.~J.,  Finoguenov A.,  Lloyd-Davies E.~J.,
  Markevitch M.,  2003, MNRAS, 340, 989

\bibitem[\protect\citeauthoryear{Sanderson \& Ponman}{Sanderson \&
  Ponman}{2003}]{sanderson03b}
Sanderson A. J.~R.,  Ponman T.~J.,  2003, MNRAS, 345, 1241

\bibitem[\protect\citeauthoryear{Sato, Akimoto, Furuzawa, Tawara, Watanabe \&
  Kumai}{Sato et~al.}{2000}]{sato00}
Sato S.,  Akimoto F.,  Furuzawa A.,  Tawara Y.,  Watanabe M.,    Kumai Y.,
  2000, ApJ, 537, L73

\bibitem[\protect\citeauthoryear{Sulentic, Rosado, Dultzin-Hacyan,
  Verdes-Montenegro, Trinchieri, Xu \& Pietsch}{Sulentic
  et~al.}{2001}]{sulentic01}
Sulentic J.~W.,  Rosado M.,  Dultzin-Hacyan D.,  Verdes-Montenegro L.,
  Trinchieri G.,  Xu C.,    Pietsch W.,  2001, AJ, 122, 2993

\bibitem[\protect\citeauthoryear{Sun, Forman, Vikhlinin, Hornstrup, Jones \&
  S.}{Sun et~al.}{2003}]{sun03}
Sun M.,  Forman W.,  Vikhlinin A.,  Hornstrup A.,  Jones C.,    S. M.~S.,
  2003, ApJ, 598

\bibitem[\protect\citeauthoryear{Tully}{Tully}{1987}]{tully87}
Tully R.~B.,  1987, ApJ, 321, 280

\bibitem[\protect\citeauthoryear{Verde, Heavens, Percival, Matarrese, Baugh,
  Bland-Hawthorn, Bridges, Cannon, Cole, Colless, Collins, Couch, Dalton,
  De~Propris, Driver, Efstathiou, Ellis, Frenk, Glazebrook, Jackson, Lahav \&
  Lewis}{Verde et~al.}{2002}]{verde02}
  Verde L.,  Heavens A.~F.,  Percival W.~J.,  Matarrese S.,  Baugh C.~M.,
  Bland-Hawthorn J.,  Bridges T.,  Cannon R.,  Cole S.,  Colless M.,  Collins
  C.,  Couch W.,  Dalton G.,  De~Propris R.,  Driver S.~P.,  Efstathiou G.,
  Ellis R.~S.,  Frenk C.~S.,  Glazebrook K.,  Jackson C.,  Lahav O.,    Lewis
  I.,  Lumsden S.,  Maddox S.,  Madgwick D.,  Norberg P.,  Peacock J.~A.,
  Peterson B.~A.,  Sutherland W.,  Taylor K.,  2002, MNRAS, 335, 432

\bibitem[\protect\citeauthoryear{Vikhlinin, Forman \& Jones}{Vikhlinin
  et~al.}{1999}]{vikhlinin99c}
Vikhlinin A.,  Forman W.,    Jones C.,  1999, ApJ, 525, 47

\bibitem[\protect\citeauthoryear{White, Jones \& Forman}{White
  et~al.}{1997}]{white97b}
White D.~A.,  Jones C.,    Forman W.,  1997, MNRAS, 292, 419

\bibitem[\protect\citeauthoryear{White, Bliton, Bhavsar, Bornmann, Burns,
  Ledlow \& Loken}{White et~al.}{1999}]{white99}
White R.~A.,  Bliton M.,  Bhavsar S.~P.,  Bornmann P.,  Burns J.~O.,  Ledlow
  M.~J.,    Loken C.,  1999, ApJ, 118, 2014

\bibitem[\protect\citeauthoryear{Wu \& Xue}{Wu \& Xue}{2002}]{wu02}
Wu X.,  Xue Y.,  2002, ApJ, 572, L19

\bibitem[\protect\citeauthoryear{Wu, Xue \& Fang}{Wu et~al.}{1999}]{wu99}
Wu X.,  Xue Y.,    Fang L.,  1999, ApJ, 524, 22

\bibitem[\protect\citeauthoryear{Xue \& Wu}{Xue \& Wu}{2000}]{xue00}
Xue Y.,  Wu X.,  2000, ApJ, 538, 65

\bibitem[\protect\citeauthoryear{Zabludoff \& Mulchaey}{Zabludoff \&
  Mulchaey}{1998}]{zabludoff98}
Zabludoff A.~I.,  Mulchaey J.~S.,  1998, ApJ, 496, 39

\bibitem[\protect\citeauthoryear{Zabludoff \& Mulchaey}{Zabludoff \&
  Mulchaey}{2000}]{zabludoff00}
Zabludoff A.~I.,  Mulchaey J.~S.,  2000, ApJ, 539, 136

\end{thebibliography}
\end{document}